\documentclass[11pt,a4paper]{article}
\pdfoutput=1

\usepackage{epsfig,amsfonts,amssymb}
\usepackage[colorlinks=0]{hyperref}

\usepackage{graphicx}
\usepackage{mathrsfs}
\usepackage{comment}
\usepackage{cite}
\usepackage{cite}
\usepackage{graphicx}
\usepackage{float}
\usepackage[normalem]{ulem}
\usepackage[utf8]{inputenc}
\usepackage[T1]{fontenc}
\usepackage{lmodern, textcomp}
\usepackage[english]{babel}
\usepackage{csquotes,cancel}

\usepackage[left=1in,right=1in,top=.8in,bottom=1in]{geometry}

\usepackage{eurosym, xspace}
\usepackage{graphicx, subfig, float}
\usepackage{verbatim}

 \usepackage{upgreek}
\usepackage{amsmath, amsfonts, amssymb,upgreek}
\usepackage{cancel}
\usepackage{mathtools}
\usepackage{slashed}

\usepackage[usenames, dvipsnames]{xcolor}
\usepackage{framed}
\usepackage[amsmath, hyperref, framed]{}
\usepackage{hhline}
\usepackage{bm}

\usepackage{array, caption}
\usepackage{array,multirow}

\newcolumntype{C}[1]{>{\centering\arraybackslash}m{#1}}

\usepackage{tikz-feynman}
\usepackage{tikz, pgf}
\usetikzlibrary{shapes.misc}
\usetikzlibrary{snakes}
\usetikzlibrary{decorations.pathmorphing}
\usetikzlibrary{arrows.meta}
\usetikzlibrary{shapes}
\usetikzlibrary{fit}

\makeatletter
\newsavebox{\@brx}
\newcommand{\llangle}[1][]{\savebox{\@brx}{\(\m@th{#1\langle}\)}%
  \mathopen{\copy\@brx\kern-0.5\wd\@brx\usebox{\@brx}}}
\newcommand{\rrangle}[1][]{\savebox{\@brx}{\(\m@th{#1\rangle}\)}%
  \mathclose{\copy\@brx\kern-0.5\wd\@brx\usebox{\@brx}}}
\makeatother

\def\ZZZ{{\hbox{ Z\kern-1.6mm Z}}}
\def\RRR{{\hbox{ R\kern-2.4mm R}}}
\def\CCC{{\hbox{ C\kern-2.0mm C}}}
\def\zzz{{\hbox{z\kern-1mm z}}}

\newcommand{\qeq}{{\hbox{=\kern-2.3mm ? \kern.5mm }}}
\renewcommand{\qeq}{=}

\newcommand{\be}{\begin{eqnarray}}
\newcommand{\ee}{\end{eqnarray}}

\newcommand{\vp}{\varphi}

\newcommand{\ben}{\begin{eqnarray}\displaystyle}
\newcommand{\een}{\end{eqnarray}}

\newcommand{\p}{\partial}

\def\one{{\hbox{ 1\kern-.8mm l}}}
\def\zero{{\hbox{ 0\kern-1.5mm 0}}}

\newcommand{\bea}[1]{\begin{eqnarray}\label{#1} }
\newcommand{\eea}{\end{eqnarray}}

\usepackage{bm}

\newcommand\non{\nonumber}

\newcommand\f{\frac}

\def\figone{

\def\JPicScale{0.8}
\ifx\JPicScale\undefined\def\JPicScale{1}\fi
\unitlength \JPicScale mm
\begin{picture}(135,80)(0,0)
\linethickness{0.3mm}
\multiput(40,80)(0.12,-0.18){167}{\line(0,-1){0.18}}
\linethickness{0.3mm}
\multiput(30,70)(0.18,-0.12){167}{\line(1,0){0.18}}
\linethickness{0.3mm}
\put(30,50){\line(1,0){30}}
\linethickness{0.3mm}
\multiput(30,30)(0.18,0.12){167}{\line(1,0){0.18}}
\linethickness{0.3mm}
\multiput(40,20)(0.12,0.18){167}{\line(0,1){0.18}}
\linethickness{0.3mm}
\put(60,50){\line(1,0){40}}
\linethickness{0.3mm}
\multiput(100,50)(0.12,0.18){167}{\line(0,1){0.18}}
\linethickness{0.3mm}
\multiput(100,50)(0.18,0.12){167}{\line(1,0){0.18}}
\linethickness{0.3mm}
\put(100,50){\line(1,0){30}}
\linethickness{0.3mm}
\multiput(100,50)(0.18,-0.12){167}{\line(1,0){0.18}}
\linethickness{0.3mm}
\multiput(100,50)(0.12,-0.18){167}{\line(0,-1){0.18}}

\put(30,80){\makebox(0,0)[cc]{$\zeta Q_B A_1^c$}}

\put(25,70){\makebox(0,0)[cc]{$A_2^c$}}

\put(25,50){\makebox(0,0)[cc]{$A_n^c$}}

\put(20,30){\makebox(0,0)[cc]{$A_1^o$}}

\put(35,20){\makebox(0,0)[cc]{$A_p^o$}}

\put(35,60){\makebox(0,0)[cc]{$\vdots$}}

\put(40,30){\makebox(0,0)[cc]{$\vdots$}}

\put(125,80){\makebox(0,0)[cc]{$B_1^c$}}

\put(135,70){\makebox(0,0)[cc]{$B_2^c$}}

\put(135,50){\makebox(0,0)[cc]{$B_m^c$}}

\put(135,30){\makebox(0,0)[cc]{$B_1^o$}}

\put(120,15){\makebox(0,0)[cc]{$B_q^o$}}

\put(115,35){\makebox(0,0)[cc]{$\vdots$}}

\put(120,57){\makebox(0,0)[cc]{$\vdots$}}

\put(66,47){\makebox(0,0)[cc]{$\psi_r^c\vp_r$}}

\put(93,47){\makebox(0,0)[cc]{$\psi_s^c\vp_s$}}

\end{picture}

}

\def\figtwo{

\def\JPicScale{0.8}
\ifx\JPicScale\undefined\def\JPicScale{1}\fi
\unitlength \JPicScale mm
\begin{picture}(135,80)(0,0)
\linethickness{0.3mm}
\multiput(40,80)(0.12,-0.18){167}{\line(0,-1){0.18}}
\linethickness{0.3mm}
\multiput(30,70)(0.18,-0.12){167}{\line(1,0){0.18}}
\linethickness{0.3mm}
\put(30,50){\line(1,0){30}}
\linethickness{0.3mm}
\multiput(30,30)(0.18,0.12){167}{\line(1,0){0.18}}
\linethickness{0.3mm}
\multiput(40,20)(0.12,0.18){167}{\line(0,1){0.18}}
\linethickness{0.3mm}
\put(60,50){\line(1,0){40}}

\put(30,80){\makebox(0,0)[cc]{$\zeta Q_B A_1^c$}}

\put(25,70){\makebox(0,0)[cc]{$A_2^c$}}

\put(25,50){\makebox(0,0)[cc]{$B_m^c$}}

\put(20,30){\makebox(0,0)[cc]{$A_1^o$}}

\put(35,20){\makebox(0,0)[cc]{$B_q^o$}}

\put(35,60){\makebox(0,0)[cc]{$\vdots$}}

\put(40,30){\makebox(0,0)[cc]{$\vdots$}}

\put(66,47){\makebox(0,0)[cc]{$\psi_r^c\vp_r$}}

\put(93,45){\makebox(0,0)[cc]{$\tilde\psi_s^c\tilde\vp^s$}}

\put(100,50){\makebox(0,0)[cc]{$\times$}}

\end{picture}

}

\definecolor{armygreen}{rgb}{0.29, 0.33, 0.13}

\begin{document}

\baselineskip 24pt

\begin{center}
{\Large \bf  Carrollian Conformal Theories in Momentum Space}

\end{center}

\vskip .6cm
\medskip

\vspace*{4.0ex}

\baselineskip=18pt

\begin{center}

{\large 
\rm  Raffaele Marotta$^a$, Arvind Shekar$^b$ and Mritunjay Verma$^{a,c}$ }

\end{center}

\vspace*{4.0ex}

\centerline{\it \small $^a$Istituto Nazionale di Fisica Nucleare (INFN), Sezione di Napoli, }
\centerline{ \it \small Complesso Universitario di Monte S. Angelo ed. 6, via Cintia, 80126, Napoli, Italy}
\centerline{ \it \small $^b$ Mathematical Sciences and STAG Research Centre, University of Southampton,}
\centerline{\it \small  Highfield, Southampton SO17 1BJ, UK}

\centerline{\it \small $^c$ Indian Institute of Technology Indore, Khandwa Road, Simrol, Indore 453552, India}
\vspace*{1.0ex}
\centerline{\small E-mail:  raffaele.marotta@na.infn.it,  A.Shekar@soton.ac.uk, 
mritunjay@iiti.ac.in }

\vspace*{5.0ex}

\centerline{\bf Abstract} \bigskip

We consider the Carrollian conformal field theories involving scalar operators in the momentum representation. The momentum space Ward identities are explicitly solved to obtain the different branches of 2 and 3 point Carrollian conformal correlators in the momentum space. These branches are symmetry-allowed structures, characterized by different analytic structures in the Carrollian energies, and are not necessarily realized simultaneously in a single consistent theory. 
For specific values of the conformal dimensions, the three-point functions in momentum space exhibit logarithmic behaviour.
This has no analogue in position space and instead originates from singularities in the Fourier transform relating position and momentum space correlators.  
We also analyze the Carrollian limit of CFT 2 and 3 point functions of scalar operators in momentum space. By taking different scalings of CFT correlators with respect to the speed of light, we obtain different branches of the Carrollian conformal correlators in the momentum space. 


\vfill

\vfill \eject

\baselineskip18pt

\tableofcontents

\section{Introduction } 

Holography is expected to be an integral feature of any theory of quantum gravity \cite{9310026,9409089}. It was mainly introduced to explain the Bekenstein–Hawking entropy–area law\cite{Bekenstein:1972tm, Bekenstein:1973ur,Hawking:1975vcx}, and it asserts that all physical information in a gravitational theory can be represented by a lower–dimensional, non-gravitational theory defined on its boundary. However, incorporating this feature into the framework of quantum gravity has proven to be a challenging task. The current efforts try to understand how holography works in certain fixed classes of spacetimes. E.g., using the insights from string theory, the holographic principle has been concretely realized in the asymptotically Anti de Sitter space times via the AdS/CFT correspondence \cite{9711200,Gubser:1998bc,9802150}. This correspondence has proven to be a powerful tool for exploring various aspects of quantum gravity.  

 \vspace*{.07in}Despite the above remarkable achievement, it remains to extend the holographic principle to more realistic spacetimes characterized by a vanishing or positive cosmological constant, namely the asymptotically flat or de Sitter spacetimes (for reviews, see \cite{Raclariu:2021zjz,Pasterski:2021rjz,Strominger:2001gp} and references therein). In fact, flat-space holography is expected to be especially rich \cite{Sen:2025oeq}. In AdS/CFT, each bulk background (specified by choices of moduli, fluxes, and compactification data) corresponds to a particular boundary theory (possibly a family related by exactly marginal couplings). Flat space behaves differently: an asymptotically flat spacetime can contain very large, weakly curved interior regions where the physics   matches different choices of moduli parameters. E.g., it has been argued that starting from type IIA string theory on $R^{3,1}\times T^6$, one can engineer large interior regions inside the flat spacetimes which resemble many other familiar vacua, such as type IIB on AdS$_5\times S^5$ or AdS$_3\times S^3\times T^4$, M theory on AdS$_4\times S^7$ or AdS$_7\times S^4$, M theory on $R^{10,1}$ as well as type IIA and IIB on $R^{9,1}$. Hence, the boundary holographic description of type IIA string theory on $R^{3,1}\times T^6$ must contain information about all these possibilities \cite{Sen:2025oeq}. This makes flat-space holography both fascinating and challenging, necessitating a dedicated study.  

\vspace*{.07in}To make progress in our understanding of flat holography, it is fruitful to study this from multiple complementary directions. In particular, some proposals for the dual boundary descriptions have been put forward for the low-energy (supergravity) sector of the bulk. For the massless bulk fields, the expected natural “location” of the dual data is the null infinity ($\mathscr{I}^+\cup \mathscr{I}^-$) of the asymptotically flat spacetime. The asymptotic symmetry group of asymptotically flat spacetimes is an infinite dimensional extension of the Poincaré group, namely, Bondi-Metzner-Sachs (BMS) group \cite{bms1, bms2}. Building on this fact, there are two closely related approaches for analyzing the physics at the null boundary of asymptotically flat spacetimes. 

\vspace*{.07in}In the first approach, known as Celestial (CFT) approach, one makes connection between the bulk scattering amplitudes and the correlators on the celestial sphere at infinity\cite{Pasterski:2017ylz, Donnay:2020guq,Pasterski:2021rjz,Donnay:2021wrk,Casali:2022fro,Mizera:2022sln,Donnay:2022wvx,Pasterski:2023ikd,Bissi:2024brf,Sleight:2023ojm}. For a $(d+1)$-dimensional bulk, the celestial sphere has dimension $(d-1)$ and hence the boundary theory is viewed as a $(d-1)$-dimensional conformal theory living on that sphere \cite{Pasterski:2021rjz}. On the other hand, in the second approach, known as Carrollian (CFT) approach \cite{Bagchi:2010zz,Bagchi:2012cy,Duval:2014uva,Bagchi:2016bcd,Donnay:2019jiz,Donnay:2022aba,Donnay:2022wvx,Bagchi:2022emh,Bagchi:2022eav,Bagchi:2022owq,Bagchi:2023fbj,Bagchi:2023cen,Saha:2023hsl,Salzer:2023jqv,2305.02884,Mason:2023mti,2406.19343,Afshar:2024llh,Nguyen:2025sqk,bagchi1,Bagchi:2025jgu,2508.06602,chen1,chen2,Nguyen:2023miw,Nguyen:2025zhg,Banerjee:2023jpi,Ara:2024fbr,deBoer:2021jej,deBoer:2023fnj}, one relates the $(d+1)$-dimensional bulk dynamics to a $d-$dimensional Carrollian conformal field theory living  on the $d$ dimensional null infinity $\mathscr{I}^+\cup \mathscr{I}^-$ \cite{ bagchi1}. Both celestial and Carrollian approaches are complementary ways to analyze the
theory at null infinity in asymptotically flat space times and have been shown to be equivalent \cite{Mason:2023mti}. 

\vspace*{.07in}In this work, we focus on the Carrollian approach. Historically, the Carrollian algebra was introduced as the vanishing speed of light $c\to0$ limit of the Poincar\'e algebra \cite{LBLL,SenGupta:1966qer}. In this limit, the Lorentzian metric becomes degenerate, and the theory becomes ultra-local. Carroll structures also arise naturally on null hypersurfaces, on which the induced metric is degenerate. This is also true for null infinities in particular. A different symmetry group implies that the properties of Carrollian theories are quite different from the standard QFTs we are used to. In particular, the standard analytic properties of Poincaré invariant QFTs do not directly apply to these theories. On the other hand, the bulk theories have well defined analytic properties. Hence, for the duality to work, we need to know how the bulk analytic properties translate to boundary analytic properties.

\vspace*{.07in}One way to approach this problem is to first note that the Carrollian conformal theories have mostly been studied in the position representation. In this representation, the analytic properties are not manifest as is also the case for usual Poincare invariant QFTs for which the analytic properties are not transparent in the position space. However, it is well known that the momentum representation makes the analytic properties of traditional QFTs quite manifest. E.g., the physical properties such as locality, causality and unitarity can be translated into some analytic behaviour of the momentum space amplitude. The locality requires some kind of polynomial boundedness of amplitudes for theories involving finite number of derivatives. The causality is ensured by behavior under the analytic continuation of the amplitudes. The unitarity shows up in various dispersion relations and cutting rules of the amplitudes.

\vspace*{.07in}This hints that for the Carrollian conformal field theories (CCFT) on the null boundaries, we can hope to get some insights about the analytic properties by analyzing them in the momentum space. The branch cuts and the poles in the momentum space correlators may tell us about these properties. E.g., the analytic structure in energy space along the null direction can tell us about the causality in the boundary Carrollian theory. Momentum space correlators carry information about energy flow direction. If under reversal of energy flow directions, some specific Carrollian correlators vanish or develop singularities, it may signal a breakdown of Carrollian causality. 

\vspace*{.07in}Another motivation for studying the Carrollian theories in momentum space comes from the fact that asymptotic symmetries of flat spacetimes have been shown to be closely related to soft theorems, memory effects and flat space S matrices (see, e.g.\cite{Strominger:2017zoo,Donnay:2023mrd}). The relations between these are most naturally expressed in terms of momenta\cite{Low:1958sn,Weinberg:1964ew,Weinberg:1965nx}. Hence, a momentum space formulation of the flat space boundary dynamics can provide the most direct bridge between these aspects of the bulk gravitational physics.  

\vspace*{.07in} With these motivations in mind, we initiate a study of Carrollian conformal field theories in momentum space. Carrollian symmetry emerges in several physical contexts, including ultra-relativistic systems such as quark–gluon plasma and certain condensed matter systems, such as fractonic phases and flat-band models (see \cite{bagchi1} and references therein). However, in the context of holography for asymptotically flat spacetimes, these theories are primarily understood at the level of symmetry and kinematics, and a complete dynamical formulation is still lacking. Our goal is, therefore,  to determine the most general class of correlation functions compatible with Carrollian conformal symmetry by solving the corresponding Ward identities in momentum space.
The classification of Carrollian correlators based on symmetry arguments has been previously studied in the literature in position space (see e.g.\ Ref.~\cite{Nguyen:2025zhg} for a review)\footnote{As far as we are aware, complete constraints on the unfixed structure of the 3-point Carrollian correlator in the position space were not worked out in the previous studies. We find additional constraints compared to \cite{Afshar:2024llh} in the position space (see equation \eqref{311we} in section \ref{sec:3pointpos}).}, where examples of theories realizing different structures have been discussed \cite{deBoer:2023fnj,bagchi1}. Our analysis extends this program to momentum space. Such a classification provides a necessary foundation for any future dynamical formulation, as it identifies the space of symmetry-allowed structures that a consistent CCFT may realize.

\vspace*{.07in}Formally, momentum-space correlators can be obtained by Fourier transforming the corresponding position-space expressions. However, as already emphasised in the CFT literature, position-space correlators are typically written for separated points and, in general, do not admit a well-defined Fourier transform prior to renormalisation \cite{Coriano:2013jba,
Isono:2018rrb,Coriano:2018bsy,Albayrak:2018tam,Farrow:2018yni,Isono:2019ihz,Coriano:2019sth,Albayrak:2019yve,Isono:2019wex,Gillioz:2019lgs,Baumann:2019oyu,Coriano:2019nkw,Lipstein:2019mpu,Albayrak:2020isk,Bzowski:2020lip,Jain:2020rmw,Jain:2020puw,Jain:2021wyn,Jain:2021qcl,Jain:2021vrv,Gillioz:2021sce,Jain:2021whr,Jain:2022ujj,Bzowski:2022rlz,Jain:2022ajd,Caloro:2022zuy,Marotta:2022jrp}.  Our approach will be to first derive the Carrollian conformal Ward identities in the momentum space. The momentum space correlation functions are solutions of these Ward identities. We shall explicitly consider the solutions of these Ward identities for 2 and 3 point functions. Unlike the standard CFTs whose 2 and 3-point functions are completely fixed by the conformal symmetry and are characterized by single terms (for scalar operators), the similar correlation functions of the Carrollian CFTs have more than one term (even for scalar operators)\footnote{We shall refer to the different terms in the Carrollian conformal correlators as different branches. These branches are characterized by different structures of Carrollian energy conserving delta functions (see, e.g., equation \eqref{hpi} for 3-point case). For 2-point functions, the two branches are also known as electric and magnetic branches.}. Further, the 3-point correlation functions are not completely fixed by the symmetry alone and additional constraints, such as operator-product expansions and consistency conditions at the level of four-point functions (including factorization and crossing symmetry), may be required. This is another feature of the Carrollian conformal theories. However, this is not unexpected. The amplitudes in the bulk theories are also not completely fixed and if the Carrollian conformal theories have to be their boundary dual, then these Carrollian conformal theories should also have some degree of non uniqueness.
These features can be understood as the momentum-space manifestation of the electric and magnetic branches of CCFTs. In particular, for two-point functions, the magnetic branch appears to be ultra-local in the
energy associated with the null direction while the electric branch exhibits a trivial dependence on the transverse $(d-1)$-dimensional momentum.
These symmetry-allowed structures are not necessarily expected to be simultaneously realized within a single theory, and may be interpreted as capturing distinct infrared and ultraviolet behaviours \cite{Cotler:2024xhb,deBoer:2023fnj}. Indeed, Carrollian theories are often classified into electric, magnetic and mixed sectors, where the mixed sector describes interacting electric and magnetic degrees of freedom, depending on how these structures are realized\cite{Duval:2014uoa,deBoer:2023fnj}.

In the present work, we do not attempt to construct explicit dynamical realizations of the Carrollian correlators allowed by the Ward identities. Instead, our aim is to provide a systematic characterization of these structures in momentum space, as a first step toward a formulation of flat-space holography in this framework.
From this perspective, it is natural to investigate the relation between Carrollian correlators and the flat, or equivalently ultra-relativistic, limit of AdS/CFT. A further step toward understanding flat holography in this setup would be to show how the different branches of Carrollian correlation functions may emerge from the asymptotic behaviour of a single bulk theory near null infinity \cite{Donnay:2023mrd,Bekaert:2024tkv, bagchi1,Nguyen:2025zhg}. Along these lines, it has been proposed that different causal structures of bulk correlators (such as Wightman or Feynman correlators) may be related to the functional freedom observed in Carrollian correlators \cite{Nguyen:2023miw}.
We leave a detailed analysis of these issues for future work.

\vspace*{.07in}The momentum space Ward identities also have logarithmic solutions for specific values of the Carrollian conformal dimensions. These solutions do not have any direct Fourier analog in the position space. Instead, it turns out that the Fourier transforms of the position space solutions are singular for specific values of the Carrollian conformal dimensions and requires regularizations. 
The logarithmic terms can arise due to the regularization process. 
This is analogous to the standard CFTs in momentum space where for special values of the conformal dimensions, the Fourier transform  may be singular and a regularization and renormalization is required \cite{Bzowski:2013sza}. The regularization gives rise to such logarithmic terms and the regularized correlators satisfy the anomalous Ward identities \cite{1510.08442, Bzowski:2017poo,Bzowski:2018fql}. 
However, what is different in Carrollian conformal theories is that the logarithmic terms obtained in momentum space satisfy the original Ward identities rather than anomalous ones. At present, we do not have a complete physical interpretation of these logarithmic solutions. Nevertheless, they are exact solutions of the momentum-space Ward identities and may arise from singularities in the Fourier transform relating position- and momentum-space correlators. Their deeper physical significance remains unclear and is left for future investigation.

 \vspace*{.07in}We shall also study the boundary Carrollian conformal theories by taking a suitable limit of Lorentzian CFT correlators. It turns out that the Carrollian approach to flat holography can emerge if we consider the flat and ultra-relativistic limits of the bulk AdS and its boundary respectively \cite{2406.19343, 2508.06602}. These two limits are closely related. In the limit where we take the AdS radius to be very large, the AdS resembles flat space. In this limit, the AdS correlators morph into the flat space S matrices \cite{9901076,Susskind:1998vk}. On the boundary side, the flat limit corresponds to a Carrollian limit where we send the speed of light to zero \cite{2406.19343}. In other words, the flat limit of the AdS geometry corresponds to an ultra-relativistic contraction of the boundary conformal algebra\cite{2406.19343}. From this perspective, Carrollian holography can be studied using the standard AdS/CFT duality. This approach has the advantage of inheriting well-understood structures from AdS/CFT. Thus, some aspects of both the bulk as well as boundary theories in the flat spacetimes can be studied by taking the flat/Carrollian limit of the AdS/CFT correspondence. In \cite{Marotta:2024sce,Farrow:2018yni,Lipstein:2019mpu,spin2}, the flat limit has been analyzed for the bulk dynamics directly using the momentum space techniques. Here, we shall consider the Carrollian limit of boundary CFT theory in the momentum space. 

\vspace*{.07in}{The approach to flat holography by using the flat limit of AdS has certain limitations. One way to see this using the boundary side is to note that Carrollian conformal theories on a null hypersurface can be obtained by taking a {I}n\"on\"u-Wigner type contraction of the standard conformal field theories where we send the speed of light $c$ to zero.} In this limit, the conformal algebra gives rise to Carrollian conformal algebra. In this Carrollian limit, the CFT correlators become Carrollian conformal correlators. The different branches of 2 and 3-point Carrollian conformal correlators are obtained by different $c\rightarrow 0$ scalings in this limit. This approach, which is tightly connected to the ultra-relativistic limit of CFT correlators, reproduces only a restricted set of Carrollian correlators on the null boundary of flat space. Relativistic CFT correlators obey the full conformal Ward identities, which are more constraining than their Carrollian counterparts, and the corresponding space of allowed parameters is therefore smaller. These restrictions persist in the $c \to 0$ limit, since the limit affects the space--time geometry but leaves the intrinsic CFT data unchanged. As a consequence, the Carrollian correlators obtained in this manner realise only a distinguished subclass of the general solutions permitted by Carrollian conformal symmetry. The limitations to this approach to flat holography can also be seen by considering the compactification of 10 dimensional theory to 4 dimensions \cite{Fontanella:2025tbs}.

\vspace*{.07in} In Lorentzian CFTs, the Wightman and time ordered correlators have different interpretations. However, in the Carrollian limit $c\rightarrow0$, the notion of time disappears. Hence, in this limit, both of these types of correlators are expected to give the same results (upto a normalization). We shall consider both the Wightman and time ordered correlators in the momentum space and consider their ultra-relativistic limit. We shall find that in this limit, they indeed give rise to same Carrollian conformal correlators which differ only by some overall constants.

 \vspace*{.07in}In the present work, we shall not discuss the connection with the bulk dynamics. This is postponed to a subsequent work \cite{bulk}. However, here we shall set up all the necessary boundary ingredients for discussing flat–space holography in momentum space.

\vspace*{.07in}The rest of the paper is organized as follows. In section \ref{sec2review3er}, we review Carrollian geometry and Carrollian conformal algebra. We also review how the Carrollian conformal algebra arises from the Inönü--Wigner contraction. In section \ref{position_rev}, we shall summarize 2 and 3 point Carrollian conformal correlators in the position space in arbitrary dimensions. We shall also clarify to what extent the Ward identities fix these correlators. In particular, we clarify what freedom is available in the 3-point correlators. In section \ref{sec4:Ward}, we derive the momentum space Carrollian conformal Ward identities and solve the resulting differential equations to obtain the 2 and 3 point correlators in the momentum space. In section \ref{sec5:inonu}, we consider the {I}n\"on\"u-Wigner contraction of the CFT 2 and 3 point functions and show that different branches of 2 and 3-point Carrollian conformal correlators can be obtained by different $c\rightarrow 0$ scalings. We end with discussions in section \ref{s4}. The appendix contains some reviews and details of some derivations. In appendix \ref{sec:appenA}, we note our conventions and some useful identities. In \ref{sec:BLor}, we review some aspects of the Lorentzian CFT correlators in the momentum space. In particular, equation \eqref{B.219df} gives a useful representation of fully time ordered 3-point function in momentum space as a convolution over two point functions. This representation will be useful for taking the Carrollian limit of time ordered 3-point function. In appendix \ref{appenc}, we review some useful delta distribution identities which are used in the analysis of momentum space Ward identities. In appendix \ref{append}, we compute the momentum space Carrollian conformal correlators directly using the Fourier transform of the position space correlators for generic values of the Carrollian conformal dimensions. Finally, in appendix \ref{appene}, we give an alternative evaluation of equation \eqref{5123ert} which is required for taking the Carrollian limit of two point function in momentum space. In the main text, it has been derived using the contour methods.

\section{Review of the Carrollian conformal group and representations}
\label{sec2review3er}
Carrollian geometry can be formulated in several equivalent ways. A geometrically transparent construction arises by considering the geometry induced on null planes in Minkowski spacetime $\mathbb{M}^{d+1}$, with coordinates $(X^0, X^i, X^d)$, where $i = 1, \dots, d{-}1$, and metric
\begin{equation}
ds^2 = \eta_{\mu\nu}\, dX^\mu dX^\nu 
= -\,du\,dv + \delta_{ij}\, dX^i dX^j,
\label{1.1a}
\qquad \mu = 0, \dots, d~.
\end{equation}
Here $\eta_{\mu\nu} = \mathrm{diag}(-1, 1, \dots, 1)$, and we have introduced the null coordinates
\begin{equation}
u = X^0 - X^d, 
\qquad 
v = X^0 + X^d~.
\end{equation}

The hypersurfaces defined by $v = \text{constant}$ form a family of null planes 
${\cal J} \simeq \mathbb{R}\times \mathbb{R}^{d-1}$, 
each parametrized by the coordinates $x^\alpha=(u, X^i)$, with $\alpha=0,1,\dots, d-1$, induced from the ambient coordinates $(X^0, X^i, X^d)$ onto the hypersurface. 
Their normal vector 
$\hat{n}_\mu = \partial_\mu v = (1, 0, \dots, 1)$ 
is itself null since $\eta^{\mu\nu} \hat{n}_\mu \hat{n}_\nu = 0$.

The metric on these hypersurfaces can be obtained by pulling back the $(d{+}1)$--dimensional Minkowski metric onto $v = \text{constant}$. 
Setting $dv = 0$ in equation \eqref{1.1a}, one finds
\begin{equation}
ds^2_{\cal{J}} \;=
\delta_{ij}\, dX^i dX^j
\;\equiv\;
q_{\alpha \beta}\, dx^\alpha dx^\beta,
\qquad 
\alpha,\beta = 0,1,\dots,d-1.
\end{equation}
The metric is degenerate being of the form $q_{\alpha\beta}=(0,\,\delta_{ij})$, and there exists a non-zero vector field $n^\beta$ satisfying
\begin{equation}
q_{\alpha\beta}\, n^\beta = 0~.
\end{equation}
This vector generates the kernel of $q_{\alpha\beta}$ and defines the null direction of the Carrollian structure. 
In the present coordinates, it is given by
\begin{equation}
n \equiv \partial_u = (1,\,0^i),
\end{equation}
that is, the tangent vector to the null plane.
The pair $(q_{\alpha\beta},\, n = \partial_u)$ thus defines the canonical flat Carrollian structure on the null hypersurface.

\vspace{0.4em}
This geometry precisely corresponds to that obtained by taking the ultra--relativistic limit $c \rightarrow 0$, 
which leads to a degenerate metric structure in which the temporal direction becomes null, 
while the spatial sections remain endowed with a non--degenerate Euclidean metric. 
Mathematically, this limit can be realized through the Inönü--Wigner contraction 
of the conformal algebra $\mathfrak{so}(2,d)$ in the ultra--relativistic limit $c \rightarrow 0$. 
To see this explicitly, let us start from the $d$--dimensional Minkowski metric
\begin{equation}
ds^2_{\mathbb{M}^d} = - (d X^0)^2 + \delta_{ij}\, dX^i dX^j,
\end{equation}
and restore the dependence on the speed of light. 
The  contraction can be implemented by rescaling the speed of light as \cite{1901.10147}
\begin{eqnarray}
c\rightarrow\epsilon c  \,~~{\rm with} ~~\epsilon\rightarrow 0~,
\label{2.7}
\end{eqnarray}
At the level of metric, we write $X^0=\epsilon\, u$ and then take the limit $\epsilon\rightarrow0$ to recover the degenerate null--plane metric
\begin{equation}
ds^2_{\mathbb{M}^d}
\;\longrightarrow\;
ds^2_{\cal J} = 0\,du^2 + \delta_{ij}\, dX^i dX^j~,
\end{equation}
which reproduces the Carrollian geometry induced on ${\cal J}$.
The algebra of conformal isometries of Minkowski space $\mathbb{M}^d$, $\mathfrak{so}(2,d)$, is given by \cite{DiFrancesco}
\begin{eqnarray}
\left[ {\cal D}, {\cal P}_\mu\right] &=& i\,{\cal P}_\mu,
\qquad 
\left[ {\cal D},\,{\cal K}_\mu\right] = -\,i\,{\cal K}_\mu,
\qquad
\left[{\cal K}_\mu, {\cal P}_\nu\right] = -\,2i\left( \eta_{\mu\nu} {\cal D}- {\cal L}_{\mu\nu}\right),
\nonumber\\[4pt]
\left[{\cal K}_\rho,\, {\cal L}_{\mu\nu}\right] &=& i\left( \eta_{\rho\mu} {\cal K}_\nu-\eta_{\rho\nu}{\cal K}_{\mu}\right),
\qquad
\left[{\cal P}_\rho, \,{\cal L}_{\mu\nu}\right] = i\left( \eta_{\rho\mu}\,{\cal P}_\nu-\eta_{\rho\nu}{\cal P}_\mu\right),
\nonumber\\[4pt]
\left[{\cal L}_{\mu\nu},\,{\cal L}_{\rho\sigma}\right] &=& i\left( 
\eta_{\nu\rho} {\cal L}_{\mu\sigma} + \eta_{\mu\sigma}{\cal L}_{\nu\rho} 
- \eta_{\mu\rho}{\cal L}_{\nu\sigma} - \eta_{\nu\sigma}{\cal L}_{\mu\rho}
\right).
\end{eqnarray}
The generators of translations, dilatations, Lorentz transformations, and special conformal transformations are represented in coordinate space, respectively, as
\begin{eqnarray}
{\cal P}_\mu &=& -\,i\,\partial_\mu,
\qquad
{\cal D} = -\,i\,X^\mu\partial_\mu,
\nonumber\\[4pt]
{\cal L}_{\mu\nu} &=& i\left( X_\mu\partial_\nu - X_\nu\partial_\mu \right),
\qquad
{\cal K}_\mu = i\left( 2X_\mu\,X^\nu\partial_\nu - X^2 \partial_\mu \right).
\end{eqnarray}
The dilatation generator is unaffected by the Inönü--Wigner contraction given in equation \eqref{2.7}, 
acting trivially on the spatial components of translations, rotations, and special conformal transformations, 
which remain finite in the limit\cite{LBLL,SenGupta:1966qer}. 
The algebra satisfied by these generators is therefore the same as the $\mathfrak{so}(2,d)$ algebra. 
For the time components, instead, one obtains
\begin{eqnarray}
&&
{\cal P}_0 \;\longrightarrow\; \frac{1}{\epsilon}\,{\cal H},
\qquad {\rm with}\quad {\cal H} = -\,i\,\partial_u,
\nonumber\\[6pt]
&&
{\cal L}_{0i} = i\left( X_0\,\partial_i - X_i\,\partial_0 \right)
\;\Rightarrow\;
i\left( \epsilon\,u\,\partial_i - \frac{1}{\epsilon}\,X_i\,\partial_u \right)
\simeq \frac{1}{\epsilon}\,{\cal B}_i,
\qquad {\rm with}\quad {\cal B}_i = -\,i\,X_i\,\partial_u,
\nonumber\\[6pt]
&&
{\cal K}_0 \;\longrightarrow\;
i\!\left[ 2\epsilon\, u\left( u\partial_u + X^i\partial_i \right)
- \left(-\,\epsilon^2 u^2 + X_i^2\right)\frac{1}{\epsilon}\,\partial_u \right]
\simeq \frac{1}{\epsilon}\,{\cal K},
\qquad {\rm with}\quad {\cal K} = -\,i\,X_i^2\,\partial_u~.\nonumber
\end{eqnarray}
These generators satisfy the Carrollian conformal algebra 
\begin{eqnarray}
\left[{\cal L}_{ij},\,{\cal B}_k\right] &=& i\left(\delta_{jk}{\cal B}_i - \delta_{ik}{\cal B}_j\right),
\qquad
\left[{\cal P}_j,\,{\cal B}_i\right] = -\,i\,\delta_{ij}\,{\cal H},
\nonumber\\[4pt]
\left[{\cal D},\,{\cal K}\right] &=& -\,i\,{\cal K},
\qquad
\left[{\cal K},\,{\cal H}\right] = 0,
\qquad
\left[{\cal K},\,{\cal P}_i\right] = 2\,i\,{\cal B}_i,
\nonumber\\[4pt]
\left[{\cal H},\,{\cal K}_i\right] &=& 2\,i\,{\cal B}_i,
\qquad
\left[{\cal D},\,{\cal H}\right] = i\,{\cal H}~~;~~\left[ {\cal K}_i,\,{\cal B}_j\right]= -i \delta_{ij}\, {\cal K}.
\end{eqnarray}
Here the time translation ${\cal H}$, the Carrollian boosts ${\cal B}_i$, and the temporal conformal generator ${\cal K}$ replace their relativistic counterparts in the contracted algebra.

\vspace*{.07in}The conformal Carroll algebra in $d$ dimensions is isomorphic to the algebra 
$\mathfrak{iso}(1,d)$ of the Poincar\'e group in $d{+}1$ dimensions. 
This can be seen by defining
\begin{eqnarray}
&&
J_{ij} = {\cal L}_{ij},
\qquad 
J_{i0} = \tfrac{1}{2}\left( {\cal P}_i + {\cal K}_i \right),
\qquad 
J_{id} = \tfrac{1}{2}\left( {\cal P}_i - {\cal K}_i \right),
\qquad 
J_{0d} = -\,{\cal D},
\nonumber\\[4pt]
&&
\hat{P}_0 = \tfrac{1}{\sqrt{2}}\left( {\cal H} + {\cal K} \right),
\qquad 
\hat{P}_i = -\,\sqrt{2}\,{\cal B}_i,
\qquad 
\hat{P}_d = \tfrac{1}{\sqrt{2}}\left( {\cal K} - {\cal H} \right).
\end{eqnarray}
The set $\{J_{MN},\,\hat{P}_M\}$ with $M,N = 0, \dots, d$, 
then satisfies the commutation relations of the $(d{+}1)$-dimensional Poincar\'e algebra. 
In other words,
\begin{eqnarray}
\mathfrak{ccarr}(d) \;\simeq\; \mathfrak{iso}(1,d)~.
\end{eqnarray}
This correspondence provides a geometrical interpretation: 
the Carrollian conformal symmetries acting on a null hypersurface in $d$ dimensions 
are isomorphic to the Poincar\'e symmetries of a $(d{+}1)$-dimensional Minkowski spacetime.
Moreover, the quadratic Casimir of the $(d{+}1)$--dimensional Poincar\'e algebra,
which is isomorphic to the Carrollian conformal algebra,
vanishes identically,
\begin{equation}
C_2 = \eta^{MN}\,\hat{P}_M \hat{P}_N = 0~,
\end{equation}
reflecting the fact that the associated momentum vector is null in  the $(d{+}1)$-dimensional Minkowski space $\mathbb{R}^{1,d}$ underlying
$\mathfrak{iso}(1,d)$.
The finite-dimensional conformal Carrollian algebra admits an infinite-dimensional enhancement, 
in which the generators ${\cal H}$, ${\cal B}_i$, and ${\cal K}$ 
are promoted to the supertranslation generator 
$M_f = f(x^i)\,\partial_t$ (see, for example, \cite{2510.21651}). 
This algebra is isomorphic to the BMS algebra and is crucial if we consider the null hypersurface to be the null boundaries of the asymptotically flat spacetimes. However, in this paper we restrict our analysis to its finite-dimensional conformal Carrollian version. 

The transformation properties of generic rank-$s$ tensor fields 
${\cal O}_{\mathbf{s}}(x^\alpha)\equiv{\cal O}_{i_1\dots i_s}(x^\alpha)$, 
forming irreducible tensor representations of the  Carrollian $SO(d-1)$ rotation group, are 
constructed by first analysing the stability subgroup of the origin and then 
extending them to arbitrary points through the action of the translation 
generators ${\cal P}_i$ and ${\cal H}$. The stability subgroup contains the 
Carrollian generators 
$\{{\cal J}_{ij},\,{\cal D},\,{\cal B}_i,\,{\cal K}_i,\,{\cal K}\}$.
The generators of spatial rotations in $\mathbb{R}^{d-1}$, namely
${\cal J}_{ij}$, and the dilatation generator ${\cal D}$ commute and can 
therefore be simultaneously diagonalized (see for example 
\cite{Mack:1969rr,2305.02884})
\begin{equation}
[{\cal J}_{ij},\,{\cal O}_{\mathbf s}(0)] =i\, \Sigma_{ij}\,{\cal O}_{\mathbf s}(0)\qquad, 
\qquad 
[{\cal D},\,{\cal O}_{\mathbf s}(0)] = i\,\Delta\,{\cal O}_{\mathbf s}(0),
\end{equation}
where $\Sigma_{ij}$ denotes the representation of the $SO(d{-}1)$ spin operator  and $\Delta$ is the Carrollian weight (or scaling dimension) of ${\cal O}_{\mathbf s}$. 
Furthermore, since the generators ${\cal{B}}_i$ and ${\cal{K}}_i$ transform as vectors under $SO(d{-}1)$, 
their non-trivial action on the representation containing ${\cal O}_{\mathbf s}(0)$ 
would produce an infinite tower of higher-rank fields. 
To ensure a finite-dimensional representation, one therefore imposes
\begin{equation}
[{\cal B}_i,\,{\cal O}_{\mathbf s}(0)] = [{\cal K}_i,\,{\cal O}_{\mathbf s}(0)] = 0~.\label{2.17}
\end{equation}
From the Carrollian algebra we also have:
\begin{eqnarray}
  \left[{\cal K},\,{\cal O}_{\mathbf s}(0)\right]= \frac{\delta^{ij}}{d-1} \left[\left[{\cal K}_i,\,{\cal O}_{\mathbf s}(0)\right],\,{\cal B}_j\right]+  
  \left[{\cal K},\,{\cal O}_{\mathbf s}(0)\right]- \frac{\delta^{ij}}{d-1} \left[\left[{\cal B}_j,\,{\cal O}_{\mathbf s}(0)\right],\,{\cal K}_j\right]=0
\end{eqnarray}
The operators ${\cal K}_\alpha = ({\cal K},\,{\cal K}_i)$ and 
${\cal P}_\alpha = ({\cal H},\,{\cal P}_i)$ satisfy the commutation relations
\begin{equation}
\left[{\cal D},\,\big[{\cal K}_\alpha,\, {\cal O}_{\mathbf s}(0)\big]\right]
= i\,(\Delta - 1)\,\big[{\cal K}_\alpha,\, {\cal O}_{\mathbf s}(0)\big],
\qquad
\left[{\cal D},\,\big[{\cal P}_\alpha,\, {\cal O}_{\mathbf s}(0)\big]\right]
= i\,(\Delta + 1)\,\big[{\cal P}_\alpha,\, {\cal O}_{\mathbf s}(0)\big].
\end{equation}
Hence, ${\cal K}_\alpha$ and ${\cal P}_\alpha$ act respectively as 
the lowering and raising operators of the Carrollian conformal dimension. 
Equation~\eqref{2.17} can thus be interpreted as defining 
the operator associated with the lowest-weight state of conformal dimension~$\Delta$. 
By analogy with standard conformal field theory, 
this condition characterizes a \emph{primary Carrollian field}. The field at a generic point of the null plane is obtained from its value at the origin 
by the action of the Carrollian translation operators ${\cal P}_\alpha$, namely,
\begin{equation}
{\cal O}_{\mathbf s}(x) = e^{\,i\,(u\,{\cal H} + x^i {\cal P}_i)}\, {\cal O}_{\mathbf s}(0)\, 
e^{-\,i\,(u\,{\cal H} + x^i {\cal P}_i)}~.
\end{equation}
The action of the translation operators on a generic element 
${\cal A}$ of the Carrollian conformal algebra, 
at the level of commutation relations, is given by
\begin{equation}
\left[  {\cal A},\, {\cal O}_{\mathbf s}(x)\right]=e^{\,i\,(u\,{\cal H} + x^i {\cal P}_i)} 
\big[e^{-\,i\,(u\,{\cal H} + x^i {\cal P}_i)}{\cal A}e^{\,i\,(u\,{\cal H} + x^i {\cal P}_i)} ,\,{\cal O}_{\mathbf s}(0)\big]\,
e^{-\,i\,(u\,{\cal H} + x^i {\cal P}_i)}~.
\end{equation}
This relation expresses how commutators involving ${\cal O}_{\mathbf s}(0)$ are mapped to those involving 
${\cal O}_{\mathbf s}(x)$ under finite Carrollian translations.
By using Hadamard’s lemma\footnote{The version of Hadamard’s lemma used here is 
\(e^{A} B e^{-A} = \sum_{n=0}^{\infty} \frac{1}{n!} [A, B]^n\), 
where \([A,B]^n \equiv [A,[A,\dots,[A,B]\dots]]\) denotes the $n$-fold nested commutator.} 
we can write
\begin{eqnarray}
\left[{\cal H},\,{\cal O}_{\mathbf s}(x)\right]
&=& e^{\,i\,(u\,{\cal H} + x^i {\cal P}_i)} 
\big[{\cal H},\,{\cal O}_{\mathbf s}(0)\big]\,
e^{-\,i\,(u\,{\cal H} + x^i {\cal P}_i)}
= -\,i\,\partial_u {\cal O}_{\mathbf s}(x)\,.
\end{eqnarray}
Similarly, one finds\footnote{In these expressions we have used the identity 
$\partial_\alpha{\cal O}_{\mathbf s}(x)= i\,e^{i x^\alpha {\cal P}_\alpha} 
\left[{\cal P}_\alpha,\,{\cal O}_{\mathbf s}(0)\right] e^{-i x^\alpha{\cal P}_\alpha}$.}
\begin{eqnarray}
\left[{\cal P}_i,\,{\cal O}_{\mathbf s}(x)\right] = -\,i\,\partial_i {\cal O}_{\mathbf s}(x)\,.
\end{eqnarray}

For the $SO(d-1)$ rotation generators and for dilatations we have
\begin{eqnarray}
e^{-\,i\,(u\,{\cal H} + x^i {\cal P}_i)}{\cal L}_{ij}e^{\,i\,(u\,{\cal H} + x^i {\cal P}_i)}    
&=&{\cal L}_{ij} +(x_i{\cal P}_j -x_{j}{\cal P}_i)\,,\nonumber\\[4pt]
e^{-\,i\,(u\,{\cal H} + x^i {\cal P}_i)}{\cal D}e^{\,i\,(u\,{\cal H} + x^i {\cal P}_i)}   
&=&{\cal D}-\,x^\alpha {\cal P}_\alpha\,.
\end{eqnarray}
These relations give
\begin{eqnarray}
\left[{\cal L}_{ij},\,{\cal O}_{\mathbf s}(x)\right]
&=& e^{i x^\alpha {\cal P}_\alpha} 
\left[{\cal L}_{ij},\,{\cal O}_{\mathbf s}(0)\right] 
e^{-i x^\alpha {\cal P}_\alpha}
- i(x_j \partial_i - x_i \partial _j){\cal O}_{\mathbf s}(x)
\nonumber\\[2pt]
&=& -\,i \left[-\,\Sigma_{ij} +(x_j \partial_i-x_i\partial _j)\right]{\cal O}_{\mathbf s}(x)\,,\nonumber\\[6pt]
\left[{\cal D},\, {\cal O}_{\mathbf s}(x)\right]
&=& e^{i x^\alpha {\cal P}_\alpha} 
\left[{\cal D},\,{\cal O}_{\mathbf s}(0)\right] 
e^{-i x^\alpha {\cal P}_\alpha}
+ i\,x^\alpha\,\partial _\alpha{\cal O}_{\mathbf s}(x)
\nonumber\\[2pt]
&=& \,i\left[\,\Delta+ u\partial_u +x^i\partial_i\right]{\cal O}_{\mathbf s}(x)\,.
\end{eqnarray}
Finally, the last commutation relations that will be used below are
\begin{eqnarray}
e^{-\,i\,(u\,{\cal H} + x^i {\cal P}_i)}{\cal K}\,e^{\,i\,(u\,{\cal H} + x^i {\cal P}_i)}
&=&{\cal K} - 2 x^i\,{\cal B}_i\,+x^i\,x_i\,{\cal H}\,,\nonumber\\[4pt]
e^{-\,i\,(u\,{\cal H} + x^i {\cal P}_i)}{\cal B}_i\,e^{\,i\,(u\,{\cal H} + x^i {\cal P}_i)}
&=&{\cal B }_i -x_i\,{\cal H}\,,\nonumber\\[4pt]
e^{-\,i\,(u\,{\cal H} + x^i {\cal P}_i)}{\cal K}_i\,e^{\,i\,(u\,{\cal H} + x^i {\cal P}_i)}
&=&{\cal K}_i + 2 u {\cal B}_i + 2 x_i{\cal D} - 2 x^j{\cal L}_{ij}
-2 u x_i {\cal H} + x^2 {\cal P}_i - 2 x^j x_i {\cal P}_j\,.
\end{eqnarray}
These relations imply
\begin{eqnarray}
\left[ {\cal K}, {\cal O}_{\mathbf s}(x)\right]&=& -\,i\,x^2\,\partial_u{\cal O}_{\mathbf s}(x)\,,\qquad
\left[ {\cal B}_i, {\cal O}_{\mathbf s}(x)\right]= i\,x_i\,\partial_u{\cal O}_{\mathbf s}(x)\,,\nonumber\\[4pt]
[\,{\cal K}_i,\,{\cal O}_{\mathbf s}(x)\,]
&=& i\Big(2x_i\,\Delta +2x_i\,u\,\partial_u +2x_i\,x^j\,\partial_j
- x^2 \partial_i - 2x^k\,\Sigma_{ik}\Big){\cal O}_{\mathbf s}(x)\,.
\end{eqnarray}

\noindent
The commutation relations obtained above determine how all Carrollian generators act on fields at a generic spacetime point $x$, once their action at the origin is specified. 
They provide the essential ingredients for realizing Carrollian conformal symmetry on both scalar and spinning fields. 
In the following sections,  we shall extensively employ them, in particular for scalar fields, to construct the CCFT correlators.

\section{Position space Carrollian correlators}
\label{position_rev}
\noindent
The study of Carrollian conformal correlators has so far been mostly carried out in position space, and primarily in specific spacetime dimensions. In arbitrary dimensions, a systematic analysis was presented in \cite{Afshar:2024llh} (also see \cite{Nguyen:2023miw}). Here, we summarize the structure of the two and three point Carrollian conformal correlators involving scalar operators in the position space. 

\vspace*{.07in}We denote a generic Carrollian conformal correlator involving spinning operators $\mathcal{O}_{\mathbf{s}}$ of spin $s$ and conformal dimension $\Delta_s$ as
\be
\mathcal{A}(u_1,\mathbf{x}_1;\cdots;u_n,\mathbf{x}_n)
=\big\langle\,\mathcal{O}_{\mathbf{s}_1}(x_1)\cdots\mathcal{O}_{\mathbf{s}_n}(x_n)\,\big\rangle,
\qquad x\equiv(u,\mathbf{x})\,.
\ee
The variation of a field under the action of a generic CCFT generator, denoted by ${\cal T}_A$, is defined as
\begin{eqnarray}
\delta {\cal O}_{\mathbf s}(x)
=\left[{\cal T}_A,\,{\cal O}_{\mathbf s} (x)\right]
=i\,T_A\, {\cal O}_{\mathbf s}(x)\,\qquad,\qquad  A=1,\dots, {\rm dim}[\mathfrak{ccarr}(d)],
\end{eqnarray}
where $T_A$ denotes the differential operator that realizes the action of ${\cal T}_A$ on the field $ {\cal O}_{\mathbf s}(x)$, as given in the previous section.

\vspace*{.07in}Assuming that the vacuum is invariant under the CCFT symmetry group, i.e.,
\begin{eqnarray}
{\cal T}_A\,|0\rangle = 0\,,
\end{eqnarray}
the action of ${\cal T}_A$ on an $n$-point function gives
\begin{eqnarray}
\langle 0|\,{\cal T}_A\,\prod_{a=1}^n \phi(x_a)\,|0\rangle
&=&\sum_{a=1}^n
\big\langle 0\big|\,\phi(x_1)\cdots[{\cal T}_A,\,\phi(x_a)]\cdots\phi(x_n)\big|0\big\rangle
=0\,.
\end{eqnarray}
Using $[{\cal T}_A, {\cal O}_{\mathbf s}(x_a)]=i\,T^{(a)}_A\,{\cal O}_{\mathbf s} (x_a)$, one obtains the Ward identities
\begin{eqnarray}
\sum_{a=1}^n T_A^{(a)}\,\big\langle\phi(x_1)\cdots\phi(x_n)\big\rangle=0\,.
\end{eqnarray}

\noindent
Specialising the CCFT algebra to the case of correlation functions leads to the Ward identities obeyed by Carrollian correlators.
In $d$ dimensional position space, they take the following form
\be
\mathcal{H}&:&
\sum_{a=1}^n \frac{\partial}{\partial u_a}\,
\mathcal{A}(u_1,\mathbf{x}_1;\cdots;u_n,\mathbf{x}_n)=0\non\\[4pt]
\mathcal{P}_i&:&
\sum_{a=1}^{n} \frac{\partial}{\partial {x}^i_a}\,
\mathcal{A}(u_1,\mathbf{x}_1;\cdots;u_n,\mathbf{x}_n)=0\non\\[4pt]
\mathcal{B}_i&:&
\sum_{a=1}^n {x}^i_a \frac{\partial}{\partial u_a}\,
\mathcal{A}(u_1,\mathbf{x}_1;\cdots;u_n,\mathbf{x}_n)=0\non\\[4pt]
\mathcal{J}_{ij}&:&
\sum_{a=1}^n\left(-\,\Sigma^{(a)}_{ij}-{x}^a_i\,\partial^a_j+{x}^a_j\,\partial^a_i\right)
\mathcal{A}(u_1,\mathbf{x}_1;\cdots;u_n,\mathbf{x}_n)=0\non\\[4pt]
\mathcal{D}&:&
\sum_{a=1}^n\left(\Delta_a+u_a\,\partial_{u_a}+{x}^i_a\,\frac{\partial}{\partial {x}^i_a}\right)
\mathcal{A}(u_1,\mathbf{x}_1;\cdots;u_n,\mathbf{x}_n)=0\non\\[4pt]
\mathcal{K}_i&:&
\sum_{a=1}^n
\Big(2{x}^a_i\,\Delta_a
-2\,{x}_a^j\,\Sigma^{(a)}_{ij}
+2u_a\,{x}^a_i\,\partial_{u_a}
+2{x}^a_i\,{x}_a^j\,\partial^a_j
-{x}_a^2\,\partial^a_i\Big)
\mathcal{A}(u_1,\mathbf{x}_1;\cdots;u_n,\mathbf{x}_n)=0\,,\nonumber\\[4pt]
\mathcal{K}&:&
\sum_{a=1}^n {x}_a^2\,\frac{\partial}{\partial u_a}\,
\mathcal{A}(u_1,\mathbf{x}_1;\cdots;u_n,\mathbf{x}_n)=0
\label{3.7wa}
\ee
where $i=1,2,\dots,d-1$ and ${x}_a^2\equiv\delta_{ij}\,{x}_a^i {x}_a^j$.
\noindent

\vspace*{.07in}We next review how to derive the two and three point correlation functions by solving the position space Ward identities. 

\subsection{Two point function}
In the two point case, the $\mathcal{H}$ and $\mathcal{P}_i$ Ward identities simply require that the correlator depends only on the differences of the coordinates,
\[
\mathcal{A}(u_1,\mathbf{x}_1;\,u_2,\mathbf{x}_2)
\equiv 
\mathcal{A}(u_{12},\,
\mathbf{x}_{12})\,,
\]
where $u_{12}= u_1 -u_2$ and $\mathbf{x}_{12}=\mathbf{x}_1-\mathbf{x}_2$. The boost Ward identity associated with $\mathcal{B}_i$ imposes
\begin{eqnarray}
({x}_1^i\partial_{u_1} + {x}_2^i\partial_{u_2})\,G^{(2)}(u_{12},\,\mathbf{x}_{12})
= {x}_{12}^i\,\partial_{u_1} G^{(2)}(u_{12},\,\mathbf{x}_{12}) = 0\,,
\end{eqnarray}
\noindent
The above equation implies that the most general two point function can only be of the form
\begin{eqnarray}
G^{(2)}(u_{12},\,\mathbf{x}_{12})
= G(\mathbf{x}_{12}) + F(u_{12})\,\delta^{(d-1)}(\mathbf{x}_{12})\,.
\end{eqnarray}
This structure naturally splits into two independent branches. 
The first term, $G(\mathbf{x}_{12})$, is entirely supported on the spatial slice $\mathbb{R}^{d-1}$ and depends only on the spatial separation $\mathbf{x}_{12}$. It is usually referred as the \emph{magnetic branch}. 
The second term, $F(u_{12})\,\delta^{(d-1)}(\mathbf{x}_{12})$, is localized in space and depends solely on the Carrollian “time” coordinate $u_{12}$. It is usually referred as the \emph{electric branch}. 
The action of the dilatation generator imposes the constraint
\begin{eqnarray}
\left(u_{12}\,\partial_{u_{12}} 
+ {{x}}_{12}^i\,\partial_{{{x}}_{12}^i}
+ \Delta_1 + \Delta_2\right)
\Big[G(\mathbf{x}_{12}) + F(u_{12})\,\delta^{(d-1)}(\mathbf{x}_{12})\Big] = 0\,.
\end{eqnarray}
Expanding this expression yields\footnote{In deriving the second identity we used the standard homogeneity property of the Dirac delta distribution,
\(
x^i\,\partial_{{x}^i}\,\delta^{(d-1)}(\mathbf{x})
= - (d-1)\,\delta^{(d-1)}(\mathbf{x})\,.
\)}
\begin{eqnarray}
&&
\Big(\Delta_1+\Delta_2+{{x}}^i_{12}\,\partial_{{{x}}^i_{12}}\Big)G({\mathbf{x}}_{12})=0 ~~~~;~~~~
\Big(\Delta_1+\Delta_2-(d-1)+u_{12}\,\partial_{u_{12}}\Big)F(u_{12})
=0\,.
\end{eqnarray}
The first equation controls the scaling behaviour of the spatial (magnetic) branch, 
while the second one constrains the temporal (electric) branch through its dependence on $u_{12}$. 
This condition ensures the full invariance of the magnetic branch under the Carrollian special conformal transformation.
One can check that the 2-point function consistent with the above Ward identities is given by \cite{Afshar:2024llh}
\be
G(\mathbf{x}_{12})=\;\f{{B_1}}{|{\mathbf{x}}_{12}|^{\Delta_1+\Delta_2}}\delta_{\Delta_1,\Delta_2} ~~~~;~~F(u_{12})=\f{{B_2}}{|u_{12}|^{\Delta_1+\Delta_2-d+1}}
\ee
Finally, note that the presence of the delta function, which enforces ${x}_1^i = {x}_2^i$, implies that 
the term $F(u_{12})\,\delta^{(d-1)}(\mathbf{x}_{12})$ automatically satisfies the 
$\mathcal{K}_i$ Ward identity for arbitrary values of $\Delta_1$ and $\Delta_2$. 
In contrast, the spatial contribution $
G(\mathbf{x}_{12})$
solves the $\mathcal{K}_i$ Ward identity only when the two operators have equal conformal dimensions. Thus, we finally arrive at following expression of two point function
\be
\bigl\langle \,\mathcal{O}_1(u_1,\textbf{x}_1) \mathcal{O}_2(u_2,\textbf{x}_2)\bigl\rangle\;=\;\f{{B_1}}{|\textbf{x}_{12}|^{\Delta_1+\Delta_2}}\delta_{\Delta_1,\Delta_2} +\f{{B_2}}{|u_{12}|^{\Delta_1+\Delta_2-d+1}}\delta^{d-1}(\textbf{x}_{12})\label{2ptposfg}
\ee

\subsection{Three point function}
\label{sec:3pointpos}
For the 3-point function also, the invariance under the $\mathcal{H}$ and $\mathcal{P}_i$ generators imply, as in the two-point case, 
that the correlator depends only on the coordinate differences. 
However, only two independent coordinate differences can be formed using three points 
since the third one is linearly dependent on the other two. E.g.,
\[
u_{13}=u_{12}-u_{23}\,, \qquad {x}^i_{13}={x}^i_{12}-{x}^i_{23}\,.
\]
Therefore, it is convenient to express the three-point correlator as a function of two independent variables. One possible choice is, for instance,
\begin{eqnarray}
\bigl\langle \,\mathcal{O}_1(u_1,\textbf{x}_1) \mathcal{O}_2(u_2,\textbf{x}_2)\mathcal{O}_3(u_3,\textbf{x}_3)\bigl\rangle
&\equiv&
\mathcal{A}_3(u_{12},{\textbf{x}}_{12};\,u_{23}, {\textbf{x}}_{23})
\label{3pt_invariance}
\end{eqnarray}
The Ward identity associated with the Carrollian boost generator ${\cal B}_i$, with the choice made in equation \eqref{3pt_invariance} can be expressed as
\begin{eqnarray}
\left(x_{12}^i\partial_{u_{12}}+x_{23}^i\partial_{u_{23}}\right)\mathcal{A}_3(u_{12},{\mathbf{x}}_{12};\,u_{23}, {\mathbf{x}}_{23})=0\label{boost_Ward_3pt}
\end{eqnarray}
\noindent
Equation~\eqref{boost_Ward_3pt} implies that the three-point correlator is either completely independent of the Carrollian temporal variables $u_{ab}$,
or any dependence on $u_{ab}$ can only appear multiplied by the corresponding spatial delta functions $\delta^{(d-1)}(\textbf{x}_{ab})$.
In other words, Carrollian boost invariance implies that the correlator can be decomposed into five distinct sectors: 
a \emph{purely magnetic branch}, depending only on the spatial separations $\textbf{x}_{ab}$;
a \emph{purely electric branch}, fully localized in space through the products of $\delta^{(d-1)}(\textbf{x}_{ab})$ and depending solely on the Carrollian ``time'' variables $u_{ab}$; and three \emph{mixed branches}, which involve the temporal differences $u_{ab}$ and are 
multiplied by the corresponding spatial delta functions $\delta^{(d-1)}(\mathbf{x}_{ab})$~\cite{Afshar:2024llh}\footnote{Ref.~\cite{Afshar:2024llh} shows that the most general solution to the ${\cal B}_i$ Ward identity also includes collinear contributions. However, as demonstrated in the same reference, once the ${\cal K}_i$ Ward identity is imposed, these terms can be consistently reabsorbed into the definition of the magnetic branch.}
\be
&&\hspace*{-.9in}\bigl\langle \,\mathcal{O}_1(u_1,\textbf{x}_1) \mathcal{O}_2(u_2,\textbf{x}_2)\mathcal{O}_3(u_3,\textbf{x}_3)\bigl\rangle\non\\
&=& G_1(u_1,u_2,u_3)\delta^{d-1}(\textbf{x}_{12})\delta^{d-1}(\textbf{x}_{23})+G_2(\textbf{x}_1,\textbf{x}_2,\textbf{x}_3) +G_3(u_{23},\textbf{x}_{12})\delta^{d-1}(\textbf{x}_{23})\non\\
&&+\;G_4(u_{31},\textbf{x}_{23})\delta^{d-1}(\textbf{x}_{31})+G_5(u_{12},\textbf{x}_{31})\delta^{d-1}(\textbf{x}_{12})\label{39we}
\ee
The remaining Ward identities fix the precise form of the functions $G_i$. We now describe these functions one by one. 

\vspace*{.07in}The function $G_1$ is not completely fixed by Carrollian conformal invariance. In particular, the Carrollian special conformal Ward identity does not impose any additional constraint not implied by other Ward identities. The Carrollian dilatation Ward identity implies that $G_1$ must be a homogeneous function of degree $ 2(d-1)-\Delta_t$ (where $\Delta_t=\Delta_1+\Delta_2+\Delta_3$), i.e., 
\be
G_1(\lambda u_1,\lambda u_2,\lambda u_3 )=\lambda^{2d-2-\Delta_t}G_1(u_1,u_2,u_3)
\ee
Any function having the form\footnote{Naively, one can assume the function $f$ to be a function of two ratios, namely, $f(\f{u_{13}}{u_{23}},\f{u_{12}}{u_{23}} )$. However, the second ratio $\f{u_{13}}{u_{23}}$ can be written in terms of the other since
\be
\f{u_{13}}{u_{23}}= \f{u_{12}}{u_{23}}+1\non
\ee}
\be
G_1(u_1,u_2,u_3)= (u_{23})^{2d-2-\Delta_t} f\left(\f{u_{12}}{u_{23}}\right),\label{311we}
\ee
where $f$ is an arbitrary function, will satisfy all the Ward identities. In particular, the dilatation Ward identity will be satisfied by \eqref{311we} since
\be
\sum_{a=1}^3 u_a\f{\p}{\p{u_a}} G_1(u_1,u_2,u_3) = (2d-2-\Delta_t)\; G_1(u_1,u_2,u_3)
\ee
and
\be
\sum_{a=1}^3 x_a^i\f{\p}{\p{x_a^i}} \delta^{d-1}(\mathbf{x}_{12})\delta^{d-1}(\mathbf{x}_{23}) = -2(d-1).
\ee
An example of a function which satisfies above equations is \cite{Afshar:2024llh}
\be
G_1(u_1,u_2,u_3)&=& \sum_{p+q+r=\Delta_1+\Delta_2+\Delta_3-2(d-1)} \f{C_1}{|u_{12}|^{p}|u_{23}|^{q}|u_{13}|^{r}} \label{314we}
\ee
However, this is only one example of a function allowed by the dilatation Ward identity. Any function consistent with the form \eqref{311we} is also acceptable and will satisfy all the Carrollian conformal Ward identities. However, in a concrete Carrollian theory, this arbitrary function would be fixed, or at least further constrained, by additional input beyond Carrollian symmetry, such as OPE data, crossing/factorization constraints from higher-point functions, regularity requirements, microscopic dynamics, or a controlled ultra-relativistic limit from a parent relativistic CFT.

\vspace*{.07in}The dilatation and ${\cal K}_i$ Ward identities for the magnetic term $G_2$ coincide with those of a Euclidean $(d-1)$-dimensional CFT. Therefore, the solution is identical to the one in that case, namely
\begin{eqnarray}
G_2(\mathbf{x}_1,\mathbf{x}_2,\mathbf{x}_3)&=&\f{C_2}{|\mathbf{x}_{12}|^{\Delta_1+\Delta_2-\Delta_3}|\mathbf{x}_{23}|^{\Delta_2+\Delta_3-\Delta_1}|\mathbf{x}_{13}|^{\Delta_1+\Delta_3-\Delta_2}}
\end{eqnarray}
For the mixed branches, for example $G_3(u_{23},\mathbf{x}_{12})$, the dilatation Ward identity takes the form
\begin{eqnarray}
&&\left(
\Delta_t 
+ u_{23}\,\frac{\partial}{\partial u_{23}}
+ x_{12}^i \,\frac{\partial}{\partial x_{12}^i}
-(d-1)\right)
G_3(u_{23},\mathbf{x}_{12}) = 0.
\end{eqnarray}
It is solved by the function
\begin{eqnarray}
G_3(u_{23},\mathbf{x}_{12})&=&\f{C_3}{|\mathbf{x}_{12}|^{a}|u_{23}|^{b}}\qquad,\qquad a+b= \Delta_t -(d-1)\label{3.54}
\end{eqnarray}
The Carrollian special conformal Ward identity instead turns out to be
\begin{eqnarray}
&&\hspace*{-.7in}\left(
x_1^{\,i}\,\Delta_1
+ x_2^{\,i}\,\Delta_2
+ x_3^{\,i}\,\Delta_3
+ (u_2 x_2^{\,i} - u_3 x_3^{\,i})\,\frac{\partial}{\partial u_{23}}
+ (x_1^{\,i} x_1^{\,j} - x_2^{\,i} x_2^{\,j})\,\frac{\partial}{\partial x_{12}^j}
- \frac{1}{2}(x_1^2 - x_2^2)\,\frac{\partial}{\partial x_{12\,i}}\right.
\nonumber\\
&&\hspace*{-.7in}\left.+ (x_2^{\,i} x_2^{\,j} - x_3^{\,i} x_3^{\,j})\,\frac{\partial}{\partial x_{23}^j}
- \frac{1}{2}(x_2^2 - x_3^2)\,\frac{\partial}{\partial x_{23\, i}}
\right)
G_3(u_{23},\mathbf{x}_{12})\,\delta^{(d-1)}(\mathbf{x}_{23})\nonumber\\[4pt]
&=& \left( \Delta_1 -\frac{a}{2} \right)x_{12}^i G_3(u_{23},\mathbf{x}_{12})\,\delta^{(d-1)}(\mathbf{x}_{23}) \non\\[4pt]
&=&0
\end{eqnarray}
This gives
\begin{eqnarray}
a=2\Delta_1\qquad,\qquad b= \Delta_2+\Delta_3 -\Delta_1-(d-1)
\end{eqnarray}
The remaining functions in \eqref{39we} are fixed in the same way by the Carrollian conformal Ward identities and are given by
\be
G_4(u_{31},\mathbf{x}_{23})&=&\f{C_4}{|\mathbf{x}_{23}|^{2\Delta_2}|u_{31}|^{\Delta_3+\Delta_1-\Delta_2-d+1}}\non\\[.3cm]
G_5(u_{12},\mathbf{x}_{31})&=&\f{C_5}{|\mathbf{x}_{31}|^{2\Delta_3}|u_{12}|^{\Delta_1+\Delta_2-\Delta_3-d+1}}
\ee
The constants $C_i$ appearing in the expressions of $G_i$ will, in general, depend upon $d$ and $\Delta_i$.

\section{Carrollian correlators from momentum space Ward identities}
\label{sec4:Ward}
As mentioned earlier, most of the results involving the Carrollian conformal correlators are only available in the position space. However, motivated by the close connection between the Carrollian conformal field theories, flat holography and 
scattering amplitudes, it is desirable to have a systematic momentum space formulation of the Carrollian theories. In this section, we obtain the 2 and 3 point Carrollian conformal correlators in the momentum space. We shall first derive the Ward identities in the momentum space. These Ward identities will then be used to find the differential equations constraining the 2 and 3 point correlations functions. These differential equations will then be solved to determine the explicit form of the 2 and 3 point correlators in the momentum space.

\subsection{Momentum space Ward identities}

\vspace*{.07in}To write the Ward identities in the momentum space, we consider the Fourier transform
\begin{eqnarray}
\mathcal{O}_{\bf s}(\omega,\textbf{p}) =\int d^{d-1}x\,du\; \mathcal{O}_{\bf s}(u,\textbf{x}) e^{i {(u \,\omega+x^ip_i)}}\, ~~;\qquad p_\mu \equiv(\omega,\,\textbf{p})
 \end{eqnarray}
We now multiply the Ward identities in \eqref{3.7wa} with $e^{i(u_a\omega_a+x^i_ap_{ia})}$ from left and integrate over the position variables. For the terms involving the derivatives, we perform an integration by parts ignoring the boundary terms and write the surviving coordinates $u_a$ and  $x^i_a$ as derivatives with respect to $\omega_a$ and $p_a^i$ respectively. Finally, using the above definition of the  transform, we get the momentum space Ward identities 
\begin{eqnarray}
{\cal H}&:&\sum_{a=1}^n\omega_a \mathcal{A}\left(\omega_1,{\bf p}_1,\dots ,\omega_n,\,{\bf p}_n\right)=0\nonumber\\
{\cal P}_i&:&\sum_{a=1}^n p_a^i\, \mathcal{A}\left(\omega_1,{\bf p}_1,\dots ,\omega_n,\,{\bf p}_n\right)=0\non\\
 {\cal B}_i&:&\sum_{a=1}^n \omega_a\frac{\partial}{\partial p_i^a} \mathcal{A}\left(\omega_1,{\bf p}_1,\dots ,\omega_n,\,{\bf p}_n\right)=0\nonumber\\
\mathcal{J}_{ij}&:&\sum_{a=1}^n\left[\Sigma_{ij}^a -i\left( p^a_j\frac{\partial}{\partial p^a_i}-p^a_i\frac{\partial}{\partial p^a_j}\right)\right] \mathcal{A}\left(\omega_1,{\bf p}_1,\dots ,\omega_n,\,{\bf p}_n\right)=0\nonumber\\
 {\cal D}&:&\sum_{a=1}^n \left[\Delta_a-d -\omega_a\frac{\partial}{\partial \omega_a}-p_a^i\frac{\partial}{\partial p_a^i}\right]\mathcal{A}\left(\omega_1,{\bf p}_1,\dots ,\omega_n,\,{\bf p}_n\right)=0\nonumber\\
  {\cal K}_i&:&\sum_{a=1}^n\left[ {\Delta_a \frac{\partial}{\partial p_{a}^i}+ \Sigma_{ij}^a\frac{\partial}{\partial p_{aj}}}-d\frac{\partial}{\partial p_a^i} -\omega_{a} \frac{\partial^2}{\partial p_a^i\partial \omega_{a}} -p_{aj} \frac{\partial^2}{\partial p_a^i\partial p_{aj}}+\frac{1}{2} p_{ai} \frac{\partial^2}{\partial p_a^j\partial p_{aj}}\right] \mathcal{A}=0\nonumber\\
{\cal K}&:&\sum_{a=1}^n \omega_a\frac{\partial^2}{\partial p_i^a\,\partial p_{ia}}\mathcal{A}\left(\omega_1,{\bf p}_1,\dots ,\omega_n,\,{\bf p}_n\right)=0
\label{3.9wa}
\end{eqnarray}
where, we have defined the momentum space correlator $\mathcal{A}\left(\omega_1,{\bf p}_1,\dots ,\omega_n,\,{\bf p}_n\right)$ by
\begin{eqnarray}
\mathcal{A}\left(\omega_1,{\bf p}_1,\dots ,\omega_n,\,{\bf p}_n\right)=\int \prod_{a=1}^n d^dx_a e^{i\sum_{b=1}^n( u_b\omega_b+x_b^i\,p_i^b)}\bigl\langle \,\mathcal{O}_{{\bf s}_1}(x_1)\dots \mathcal{O}_{{\bf s}_n}(x_n)\bigl\rangle
\end{eqnarray}
Note that the correlators appearing in the Ward identities \eqref{3.9wa} include the energy momentum conserving delta functions. This is in contrast to the momentum space Ward identities one considers in usual CFTs where the overall energy momentum conserving delta functions are brought outside the Ward identity generators \cite{Bzowski:2013sza}. The reason for this is that in the Carrollian case, the Ward identities allow for more general solutions where we may not necessarily have the usual overall energy and momentum conserving delta functions as we shall see below.

\subsection{ Ward identity constraints}

\subsubsection{2-point function}
We start by finding the consequences of different Ward identities on the 2-point correlation function  of scalar fields. The $\mathcal{P}_i$ Ward identity implies that the function $\mathcal{A}$ must take the form
\be
\mathcal{A}_2(\omega_1,{\bf p}_1,\omega_2,\,{\bf p}_2)\;=\; \delta^{d-1}({\bf p}_1+\,{\bf p}_2)\tilde{\mathcal{A}}_2(\omega_1,{\bf p}_1,\omega_2,\,{\bf p}_2)
\ee
On the other hand, the $\mathcal{H}$ Ward identity implies
\be
\mathcal{A}_2(\omega_1,{\bf p}_1,\omega_2,\,{\bf p}_2)\;=\; \delta^{d-1}({\bf p}_1+\,{\bf p}_2)\Bigl[ \delta(\omega_1+\omega_2) F_1(\omega_1,{\bf p}_1,\omega_2,\,{\bf p}_2) +\delta(\omega_1)\delta(\omega_2) F_2(\omega_1,{\bf p}_1,\omega_2,\,{\bf p}_2)\Bigl]\non
\ee
Note that in the second term, the product of delta functions $\delta(\omega_1)\delta(\omega_2)$ sets the $\omega_1$ and $\omega_2$ to zero and hence we can write
\be
F_2(\omega_1,{\bf p}_1,\omega_2,\,{\bf p}_2) = F_2({\bf p}_1,\,{\bf p}_2)
\ee
Ultra-local contact terms supported at vanishing momenta, $\delta^{d-1}(\mathbf{p}_1)\delta^{d-1}(\mathbf{p}_2)$  are in principle compatible with translation invariance, but are ruled out by the full set of Carrollian Ward identities.

Next, we consider the consequences of $\mathcal{B}_i$ Ward identity 
\be
0
&=&\left(\omega_1\f{\p}{\p p_i^1} + \omega_2\f{\p}{\p p_i^2} \right)\biggl[\delta^{d-1}({\bf p}_1+\,{\bf p}_2)\Bigl( \delta(\omega_1+\omega_2) F_1(\omega_1,{\bf p}_1,\omega_2,\,{\bf p}_2) +\delta(\omega_1)\delta(\omega_2) F_2({\bf p}_1,\,{\bf p}_2)\Bigl)\biggl]\non\\
&=&\delta(\omega_1+\omega_2)\left(\omega_1\f{\p}{\p p_i^1} + \omega_2\f{\p}{\p p_i^2} \right)\biggl[\delta^{d-1}({\bf p}_1+\,{\bf p}_2) F_1(\omega_1,{\bf p}_1,\omega_2,\,{\bf p}_2) \biggl]\non\\
&=&2\delta(\omega_1+\omega_2)\delta^{d-1}({\bf p}_1+\,{\bf p}_2)\omega_1\f{\p}{\p p_i^1} F_1(\omega_1,\textbf{p}_1,\omega_2)
\ee
This can be satisfied if $F_1$ is chosen to be independent of the momenta along the $\mathbb{R}^{d-1}$ directions, i.e., 
\be
\f{\p}{\p p_i^1} F_1(\omega_1,\textbf{p}_1,\omega_2)=0\quad\implies\qquad F_1 = F_1(\omega_1,\omega_2)
\ee
Thus, the 2-point function takes the form
\be
\mathcal{A}_2(\omega_1,{\bf p}_1,\omega_2,\,{\bf p}_2)\;=\; \delta^{d-1}({\bf p}_1+\,{\bf p}_2)\Bigl[ \delta(\omega_1+\omega_2) F_1(\omega_1,\omega_2) +\delta(\omega_1)\delta(\omega_2) F_2({\bf p}_1,\,{\bf p}_2)\Bigl]\label{446tyr}
\ee
The $\mathcal{J}_{ij}$ Ward identity can be satisfied if we assume that the correlator depends only upon the magnitude of the momenta $p^a_i$ along the $R^{d-1}$ directions. Let's denote the magnitude of the vector $p_a^i$ as
 \begin{eqnarray}
k_a&\equiv&|{\bf p}_a|=\sqrt{\sum_{i=1}^{d-1} (p^i_a)^2} 
\end{eqnarray}
For later applications, we note the following chain rule identities 
\begin{eqnarray}
\frac {\partial}{\partial p^i}&=&  \frac{p_i}{k} \frac{\partial}{\partial k}\quad;\qquad\frac{\partial^2}{\partial p^i\partial p^j} \;\;=\;\; \left[\frac{\delta_{ij}}{k} -\frac{p_i\,p_j}{k^3}\right]\frac{\partial}{\partial k}+\frac{p_i\,p_j}{k^2} \frac{\partial^2}{\partial k^2}\label{5.96}
\end{eqnarray}
\begin{eqnarray}
p^j\,\frac{\partial^2}{\partial p^i\partial p^j}&=& p_i\frac{\partial^2}{\partial k^2}\quad;\qquad p_i\frac{\partial^2}{\partial p_j\partial p^j}\;\;=\;\; \frac{(d-2)p_i}{k}\frac{\partial}{\partial k} +p_i\frac{\partial^2}{\partial k^2} 
\end{eqnarray}
Now, acting with the $\mathcal{J}_{ij}$ operator and using the identity \eqref{4210}, we see that $\mathcal{J}_{ij}$ trivially annihilates the first term of \eqref{446tyr}. Similarly, the action of $\mathcal{J}_{ij}$ on the second term gives
\begin{eqnarray}
&&\hspace*{-.9in}\sum_{a=1}^2\left[-i\left( p^a_j\frac{\partial}{\partial p^a_i}-p^a_i\frac{\partial}{\partial p^a_j}\right)\right]  \delta^{d-1}(\textbf{p}_1+\textbf{p}_2)F_2(k_1)\non\\
&=&-i \left[\frac{\partial }{\partial P^i}\delta^{d-1}(\textbf{P})\sum_{a=1}^2 p_a^j-\frac{\partial }{\partial P^j}\delta^{d-1}(\textbf{P})\sum_{a=1}^2 p_a^i\right]F_2(k_1)\nonumber\\
&&-i \delta^{d-1}(\textbf{p}_1+\textbf{p}_2)\sum_{a=1}^2\left[-i\left( p^a_j\,p^a_i-p^a_i\,p^a_j\right)\right] \frac{1}{k_1}\frac{\partial}{\partial k_1} F_2(k_1)\non\\
&=&0
 \end{eqnarray}
with $P^i\equiv \sum_{a=1}^{n}p_a^i$ and $n=2$.

\vspace*{.07in}The ${\cal K}$-Ward identity is trivially satisfied. The first term in \eqref{446tyr} is annihilated due to the identity \eqref{a94} while the 2nd term is annihilated since we have $\omega_a=0$ due to the delta functions. 

\vspace*{.07in}Next, we apply the dilation Ward identity. Acting with the dilatation generator, we get
 \begin{eqnarray}
0&=& \sum_{a=1}^2 \left[\Delta_a-d -\omega_a\frac{\partial}{\partial \omega_a} -p_a^i\frac{\partial }{\partial p_a^i}\right]\mathcal{A}_2(\omega_1,{\bf p}_1,\omega_2,\,{\bf p}_2)\nonumber\\
 &=& (\Delta_1+\Delta_2-2d) \delta^{d-1}\left(\sum_{b=1}^2{\bf p}_b\right)\left[ \delta\left(\omega_1+ \omega_2\right)F_1(\omega_1, \omega_{2})+\delta(\omega_1)\delta(\omega_2)\, F_2({\bf p}_1,{\bf p}_{2})\right]\nonumber\\
 &&-\delta(\omega_1)\delta(\omega_2)\delta^{d-1}(\textbf{P})
 \left[-(1+d) F_2({\bf p}_1,{\bf p}_{2})+\sum_{a=1}^2p_a^i\frac{\partial}{\partial p_a^i}F_2({\bf p}_1, {\bf p}_{2})\right] \nonumber\\
 &&- \delta^{d-1}(\textbf{P}) \delta\left(\omega_1+ \omega_2\right) \left[- dF_1(\omega_1, \omega_{2}) +\sum_{a=1}^2\omega_a\frac{\partial}{\partial \omega_a}F_1(\omega_1,\omega_{2}) \right]
 \end{eqnarray}
The above Ward identity is satisfied if the functions $F_1$ and $F_2$ satisfy 
 \begin{eqnarray}
   \left[\Delta_1+\Delta_2 -d - \omega_1\frac{\partial}{\partial \omega_1}\right]F_1(\omega_1)\;=\;0\quad;\qquad 
 \left[\Delta_1+\Delta_2 -(d-1) -p_1^i\frac{\partial}{\partial p_1^i}\right]F_2({\bf p}_1)\;=\;0\label{5.87v2}
 \end{eqnarray}
The solutions of those differential equations are given by 
\begin{eqnarray}
F_1(\omega_1)=\omega_1^{\Delta_1+\Delta_2 -d}\quad;\qquad F_2({\bf p}_1)=|{\bf p}_1|^{\Delta_1+\Delta_2 -d+1}\label{2points}
\end{eqnarray}
Finally, we need to consider the special conformal Ward identity $\mathcal{K}_i$. The action of $\mathcal{K}_i$ generator on the 2-point function is given by
\begin{eqnarray}
&&\sum_{a=1}^2 \left[(\Delta_a-d) \frac{\partial}{\partial p_a^i} -\omega_a\frac{\partial^2}{\partial p_a^i\partial \omega_a}
-p_{aj}\frac{\partial}{\partial p_a^i\partial p_{aj}} +\frac{1}{2}p_{a i} \frac{\partial^2}{\partial p_a^j\partial p_{ai}}\right]\mathcal{A}_2(\omega_1,{\bf p}_1,\omega_2,{\bf p}_2)\non\\[.2cm]
&=& \delta^{d-1}(\textbf{p}_1+\textbf{p}_2)\delta(\omega_1+\omega_2) \sum_{a=1}^2\left[(\Delta_a-d) \frac{\partial}{\partial p_a^i} -p_{aj}\frac{\partial^2}{\partial p_{aj}\partial \omega_a}-p_{aj} \frac{\partial^2}{\partial p_a^i\partial p_{aj}}+\frac{1}{2}p_{ai}\frac{\partial^2}{\partial p_a^j\partial p_{a_j}}\right]F_1\non\\
&&+\delta^{d-1}(\textbf{p}_1+\textbf{p}_2)\delta(\omega_1)\delta(\omega_2)\sum_{a=1}^2\left[\left(\Delta_a-(d-1)\right) \frac{\partial}{\partial p_a^i}
-p_{aj} \frac{\partial^2}{\partial p_a^i\partial p_a^j} +\frac{1}{2}p_{ai} \frac{\partial^2}{\partial p_a^j \partial p_{aj} }\right]F_2\nonumber
\end{eqnarray}
where, we have again used the delta function identities to commute the delta functions across the special conformal generator. The first term involving $F_1$ now vanishes since it does not depend upon the momenta $p^i_a$. From the 2nd term, we get
\begin{eqnarray}
p_{1i}\left[\left(\Delta_1 -\frac{d}{2} \right)\frac{1}{k_1}\frac{\partial}{\partial k_1}-\frac{1}{2}\,\frac{\partial^2}{\partial k_1^2} \right]F_2({\bf p}_1, {\bf p}_{2})=0\label{2nd}
\end{eqnarray}
This is the special conformal generator for a Euclidean CFT in $d-1$ dimensions and as is well known, it imposes the condition $\Delta_1=\Delta_2$.

\subsubsection{3-point function}
For the 3-point correlator, the $\mathcal{P}_i$ Ward identity again imposes the overall spatial momentum conservation
\be
\mathcal{A}_3(\omega_1,{\bf p}_1,\omega_2,\,{\bf p}_2,\omega_3,\,{\bf p}_3)\;=\; \delta^{d-1}({\bf p}_1+\,{\bf p}_2+\,{\bf p}_3)\tilde{\mathcal{A}}_3(\omega_1,{\bf p}_1,\omega_2,\,{\bf p}_2,\omega_3,\,{\bf p}_3)
\ee
On the other hand, the $\mathcal{H}$ Ward identity implies
\be
&&\hspace*{-.7in}\mathcal{A}_3(\omega_1,{\bf p}_1,\omega_2,\,{\bf p}_2,\omega_3,\,{\bf p}_3)\non\\
&=& \delta^{d-1}({\bf p}_1+\,{\bf p}_2+\,{\bf p}_3)\Bigl[ \delta(\omega_1+\omega_2+\omega_3) F_1 +\delta(\omega_1)\delta(\omega_2) \delta(\omega_3) F_2\non\\
&&+\delta(\omega_1)\delta(\omega_2+\omega_3)  F_3\;+\;\delta(\omega_2) \delta(\omega_1+\omega_3) F_4+\delta(\omega_3) \delta(\omega_1+\omega_2) F_5 \Bigl]
\label{hpi}
\ee
Thus, we have a total of 5 independent functions appearing in the 3-point Carrollian correlator of scalar operators. Each of these functions depend upon the 6 variables $\omega_i$ and $\textbf{p}_i$ (for $i=1,2,3$) to begin with. For further restricting the form of the functions $F_i$, we need to consider the rest of the Ward identities. We consider them one by one. 
\subsubsection*{$\mathcal{B}_i$ Ward identity}
The functions $F_i$ appearing in the 3-point function are all independent. Hence, each term in \eqref{hpi} must be annihilated by the $\mathcal{B}_i$ Ward identity individually. For the $F_1$ term, this means
\begin{eqnarray}
0&=&\delta(\omega_1+\omega_2+\omega_3)\left(\omega_1\f{\p}{\p p_i^1} + \omega_2\f{\p}{\p p_i^2}+ \omega_3\f{\p}{\p p_i^3}  \right)\delta^{d-1}({\bf p}_1+\,{\bf p}_2+\,{\bf p}_3) F_1\non\\
&=&\frac{\partial}{\partial P^i} \delta^{d-1}(\textbf{P}) (\omega_1+\omega_2+\omega_3)\delta(\omega_1+\omega_2+\omega_3)F_1\nonumber\\
&&+
\delta(\omega_1+\omega_2+\omega_3)\delta^{d-1}({\bf p}_1+{\bf p}_2+{\bf p}_3)\left[\omega_1\left( \frac{\partial}{\partial p_1^i}-{\frac{\partial}{\partial p_3^i}}\right)
 +\omega_2\left(\frac{\partial}{\partial p_2^i}-{\frac{\partial}{\partial p_3^i}}\right)\right]F_1(\omega_1,\,{\bf p}_1,\,\omega_2,\,{\bf p}_2)\non\\
&=&0+\delta(\omega_1+\omega_2+\omega_3)\delta^{d-1}({\bf p}_1+{\bf p}_2+{\bf p}_3)\left[\omega_1 \frac{\partial}{\partial p_1^i}
 +\omega_2\frac{\partial}{\partial p_2^i}\right]F_1(\omega_1,\,{\bf p}_1,\,\omega_2,\,{\bf p}_2)
 \end{eqnarray}
In going from the first to second line, we have used the identity \eqref{a88}. In going from the second to third line, we have used the fact that $F_1$ can be made independent of $\textbf{p}_3^i$ (due to the delta function) and hence its' derivative with respect to $\textbf{p}_3^i$ vanishes. The above equation now implies
 \begin{eqnarray}
 \frac{\partial}{\partial p_1^i}F_1(\omega_1,\,{\bf p}_1,\,\omega_2,\,{\bf p}_2)\;\;=\;\;0\;\;=\;\; \frac{\partial}{\partial p_2^i}F_1(\omega_1,\,{\bf p}_1,\,\omega_2,\,{\bf p}_2)\;\;\Rightarrow\;\; F_1\equiv F_1(\omega_1,\omega_2)
\end{eqnarray}
The $\mathcal{B}_i$ Ward identity on $F_2$ term is trivially satisfied due to the presence of the product of delta function $\delta(\omega_1)\delta(\omega_2)\delta(\omega_3)$. On the $F_3$ term, we get 
\begin{eqnarray}
0&=&\delta(\omega_1)\delta(\omega_2+\omega_3)  \left( \omega_2\f{\p}{\p p_i^2}+ \omega_3\f{\p}{\p p_i^3}  \right)\delta^{d-1}({\bf p}_1+\,{\bf p}_2+\,{\bf p}_3) F_3\non\\
&=&\delta(\omega_1)\,\delta(\omega_2+\omega_3 ) \delta^{d-1} (\textbf{p}_1+\textbf{p}_2+\textbf{p}_3) \omega_2\left( \frac{\partial}{\partial p_2^i}-\frac{\partial}{\partial p_3^i}\right)F_3({\bf p}_1;\,\omega_2,\,{\bf p}_2;\,\omega_3,\,{\bf p}_3)\non
\end{eqnarray}
The above equation will be satisfied provided
\begin{eqnarray}
\frac{\partial}{\partial p_2^i}F_3({\bf p}_1;\,\omega_2,\,{\bf p}_2;\,\omega_3,\,{\bf p}_3)=\frac{\partial}{\partial p_3^i}F_3({\bf p}_1;\,\omega_2,\,{\bf p}_2;\,\omega_3,\,{\bf p}_3)
\ee
This gives
\be
F_3({\bf p}_1;\,\omega_2,\,{\bf p}_2;\,\omega_3,\,{\bf p}_3)=F_3(\omega_2,\,{\bf p}_2+{\bf p}_3)\equiv  F_3(\omega_2,\,{\bf p}_1)\equiv F_3(\omega_3,\,{\bf p}_1)
\end{eqnarray}
where we have discarded constant contributions, independent of 
$\mathbf{p}_2$ and $\mathbf{p}_3$, that carry no dynamical information.
In writing the above expression, we have implicitly made use of the delta functions $\delta(\omega_1)\,\delta(\omega_2+\omega_3 ) \delta^{d-1} (\textbf{p}_1+\textbf{p}_2+\textbf{p}_3)$ which multiplies $F_3$ term. The analysis for $F_4$ and $F_5$ follows exactly as for $F_3$ implying
\be
F_4= F_4(\omega_1,\,{\bf p}_2)\quad;\qquad F_5 = F_5(\omega_1,\,{\bf p}_3)
\end{eqnarray}
Thus, the $\mathcal{B}_i$ generator restricts the correlator to have the form
\begin{eqnarray}
&&\hspace*{-.5in}\mathcal{A}_3(\omega_1,\,{\bf p}_1;\,\omega_2,\,{\bf p}_2,\,\omega_3,\,{\bf p_3})\non\\[.2cm]
&=&\delta^{d-1}({\bf p}_1+{\bf p}_2+{\bf p}_3)\Bigl[ \delta(\omega_1+\omega_2+\omega_3)F_1(\omega_1,\,\omega_2)+\delta(\omega_1)\delta(\omega_2)\delta(\omega_3)F_2({\bf p}_1,\,{\bf p}_2)\nonumber\\[.2cm]
&&
+\delta(\omega_1)\delta(\omega_2+\omega_3)F_3(\omega_2,\,{\bf p}_1)+\delta(\omega_2)\delta(\omega_1+\omega_3)F_4(\omega_1,\,{\bf p}_2)+\delta(\omega_3)\delta(\omega_1+\omega_2)F_5(\omega_1,\textbf{p}_3)\Bigl]\non
\end{eqnarray}

\subsubsection*{$\mathcal{K}$ Ward identity}

The action of ${\cal K}$ generator on the $F_1$ term is given by 
\begin{eqnarray}
0&=&\delta(\omega_1+\omega_2+\omega_3)F_1(\omega_1,\omega_2)\sum_{a=1}^3 
\omega_a\frac{\partial^2}{\partial p_a^i\partial p_{ia}}\delta^{d-1}(\textbf{P})\non\\
&=&\delta(\omega_1+\omega_2+\omega_3) F_1(\omega_1,\omega_2)\sum_{a=1}^3 \omega_a\frac{\partial^2}{\partial P^i\partial P_{i}}\delta^{d-1}(\textbf{P)}
\end{eqnarray}
The last equation is trivially satisfied. The term involving $F_2$ is also trivially annihilated by $\mathcal{K}$ generator due to the delta function over each $p_a$. For $F_3$, we have 
\begin{eqnarray}
&&\hspace*{-.2in} \sum_{a=1}^3 
\omega_a\frac{\partial^2}{\partial p_a^i\partial p_{ia}}\delta^{d-1}({\bf p}_1+{\bf p}_2+{\bf p}_3)\delta(\omega_1)\delta(\omega_2+\omega_3)F_3({\omega}_2,\,{\bf p}_1)\nonumber\\
&=&\delta(\omega_1)\delta(\omega_2+\omega_3)F_3({\omega}_2,\,{\bf p}_1)\, \omega_2\,{\left( \frac{\partial^2}{\partial P^i\partial P_{i}}-\frac{\partial^2}{\partial P^i\partial P_{i}}\right)\delta^{d-1}(\textbf{P})}\nonumber\\
&&+\delta(\omega_1)\delta(\omega_2+\omega_3)\delta^{d-1}({\bf p}_1+{\bf p}_2+{\bf p}_3)F_3({\omega}_2,\,{\bf p}_1) \omega_2\left( \frac{\partial^2}{\partial p_2^i\partial p_{i2}}-\frac{\partial^2}{\partial p_3^i\partial p_{i3}}\right)F_3({\omega}_2,\,{\bf p}_1)\nonumber
\end{eqnarray}
The first term in the RHS of above expression identically vanishes. The second term also vanishes if we note that we can write $F_3(\omega_2, \textbf{p}_1)=F_3(\omega_2, -\textbf{p}_2-\textbf{p}_3) $ due to the presence of delta function over spatial momenta. The analysis of the action of $\mathcal{K}$ generators on the $F_4$ and $F_5$ terms is exactly identical to the $F_3$ case. Thus, we see that $\mathcal{K}$ Ward identity does not impose any constraint on $F_i$.

\subsubsection*{$\mathcal{J}_{ij}$ Ward identity}

The $\mathcal{J}_{ij}$ Ward-identity imposes the $SO(d-1)$ invariance along $R^{d-1}$. It implies that the correlator can depend only upon the magnitude of the momenta along these directions. This can be shown explicitly by observing that the generator 
${\cal J}_{ij}$ commutes with 
$\delta^{(d-1)}(\mathbf{p}_1+\mathbf{p}_2+\mathbf{p}_3)$ 
and by using the identities given in equation \eqref{5.96}. Thus, denoting the magnitude of the momenta ${p}^i_a$ by $k_a$, the $\mathcal{J}_{ij}$ Ward identity implies 
\begin{eqnarray}
&&\hspace*{-.5in}\mathcal{A}_3(\omega_1,\,{\bf p}_1;\,\omega_2,\,{\bf p}_2,\,\omega_3,\,{\bf p_3})\non\\[.2cm]
&=&\delta^{d-1}({\bf p}_1+{\bf p}_2+{\bf p}_3)\Bigl[ \delta(\omega_1+\omega_2+\omega_3)F_1(\omega_1,\,\omega_2)+\delta(\omega_1)\delta(\omega_2)\delta(\omega_3)F_2({ k}_1,\,{k}_2)\nonumber\\[.2cm]
&&
+\delta(\omega_1)\delta(\omega_2+\omega_3)F_3(\omega_2,\,{k}_1)+\delta(\omega_2)\delta(\omega_1+\omega_3)F_4(\omega_1,\,{k}_2)+\delta(\omega_3)\delta(\omega_1+\omega_2)F_5(\omega_1,k_3)\Bigl]\non
\end{eqnarray}

\subsubsection*{Dilatation Ward identity}
Again, each term in the 3-point function must be annihilated by the dilatation generator. On the first term, it gives
\begin{eqnarray}
0&=&\delta^{d-1}({\bf{p}}_1+{\bf{p}}_2+{\bf{p}}_3)\delta(\omega_1+\omega_2+\omega_3)\left[ \Delta_t - 3\,d +1 +(d-1) -\sum_{a=1}^2 \omega_a\frac{\partial}{\partial \omega_a}\right]F_1(\omega_1,\omega_2)\nonumber
\end{eqnarray}
where we have defined $\Delta_t=\Delta_1+\Delta_2+\Delta_3$ and used the delta function identities \eqref{ident1}. 

\vspace*{.07in}The action of dilatation generator on $F_2$ gives
\begin{eqnarray}
0&=&\left(\Delta_t -3\,d\right)\prod_{a=1}^3 \delta(\omega_a) \delta^{d-1}({\bf p}_1+{\bf p}_2+{\bf p}_3)F_2-\delta^{d-1}({\bf p}_1+{\bf p}_2+{\bf p}_3)F_2\sum_{a=1}^3\omega_a\frac{\partial}{\partial \omega_a}\prod_{b=1}^3 \delta(\omega_b)\nonumber\\
&&-\prod_{b=1}^3 \delta(\omega_b)F_2\sum_{a=1}^3p_a^i\frac{\partial}{\partial p_a^i}\delta^{d-1}({\bf p}_1+{\bf p}_2+{\bf p}_3)-\prod_{b=1}^3 \delta(\omega_b) \delta^{d-1}({\bf p}_1+{\bf p}_2+{\bf p}_3)\sum_{a=1}^2p_a^i\frac{\partial}{\partial p_a^i}F_2\nonumber\\
&=&\delta(\omega_1)\delta(\omega_2)\delta(\omega_3) \delta^{d-1}({\bf p}_1+{\bf p}_2+{\bf p}_3)\left[\Delta_t -3\,d+ 3+(d-1) -\sum_{a=1}^3 k_a\frac{\partial}{\partial k_a}\right]F_2
\end{eqnarray}
where we have used the identity
\begin{eqnarray}
\sum_{a=1}^3\omega_a\frac{\partial}{\partial \omega_a}\prod_{b=1}^3 \delta(\omega_b)=\sum_{ a=1}^3\omega_a\frac{\partial}{\partial \omega_a} \delta(\omega_a) \prod_{b\neq a} \delta(\omega_b) =-\sum_{a=1}^3 \delta(\omega_a) \prod_{b\neq a} \delta(\omega_b)=-3 \prod_{a=1}^3\delta(\omega_a)
\end{eqnarray}
Next, the action of dilatation generator on $F_3$ gives
\begin{eqnarray}
0=\delta(\omega_1) \delta(\omega_2+\omega_3)\delta^{d-1}({\bf p}_1+{\bf p}_2+{\bf p}_3)\left[ \Delta_t - 2 d+ 1  -\omega_2\frac{\partial}{\partial \omega_2} -k_1\frac{\partial}{\partial k_1}\right] F_3(\omega_2,k_1) \label{339uyt}
\end{eqnarray}
where, we have used the identity
\begin{eqnarray}
\sum_{a=1}^3p_a^i\frac{\partial}{\partial p_a^i}F_3(\omega_2,\,k_1)
&=& k_1 \frac{\partial}{\partial k_1}F_3(\omega_2,k_1)
\end{eqnarray}
and 
\begin{eqnarray}
\sum_{a=1}^3 \omega_a\frac{\partial}{\partial \omega_a} \delta(\omega_1)\delta(\omega_2+\omega_3)&=& \delta(\omega_2+\omega_3){ \omega_1\frac{\partial}{\partial \omega_1} \delta(\omega_1)} +\delta(\omega_1){\sum_{a=2}^3\omega_a \frac{\partial}{\partial \omega_a}\delta(\omega_2+\omega_3)}\nonumber\\
&=& -\delta(\omega_2+\omega_3)\delta(\omega_1) -\delta(\omega_1)\delta(\omega_2+\omega_3)\nonumber\\
&=&- 2\delta(\omega_1) \delta(\omega_2+\omega_3)
\end{eqnarray}
The analysis for $F_4$ and $F_5$ is identical to the case of $F_3$. Hence, we can write the resulting condition for them by changing the momentum labels in \eqref{339uyt}. Collecting all the results, the dilatation Ward identity gives following differential equations for $F_i$ 
\begin{eqnarray}
\left[\Delta_t -2\,d -\sum_{a=1}^2 \omega_a\frac{\partial}{\partial \omega_a}\right]F_1(\omega_1,\omega_2)&=&0\label{6.42}\\[.2cm]
\left[\Delta_t -2d+2-\sum_{a=1}^3 k_a\frac{\partial}{\partial k_a}\right]F_2(k_1,k_2)&=&0\label{4450}\\[.2cm]
\left[ \Delta_t - 2 d+1 -\omega_2\frac{\partial}{\partial \omega_2} -k_1\frac{\partial}{\partial k_1}\right] F_3(\omega_2,k_1) &=&0\label{4455}\\[.2cm]
\left[ \Delta_t -  2 d+1 -\omega_1\frac{\partial}{\partial \omega_1} -k_2\frac{\partial}{\partial k_2}\right] F_4(\omega_1,k_2) &=&0\\[.3cm]
\left[ \Delta_t -  2 d+1 -\omega_1\frac{\partial}{\partial \omega_1} -k_3\frac{\partial}{\partial k_3}\right] F_5(\omega_1,k_3) &=&0
\end{eqnarray}

\subsubsection*{Carrollian special conformal Ward identity $\mathcal{K}_i$}
Finally, we consider the effect of the carrollian special conformal Ward identity $\mathcal{K}_i$ on the 3-point function. As before, the corresponding generator acts on each $F_i$ independently. For the $F_1$ term, we have (denoting $\textbf{P}={\bf p}_1+{\bf p}_2+{\bf p}_3 $ as before)
\begin{eqnarray}
0&=&\sum_{a=1}^3\left[ (\Delta_a-d) \frac{\partial}{\partial p_{ai}} -\omega_{a} \frac{\partial^2}{\partial p_a^i\partial \omega_{a}} -p_{aj} \frac{\partial^2}{\partial p_a^i\partial p_{aj}}+\frac{1}{2} p_{ai} \frac{\partial^2}{\partial p_a^j\partial p_{aj}}\right]\delta(\omega_1+\omega_2+\omega_3)\,\delta^{d-1}({\bf P})F_1\non\\
&=&\delta(\omega_1+\omega_2+\omega_3) \frac{\partial}{\partial P^i}\delta^{d-1}(\textbf{P}) \left[\Delta_t -2d-\sum_{a=1}^3 \omega_a\frac{\partial}{\partial \omega_a} \right]F_1(\omega_1,\omega_2)
\end{eqnarray}
where we have used identities \eqref{ident1}, \eqref{C.1} and \eqref{C.5} in going from the first to second line. The differential operator in the above expression is same as the one appearing in the dilatation Ward identity \eqref{6.42} on $F_1$. Hence, the Carrollian special conformal Ward identity does not impose any additional constraint on $F_1$. 

\vspace*{.07in}Next, the action of $\mathcal{K}_i$ generator on $F_2$ is given by 
\begin{eqnarray}
0&=&\sum_{a=1}^3\left[ (\Delta_a-d) \frac{\partial}{\partial p_{ai}} -\omega_{a} \frac{\partial^2}{\partial p_a^i\partial \omega_{a}} -p_{aj} \frac{\partial^2}{\partial p_a^i\partial p_{aj}}+\frac{1}{2} p_{ai} \frac{\partial^2}{\partial p_a^j\partial p_{aj}}\right]\,\delta^{d-1}({\bf P}) \prod_{b=1}^3\delta(\omega_b) \,F_2(k_1,\,k_2,\,k_3)\non\\[.3cm]
&=&  \prod_{b=1}^3\delta(\omega_b)  \frac{\partial}{\partial P^i}\delta^{d-1}({\bf P})\Bigg[\sum_{a=1}^3 (\Delta_a-d) +3+d -\sum_{a=1}^3 p_a^j\frac{\partial}{\partial p_a^j}-1\Bigg]F_2\nonumber\\
&&+ \prod_{b=1}^3\delta(\omega_b)  \frac{\partial}{\partial P^j}\delta^{d-1}({\bf P})\Bigg[-\sum_{a=1}^3\left( p_a^j\frac{\partial}{\partial p_a^i} - p_{ai} \frac{\partial}{\partial p_a^j}\right)\Bigg]F_2\nonumber\\
&&+  \prod_{b=1}^3\delta(\omega_b) \delta^{d-1}({\bf P})\Bigg[\sum_{a=1}^3(\Delta_a -d+1) \frac{\partial}{\partial p_a^i}-p_a^j\frac{\partial^2}{\partial p_a^i\partial p_a^j} +\frac{1}{2} p_{ai}\frac{\partial^2}{\partial p_a^j\partial p_{aj}}\Bigg]F_2
\end{eqnarray} 
The first line of the above expression vanishes by the dilatation Ward identity \eqref{4450} and the second line vanishes by $\mathcal{J}_{ij}$ Ward identity. The 3rd line is the standard special conformal Ward identity in an Euclidean CFT in $(d-1)$ dimensions (see for example \cite{Coriano:2020ees,Bzowski:2013sza}). Expressing the operator in the last line in terms of the derivatives with respect to the magnitudes of momenta, we get the equation satisfied by $F_2$ to be 
\begin{eqnarray}
\sum_{a=1}^3 p^i_{a} \left[ \left( \Delta_a -\frac{d}{2} \right) \frac{1}{k_a}\frac{\partial}{\partial k_a} -\frac{1}{2} \frac{\partial^2}{\partial k_a^2}\right]F_2\;\;=\;\;0
\end{eqnarray}
Next, we consider the action of the $\mathcal{K}_i$ generator on the $F_3$ term. We shall evaluate the action of the different terms in $\mathcal{K}_i$ generator one by one. The first term is given by
\begin{eqnarray}
&&\sum_{a=1}^3 \left( \Delta_a -d \right)\,\frac{\partial}{\partial p_a^i} \delta(\omega_1)\,\delta(\omega_2+\omega_3) \delta^{d-1}(\textbf{P})\, F_3(\omega_2, k_1)\nonumber\\
&&= \delta(\omega_1) \delta(\omega_2+\omega_3)\Bigg\{\left(\Delta_t-3\,d\right)F_3\frac{\partial}{\partial P^i}\delta^{d-1}(\textbf{P}) {+ (\Delta_2+\Delta_3-2d)\delta^{d-1}(\textbf{P})\f{p_2^i+p_3^i}{k_1}\frac{\partial}{\partial k_1} F_3(\omega_2,\,k_1)}\Bigg\}\non
\end{eqnarray} 
The second term gives
\begin{eqnarray}
&&\hspace*{-.4in}-\sum_{a=1}^3 \omega_a\frac{\partial^2}{\partial \omega_a\partial p_a^i} \delta(\omega_1) \delta(\omega_2+\omega_3)\, \delta^{d-1}(\textbf{P})\, F_3(\omega_2,\,k_1)\nonumber\\
&=&\delta(\omega_1) \,\delta(\omega_2+\omega_3)\left[\frac{\partial}{\partial P^i}\delta^{d-1}(\textbf{P})\left(2 -\omega_2\frac{\partial}{\partial \omega_2}\right)F_3 +{  \delta^{d-1}(\textbf{P}) (p_2^i+p_3^i)\biggl(\f{1}{k_1}\frac{\partial}{\partial k_1}-\f{\omega_2}{k_1}\frac{\partial^2}{\partial k_1\p \omega_2}\biggl) F_3}\right]\non
\end{eqnarray}
where we have used the first identity in \eqref{ident1} and used $\f{\p}{\p \omega_3}F_3=0$. 

The 3rd term of the special conformal generator acting on $F_3$ gives
\begin{eqnarray}
&&- \sum_{a=1}^3 p_{aj} \frac{\partial^2}{\partial p_{aj}\partial p_a^i}\delta(\omega_1)\delta(\omega_2+\omega_3)\,\delta^{d-1}(\textbf{P}) F_3(\omega_2,\,{ k}_1)\nonumber\\
&=&-\delta(\omega_1)\delta(\omega_2+\omega_3)\,\Bigg[\frac{\partial}{\partial P^i}\delta^{d-1}(\textbf{P})\Bigl(-d +k_{1}\frac{\partial}{\partial k_{1}}\Bigl)F_3+\frac{\partial}{\partial P^j}\delta^{d-1}(\textbf{P})\f{(p_2+p_3)^i(p_2+p_3)^j}{k_1}\frac{\partial}{\partial k_1}F_3\non\\
&&+\delta^{d-1}(\textbf{P})(p_{2}+p_{3})_i \,\frac{\partial^2}{\partial k_1^2}F_3\Bigg]\non
\end{eqnarray}
where we have used
\be
\sum_{a=1}^3p_a^j\f{\p F_3}{\p p_a^i}&=& (p_2^j+p_3^j)\f{p_2^i+p_3^i}{k_1} \f{\p F_3}{\p k_1}\quad;\qquad
\sum_{a=1}^3p_a^j\f{\p F_3}{\p p_a^j}=k_1\f{\p F_3}{\p k_1}
\ee
\begin{eqnarray}
\mbox{and},\qquad\qquad \sum_{a=1}^3 p_a^j\frac{\partial^2}{\partial p_a^j\partial p_a^i} F_3(p_2, \,k_1)
 =({ p}_2+{ p}_3)_i\,\frac{\partial^2}{\partial k_1}F_3(p_2,\,k_1)
\end{eqnarray}
The 4th term of the $\mathcal{K}_i$ generator acting on $F_3$ gives
  \begin{eqnarray}
&&\frac{1}{2}\sum_{a=1}^3 p_{ai}\frac{\partial^2}{\partial p_a^j \partial p_{aj}}\delta(\omega_1)\delta(\omega_2+\omega_3)\,\delta^{d-1}(\textbf{P}) F_3(\omega_2,\,k_1)\nonumber\\
&=&\frac{1}{2}\delta(\omega_1)\delta(\omega_2+\omega_3)\,\Bigg[-2F_3 \frac{\partial}{\partial P^i} \delta^{d-1}(\textbf{P})+2\frac{\partial}{\partial P^j} \delta^{d-1}(\textbf{P})   \frac{({ p}_2+{ p}_3)_i\,({ p}_2+{p}_3)^j}{k_1}\frac{\partial}{\partial k_1}F_3\non\\
&&+\delta^{d-1}(\textbf{P})({ p}_2+{ p_3})_i  \Big(\frac{d-2}{k_1} \frac{\partial}{\partial k_1} +\frac{\partial^2}{\partial k_1^2}\Big)F_3(\omega_2,k_1)\Bigg]\non
\end{eqnarray} 
where we have used 
\begin{eqnarray}
\sum_{a=1}^3 p_{ai}\frac{\partial}{\partial p_{aj}}F_3(\omega_2,k_1)&=& 
   \frac{({ p}_2+{ p}_3)_i\,({ p}_2+{ p}_3)^j}{k_1}\frac{\partial}{\partial k_1}F_3(\omega_2,k_1)\non\\
  \sum_{a=1}^3 p_{ai} \frac{\partial^2}{\partial p_a^j\partial p_{aj}}F_3(\omega_2, \,k_1) 
  &=&({ p}_2+{ p_3})_i \Big[\frac{d-2}{k_1} \frac{\partial}{\partial k_1} +\frac{\partial^2}{\partial k_1^2}\Big]F_3(\omega_2,k_1)
  \end{eqnarray}
Combining all the terms, the action of Carrollian special conformal Ward identity on $F_3$ gives
\begin{eqnarray}
&&\delta(\omega_1)\delta(\omega_2+\omega_3) \Biggl[\frac{\partial}{\partial P^i}\delta^{d-1}(\textbf{P}) \Bigg(\Delta_t - 2d + 1 -\omega_2\frac{\partial}{\partial \omega_2} -k_{1}\frac{\partial}{\partial k_{1}}\Bigg)F_3(\omega_2,\,k_1)\nonumber\\
&&+\delta^{d-1}(\textbf{P})({ p}_2+{ p_3})_i  \Bigg(\Bigl(\Delta_2+\Delta_3-\f{3d}{2} \Bigl)\f{1}{k_1}\frac{\partial}{\partial k_1}-\f{\omega_2}{k_1} \frac{\partial^2}{\partial \omega_2\partial k_1} -\frac{1}{2}\frac{\partial^2}{\partial k_1^2}\bigg)F_3(\omega_2,\,k_1)\Bigg]=0
\end{eqnarray}
The first line vanishes by the dilatation Ward identity \eqref{4455}. The action of $\mathcal{K}_i$ on the remaining $F_4$ and $F_5$ terms can be similarly obtained. Thus, collecting all the results, the Carrollian special conformal Ward identity $\mathcal{K}_i$ imposes the following constraints 
\begin{eqnarray}
\sum_{a=1}^3 p^i_{a} \left[ \left( \Delta_a -\frac{d}{2} \right) \frac{1}{k_a}\frac{\partial}{\partial k_a} -\frac{1}{2} \frac{\partial^2}{\partial k_a^2}\right]F_2(k_1,k_2)&=&0\\[.2cm]
\Bigg[\left(\Delta_2+\Delta_3-\f{3d}{2} \right)\f{1}{k_1}\frac{\partial}{\partial k_1}-\f{\omega_2}{k_1} \frac{\partial^2}{\partial \omega_2\partial k_1} -\frac{1}{2}\frac{\partial^2}{\partial k_1^2}\bigg]F_3(\omega_2,\,k_1)&=&0\label{f3sct}\\[.2cm]
\Bigg[\left(\Delta_1+\Delta_3-\f{3d}{2} \right)\f{1}{k_2}\frac{\partial}{\partial k_2}-\f{\omega_1}{k_2} \frac{\partial^2}{\partial \omega_1\partial k_2} -\frac{1}{2}\frac{\partial^2}{\partial k_2^2}\bigg]F_4(\omega_1,\,k_2)&=&0\\[.2cm]
\Bigg[\left(\Delta_1+\Delta_2-\f{3d}{2} \right)\f{1}{k_3}\frac{\partial}{\partial k_3}-\f{\omega_1}{k_3} \frac{\partial^2}{\partial \omega_1\partial k_3} -\frac{1}{2}\frac{\partial^2}{\partial k_3^2}\bigg]F_5(\omega_1,\,k_3)&=&0
\end{eqnarray}

\subsection{Solutions of Ward identities}
\subsubsection{2-point function}
The solution to the differential equations \eqref{5.87v2} and \eqref{2nd},
which constrain the two-point function, has already been provided in
equation \eqref{2points}.  
For completeness, we recall here its full expression, giving the general
form of the Carrollian two-point function in momentum space 
\begin{eqnarray}
&&\Bigl\langle \mathcal{O}_{\Delta_1}(\omega_1, {\bf p}_2)\,  \mathcal{O}_{\Delta_2}(\omega_2,\, {\bf p}_2)\Bigl\rangle=\delta^{d-1}({\bf p}_1+{\bf p}_2)\biggl[ \delta(\omega_1+\omega_2)\; b_1\;\omega_1^{\Delta_1+\Delta_2 -d}+\delta(\omega_1)\delta(\omega_2)\; b_2\;{|\bf p_1|}^{2\Delta_1 -d+1}\biggl]\non\\
&&\label{3.55}
\end{eqnarray}
where $b_1$ and $b_2$ are arbitrary constants which may depend upon $d$ and $\Delta_i$. 

\subsubsection{3-point function}
A generic $d$ dimensional Carrollian CFT 3-point function of scalar operators in the momentum space takes the form 
\begin{eqnarray}
&&\hspace*{-.5in}\mathcal{A}_3(\omega_1,\,{\bf p}_1;\,\omega_2,\,{\bf p}_2,\,\omega_3,\,{\bf p_3})\non\\[.2cm]
&=&\delta^{d-1}({\bf p}_1+{\bf p}_2+{\bf p}_3)\Bigl[ \delta(\omega_1+\omega_2+\omega_3)F_1(\omega_1,\,\omega_2)+\delta(\omega_1)\delta(\omega_2)\delta(\omega_3)F_2({ k}_1,\,{k}_2)\nonumber\\[.2cm]
&&
+\delta(\omega_1)\delta(\omega_2+\omega_3)F_3(\omega_2,\,{k}_1)+\delta(\omega_2)\delta(\omega_1+\omega_3)F_4(\omega_1,\,{k}_2)+\delta(\omega_3)\delta(\omega_1+\omega_2)F_5(\omega_1,k_3)\Bigl]\non
\end{eqnarray}
We consider the general solutions of each $F_i$ one by one. Starting with $F_1$, we note that it is the solution of equation \eqref{6.42}. Now, the equation \eqref{6.42} imposes the scale invariance and only tells us that $F_1$ is a homogeneous function of degree $ \Delta_t-2d$.  Hence, the most general solution must take the form
\be
F_1(\omega_1,\omega_2) &=& \omega_1^{\Delta_t-2d}\;f\left(\f{\omega_2}{\omega_1}\right)\label{3.57}
\ee
where $f$ can be any arbitrary function\footnote{ Due to the delta function over the ``energies'', 
$\omega_3 = -\omega_2 - \omega_1$, there is only one independent
energy ratio. Indeed, $
\frac{\omega_3}{\omega_1} = -1 - \frac{\omega_2}{\omega_1}\,
$.}. As in position space, in a concrete theory this function is expected to be fixed, or at least restricted, by additional dynamical or consistency conditions beyond the Ward identities.

\vspace*{.07in}Next, we consider $F_2$. It is given by the solutions of the differential equations \eqref{4450} and \eqref{f3sct}. These two equations are recognized to be the momentum space dilatation and special conformal Ward identities of 3 point function involving scalar operators in $d-1$ dimensional Euclidean CFT \cite{Coriano:2020ees,Bzowski:2013sza}. Hence, its solution can be immediately written down in terms of the triple K integrals  
\be
F_2(k_1,k_2,k_3)&=&c_2\int_0^\infty dy\; y^{\f{d-3}{2}}\prod_{a=1}^3k_a^{\Delta_a-\f{d-1}{2}}K_{\Delta_a-\f{d-1}{2}}(k_ay)\label{3.58}
\ee
For special values of $\Delta_a$, the above expression diverges and needs regularization. The analysis for this follows exactly the case of standard CFTs. For more details on this, see, e.g., \cite{1510.08442}. 

\vspace*{.07in}Next, $F_3$ is given by the solutions of \eqref{4455} and \eqref{f3sct}. For generic values of $\Delta_a$, the solution is given by 
\be
F_3(\omega_2,k_1) \;=\; c_3\; (k_1)^{2\Delta_1-d+1}(\omega_2)^{\Delta_2+\Delta_3-\Delta_1-d} \;+\; d_3\; (\omega_2)^{\Delta_1+\Delta_2+\Delta_3-2d+1}\label{4110fg}
\ee
In the special case $\Delta_1= \f{d-1}{2}$, the two solutions above coincide and an additional solution involving logarithm emerges
\be
F_3(\omega_2,k_1)&=& e_3\;(\omega_2)^{\Delta_1+\Delta_2+\Delta_3-2d+1} \ln\left(\f{k_1}{\omega_2} \right)\quad;\qquad \Delta_1 = \f{d-1}{2}\label{4.111}
\ee
Note that the above logarithmic solution is valid for arbitrary values of $\Delta_2$ and $\Delta_3$. It only requires $\Delta_1$ to be specific.  

Finally, we consider the solutions of $F_4$ and $F_5$. Since the differential equations for these are similar to $F_3$, their solutions can be written down immediately by permuting the indices. Hence, we have 
\be
F_4(\omega_1,k_2) &=& c_4\; (k_2)^{2\Delta_2-d+1}(\omega_1)^{\Delta_1+\Delta_3-\Delta_2-d} \;+\;d_4\;(\omega_1)^{\Delta_1+\Delta_2+\Delta_3-2d+1}\;  \non\\[.3cm]
F_5(\omega_1,k_3) &=& c_5\; (k_3)^{2\Delta_3-d+1}(\omega_1)^{\Delta_1+\Delta_2-\Delta_3-d}+ \;d_5\; (\omega_1)^{\Delta_1+\Delta_2+\Delta_3-2d+1}
\ee
Again for $\Delta_2$ and $\Delta_3$ equal to $\f{d-1}{2}$, the two solutions coincide and we have additional solutions
\be
F_4(\omega_1,k_2)&=& e_4\;(\omega_1)^{\Delta_1+\Delta_2+\Delta_3-2d+1} \ln\left(\f{k_2}{\omega_1} \right)\quad;\qquad \Delta_2 = \f{d-1}{2}\non\\[.3cm]
F_5(\omega_1,k_3)&=& e_5\;(\omega_1)^{\Delta_1+\Delta_2+\Delta_3-2d+1} \ln\left(\f{k_3}{\omega_1} \right)\quad;\qquad \Delta_3 = \f{d-1}{2}\ee

A final remark concerns the origin and interpretation of the logarithmic terms that appear when solving the Ward identities in momentum space. These logarithmic solutions do not correspond to additional branches of the correlator in position space. Instead, they arise at special loci in the parameter space $\{\Delta_a, d\}$ where the  transform of the position-space correlators becomes singular and requires regularization. This can be seen explicitly by considering the  transform of the $|\mathbf{x}_{12}|^{-2\Delta_1}$ factor appearing in equation \eqref{3.54}
\begin{equation}
\int d^{d-1}x \, 
\frac{e^{i {\bf p}_1\cdot \mathbf{x}_{12}}}{|\mathbf{x}_{12}|^{2\Delta_1}}
=
\frac{{2^{d-1-2\Delta_1}}\,\pi^{\frac{d-1}{2}}}{\Gamma(\Delta_1)}\,
\Gamma\!\left(\frac{d-1-2\Delta_1}{2}\right)\,
k_1^{\,2\Delta_1-(d-1)} .
\end{equation}
The prefactor is singular when $\Delta_1=\tfrac{d-1}{2}$, i.e.\ when the exponent of $k_1^{2\Delta_1-(d-1)}$ reaches its critical value. In this limit,  we get (denoting $\epsilon = d-1-2\Delta_1$)
\begin{equation}
\lim_{\epsilon\rightarrow0}\; \Gamma\left(\frac{\epsilon}{2}\right)\,k_1^{-\epsilon}
= \frac{2}{\epsilon} -2 \log k_1 + O(\epsilon)
\end{equation}
The above equation shows that the regularized  transform acquires a $\log k_1$ contribution, as in equation \eqref{4.111}.

A completely analogous mechanism affects the  transform of the $u_{23}$-dependent factor in equation \eqref{3.54}. A logarithmic dependence on $\omega_2$ as in \eqref{4.111} emerges only when an additional constraint relating the parameters $\Delta_a$ are satisfied. These constraints do not directly emerge from solving the Ward identities in 
position space. This mismatch indicates that the regularization of the  transform must be handled with care. These issues deserve further investigation and we leave these for a future work. 

We also remark that the logarithmic term in \eqref{4.111} are divergent as $k_1\rightarrow 0$ or $ \infty$ for fixed values of $\omega_2$. These singularities can be regulated by introducing a cutoff, and in this sense the variable $\omega_2$ naturally plays the role of an effective IR/UV regulator as we span the spatial momentum space $\mathbf{p}_1$. Similarly, the logarithmic term also diverges as $\omega_2\rightarrow 0$ or $ \infty$ for fixed values of $k_1$. These singularities can be regulated by introducing a cutoff, and hence the variable $k_1$ can be thought as an effective IR/UV regulator as we span the Carrollian energy space $\omega_2$.

\section{Carrollian limit of CFT correlators}
\label{sec5:inonu}

The Carrollian limit of CFT correlators has been considered in position space for some branches \cite{2406.19343,2508.06602}. However, as emphasized earlier, the momentum space provides a natural language for the flat-space amplitudes. Therefore, it is useful to have a control over the Carrollian limit of CFT correlators directly in the momentum space along with the position space. 

\vspace*{.07in}In this section, we shall consider the Carrollian limit of the standard CFT correlators.  We shall show how, in this 
 limit, the CFT 2 and 3 point functions give rise to 2 and 3 point Carrollian conformal correlators. We will keep the conformal dimensions of CFT-operators  arbitrary;  
under this assumption, we do not expect to reproduce, in the $c\rightarrow 0$ limit,  the logarithmic behavior discussed in the previous section.  The main strategy would be to explicitly keep track of the factors of speed of light $c$ in the correlators and send it to zero. By rescaling the correlators with different powers of $c$, we shall obtain different branches of  2 and 3 point Carrollian conformal correlators.

\vspace*{.07in}In the CFT correlators, the speed of light $c$ always appears with time coordinate, so the conventional notion of time ceases to apply when we send $c$ to zero. As a consequence, the correlators lose their standard causal interpretation for spatially separated points. Due to this, the Carrollian limit of Wightman and time ordered correlators requires seperate treatments. We shall consider both Wightman as well as time-ordered CFT correlators in our analysis.

\subsection{ 2-point function}

 We begin this section by reviewing the Carrollian limit of the two-point function in position space. This limit amounts to taking $
(c\,u,\,{\bf x}) \;\rightarrow\; (0,\,{\bf x})$ 
which is equivalent to considering the ultra-relativistic limit of vanishing light velocity, $c \to 0$. In this limit, as discussed previously, the conformal group $SO(2,d)$ of Minkowski space $\mathbb{M}^{1,d-1}$ reduces to the massless Poincaré group $ISO(1,d)$, which acts as the group of conformal isometries of the null plane ${\cal J}=\mathbb{R}\times\mathbb{R}^{d-1}$. The natural starting point for studying the Carrollian limit are  therefore the CFT correlation functions in Lorentzian signature.  

\vspace*{.07in}We start by reviewing the Carrollian limit of time ordered 2-point function in the position space following \cite{2406.19343,2508.06602}. As reviewed in appendix \ref{sec:BLor}, the Lorentzian time-ordered CFT two-point function takes the form (see, e.g.,~\cite{2406.19343})
\begin{eqnarray}
\left\langle T\Bigl( {\cal O}(t_1,{\bf x}_1)\,{\cal O}(t_2,{\bf x}_2)\Bigl)\right\rangle
= \frac{C_2(\Delta)}{\left[-c^2 t_{12}^2+|{\bf x}_{12}|^2+i\epsilon \right]^\Delta}\,.
\end{eqnarray}
In the Carrollian limit $c\rightarrow0$, two distinct behaviors of the two-point function emerge. For ${\bf x}_{12}\neq 0$, the two-point simply reduces to $|{\bf x}_{12}|^{-2\Delta}$ in the limit $c\rightarrow0$. The case ${\bf x}_{12}=0$ is instead singular. Combining both situations, we can schematically write \cite{2508.06602}
\begin{eqnarray}
\left\langle T\Bigl( {\cal O}(t_1,{\bf x}_1)\,{\cal O}(t_2,{\bf x}_2)\Bigl)\right\rangle
\xrightarrow[c\to 0]{} \;
\frac{C_2(\Delta)}{|{\bf x}_{12}|^{2\Delta}}
\;+\;\frac{1}{c^\alpha}\,g(\Delta,\,t)\,\delta^{(d-1)}({\bf x}_{12})\,.
\end{eqnarray}
where the function $g(\Delta,t)$ and the constant $\alpha$ need to be fixed. The first and second terms in the RHS of above expression correspond respectively to the \emph{magnetic} and \emph{electric} branches of the Carrollian two-point function in position space.
The function $g(\Delta,t)$ appearing in the electric branch is determined by multiplying the time-ordered two-point function by $c^\alpha$ (which in the limit $c\rightarrow0$ suppresses the magnetic contribution) and then integrating over the spatial coordinates for removing the delta-function.
The result of the integration gives \cite{2508.06602}
\begin{eqnarray}
g(\Delta,\,t_{12})=  \frac{\pi^{\frac{d-1}{2}}\,\Gamma[\Delta-\frac{d-1}{2}]}{\Gamma[\Delta]}   \frac{C_2(\Delta)}{(-t_{12}^2+i\epsilon)^{\Delta-\frac{d-1}{2}}}~~;~~\alpha= 2\Delta-d+1
\end{eqnarray}
We shall follow a similar procedure for taking the Carrollian limit of 2-point functions in the momentum space. We start with the Wightman 2-point function.

\subsubsection{Wightman}

The Wightman (Lorentzian) two–point function in momentum space for a scalar primary operator
\(\mathcal O\) of conformal dimension \(\Delta\) is given by (see, e.g., \cite{1807.07003})
\begin{equation}
G_2^W(p)\;\equiv\;\langle\!\langle \mathcal O(p)\,\mathcal O(-p)\rangle\!\rangle
=\frac{\pi^{\frac{d}{2}+1}\,  \Theta\!\big(p^0 - |\textbf{p}|\big)  }{2^{\,2\Delta - d - 1}\,\Gamma(\Delta)\,
\Gamma\!\big(\Delta - \tfrac{d}{2} + 1\big)}\;
\;\frac{1}{c^{2\Delta -d} (\omega^2- c^2\,|{\bf p}|^2)^{\frac{d}{2} - \Delta}}
\label{eq:GW}
\end{equation}
where \(\Theta\) is the Heaviside step function (\(\Theta(x)=1\) for \(x>0\), \(\Theta(x)=0\) for \(x\leq 0\)), and we have written
\begin{equation}
-p^2= (p^0)^2-|{\bf p}|^2~~;~~ \;p^0=\; \frac{\omega}{c}.
\end{equation}
The Heaviside factor enforces support on the future timelike (or null) cone,
\(p^0=\frac{\omega}{c} \ge |\textbf{p}| \ge 0\).
In the Carrollian limit, with fixed $\omega$,
\begin{equation}
\lim_{c\to 0^+}\,\Theta\!\left(\frac{\omega}{c}-|\textbf{p}|\right)\;\equiv\;\Theta(\omega)=1.
\end{equation}
Moreover, taking the limit \(\omega\to 0^{+}\) (from the positive side) one still preserves the condition 
\(\Theta(\omega)\to 1\). As in position space, the Carrollian limit of the two–point function can be expressed as
\begin{eqnarray}
G_2^W(p) \xrightarrow{c \rightarrow 0} \frac{\pi^{\frac{d}{2}+1} }{2^{2\Delta -d-1}\,\Gamma[\Delta -\frac{d}{2} +1]\,\Gamma(\Delta)}\frac{1}{c^{2\Delta-d}}\Bigg[\,\omega^{2\Delta-d}+\frac{1}{c^\alpha}\,\delta(\omega) g(|{\bf p}|)\Bigg]~~;~~
\end{eqnarray}
The first term captures the regular contribution for non-vanishing energy, while the second term, proportional to the delta function, accounts for the singular contribution originating from the threshold region $\omega\simeq c|{\bf p}|\rightarrow 0$. This contact term arises only for  $\Delta<\frac{d}{2}$ where the two points develop a divergence in the Carrollian limit.
At this stage, there are two possible Carrollian limits depending on the value of the scaling exponent $\alpha$. 
The function $ g(|{\bf p}|)$ is fixed by integrating\footnote{Here, we have used the identity $\int_{c|{\bf p}|}^\infty d\omega\,(\omega^2-c^2|{\bf p}|^2)^{\Delta-\frac{d}{2}}= (2 c |{\bf p}|)^{2\Delta -d+1} B(\Delta-\frac{d}{2}+1,\,-2\Delta +d-1)$.}
\begin{eqnarray}
 \lim_{c\rightarrow 0}c^{2\Delta -d+\alpha}\,\int_{-\infty}^{\infty} d\omega\, \, {G_2^W({p})}= \frac{\pi^{\frac{d}{2}+1}\,\Gamma[d-2\Delta-1]}{4\,\Gamma[-\Delta +\frac{d}{2} +1]\,\Gamma(\Delta)}\,\frac{c^\alpha}{c^{d-2\Delta -1}}\, |{\bf p}|^{2\Delta-d+1} 
\end{eqnarray}
that fixes $\alpha=d-2\Delta-1$ and gives
\begin{eqnarray}
c^\beta\,G_2^W(p) \xrightarrow{c \rightarrow 0}\left\{\begin{array}{ll}  
\frac{\pi^{\frac{d}{2}+1} }{2^{2\Delta -d-1}\,\Gamma[\Delta -\frac{d}{2} +1]\,\Gamma(\Delta)}\,\omega^{2\Delta-d}& \beta= 2\Delta-d\\[.4cm]
\frac{\pi^{\frac{d}{2}+1}\,\Gamma[d-2\Delta -1]}{4\, \Gamma[\frac{d}{2} -\Delta+1]\,\Gamma[\Delta]}\,|{\bf p}|^{2\Delta -d+1}&\beta= -1\end{array}\right.\label{carlw}
\end{eqnarray}
This expression is consistent with Eq. \eqref{3.55}, evaluated in the electric branch, with $\Delta_1=\Delta_2\equiv\Delta$.
\begin{figure}[t!]
 	\centering
 	\begin{tikzpicture}[scale=1]
 	\draw[->] (-5, 0) -- (5, 0) node[below right] {${\rm Re} (s)$};
 	\draw (0,-2) -- (0,2) node[left] {${\rm Im}( s)$};
 	\draw[very thick, black, decorate, decoration={snake,amplitude=0.6mm, segment length=2mm} ] (-5, 0.5) -- (0,0.5);	
    \node at (2,-1){$s=-\omega^2 +c^2|{\bf p}|^2$};
 	\fill (0,0.5) circle (2pt);
 	\node at (0.4, 0.5) {$i\,\epsilon$};
 	\end{tikzpicture}
 	\caption{The branch cut in the $s \equiv p^2c^2$ plane for the time ordered 2-point function in equation \eqref{5.116to}. }
\label{Fig1}
 \end{figure}

\subsubsection{Time ordered}
\label{Time-ordered 2point}
The time ordered two point CFT correlation function in the Lorentzian signature takes the form \cite{1807.07003}
\begin{eqnarray}
G_2^T(p)=-i\frac{\pi^{\frac{d}{2}}\,\Gamma[\frac{d}{2}-\Delta]}{2^{\Delta-d}\,\Gamma[\Delta]}\frac{1}{c^{2\Delta-d}(-\omega^2+c^2|{\bf p}|^2-i\epsilon)^{\frac{d}{2}-\Delta}} \quad;\qquad \epsilon >0\label{5.116to}   
\end{eqnarray}
Compared to the Wightman two-point function \eqref{eq:GW}, it differs only by an overall factor and the $i\epsilon$ prescription (see Fig.\ref{Fig1}). 

 The analysis is the same as for the Wightman 2-point function. Hence, we write
 \begin{eqnarray}
G_2^T(p) \xrightarrow{c \rightarrow 0} -i\frac{\pi^{\frac{d}{2}}\,\Gamma[\frac{d}{2}-\Delta]}{2^{\Delta-d}\,\Gamma[\Delta]}\,\frac{1}{c^{2\Delta-d}}\Bigg[\,\omega^{2\Delta-d}+\frac{1}{c^\alpha}\,\delta(\omega) \tilde g(|{\bf p}|)\Bigg].\label{5124ewr}
\end{eqnarray}
Again, to determine the function $\tilde g(|\textbf{p}|)$ and the coefficient $\alpha$, we can integrate both sides over $\omega$. The integration over the left side involves the integral
\begin{eqnarray}
\int_{-\infty}^\infty \frac{d\omega}{[(-\omega+c\,|{\bf p}|-i\epsilon)(\omega+c|{\bf p}| -i\epsilon)]^{\frac{d}{2} -\Delta}}\label{5123ert}
\end{eqnarray}
This can be evaluated directly in a brute force way carefully keeping track of the phase factors in the denominators. Alternatively, we can evaluate it by a contour method. Here we give the argument based on contour deformation while providing the details of direct evaluation in the appendix. Both approaches, of course, give the same result. In the contour method, we consider the integrand in \eqref{5123ert} evaluated along the contour shown in figure \ref{Fig2} in the complex $\omega$ plane. For arbitrary $\Delta$, the branch cuts start from the $\pm c|{\bf p}|\mp i\epsilon$ points and extend to $\pm \infty\mp i\epsilon$. 
\begin{figure}[t!]
	\centering
	\begin{tikzpicture}[scale=1]
	\draw[thick, Black,->] (-6, 0) -- (6, 0) node[below right] {${\rm Re}( \omega)$};
	\draw[thick, Black,->] (0,-1) -- (0,6) node[left] {${\rm Im}( \omega)$};

	\draw[thin, red!80, decorate, decoration={snake,amplitude=0.6mm, segment length=2mm} ] (-5, 0.5) -- (-1.5,0.5);
	\draw[thin, red, decorate, decoration={snake,amplitude=0.6mm, segment length=2mm} ] (5, -0.5) -- (1.5,-0.5);
	\draw[thick, blue] 
	(5,0.0) arc (-0.5:172.5:5) node[midway, above] {$\Gamma$};
	\draw[thick, blue] 
	(-5,0.3) arc (176.4:180:5) node[midway, above] {$$};
	
	\draw[thick, blue]
	(-4.96,0.7) -- (-1.5,0.7) node[midway, above] {$$};
	
	\draw[thick, blue]
	(-1.5,0.5) ++(0.0,-0.2) 
	arc (-90:90:0.2) 
	node[right] {$$};
	
	\draw[thick, blue]
	(-1.5, 0.3) -- (-5, 0.3) node[midway, below] {};

\draw[thick, blue,->]
	(-5, 0) -- (5, 0) node[midway, below] {};
    
	\node at (-1.5, -0.3) {$-c|{\bf p}|$};
	\fill (-1.5,0) circle (2pt);
	\fill (0,0.5) circle (2pt);
	\node at (0.4, 0.5) {$i\,\epsilon$};
	\node at (1.5, 0.3) {$c|{\bf p}|$};
	\fill (1.5,0) circle (2pt);
	\fill (0,-0.5) circle (2pt);
	\node at (0.4, -0.5) {$-i\,\epsilon$};
	\node at (-0.8, 0.5) {$C_\epsilon$};
	\end{tikzpicture}
	\caption{The contour choice (shown in blue) for performing the $\omega$ integration in equation \eqref{5123ert}. The red wiggly lines represent the two branch cuts due to the factors in denominator. }
	\label{Fig2}
\end{figure}
Assuming $\frac{d}{2}-\Delta>0$, the contributions from the arc at infinity vanish. In addition, the integral over the semicircle $C_\epsilon$ centered at the branch point $-c|{\bf p}|+i\epsilon$ also vanishes. This implies that the desired integral along the real axis is given by the discontinuity across the branch cut in the second quadrant leaving us with the identity
\begin{eqnarray}
&&\hspace*{-.3in}\int_{-\infty}^\infty \frac{d\omega}{[(-\omega+c\,|{\bf p}|-i\epsilon)(\omega+c|{\bf p}| -i\epsilon)]^{\frac{d}{2} -\Delta}}\nonumber\\[.2cm]
&=& -\left[e^{-i\pi (\frac{d}{2}-\Delta)}\int_{-\infty}^{-c|{\bf p}|}+e^{i\pi (\frac{d}{2}-\Delta)}\int^{-\infty}_{-c|{\bf p}|}\right]\frac{d\omega}{[(-\omega+c\,|{\bf p}|)(\omega+c|{\bf p}| )]^{\frac{d}{2} -\Delta}}
\end{eqnarray}
The evaluation of the two integrals in the right hand side follows the same method as in the case of Wightman two-point function, yielding
\begin{eqnarray}
\int_{-\infty}^\infty \frac{d\omega}{[(-\omega+c\,|{\bf p}|-i\epsilon)(\omega+c|{\bf p}| -i\epsilon)]^{\frac{d}{2} -\Delta}}
&=& \f{2i\pi\Gamma\left(d-2\Delta-1\right)}{\Gamma^2\left(\f{d}{2}-\Delta\right)}(2c|\textbf{p}|)^{2\Delta-d+1}\label{4.77}
\end{eqnarray}
This fixes the function $\tilde g$ in \eqref{5124ewr} giving the Carrollian limit of the time ordered CFT 2-point function to be
\begin{eqnarray}
c^\beta\,G_2^T(p) \xrightarrow{c \rightarrow 0}\left\{\begin{array}{ll}  
-i\frac{\pi^{\frac{d}{2}}\,\Gamma[\frac{d}{2}-\Delta]}{2^{\Delta-d}\,\Gamma[\Delta]}\,\,{\omega^{2\Delta-d}}& \beta= 2\Delta-d\\[.4cm]
\frac{2^\Delta\pi^{\frac{d}{2}+1}\,\Gamma[d-2\Delta -1]}{4\, \Gamma[\Delta-\frac{d}{2} ]\,\Gamma[\Delta]}\,|{\bf p}|^{2\Delta -d+1}&\beta= -1\end{array}\right.\label{carlt}
\end{eqnarray}
 {Comparing \eqref{carlw} and \eqref{carlt}, we see that the Carrollian limit of Wightman and time ordered Lorentzian CFT 2-point functions differ only in the overall constants. 
 
We conclude by noting that, although the CCFT algebra is obtained from the
CFT algebra via an ultrarelativistic contraction, the same limit applied
to CFT correlators generates only a subset of the full Carrollian
solution space. The reason is that relativistic CFT two-point functions
are already constrained by the full conformal symmetry to satisfy
$\Delta_1=\Delta_2$, a condition that follows from invariance under the
algebra $\mathfrak{so}(2,d)$. Consequently,
the ultra-relativistic limit inherits this restriction and cannot access
the most general Carrollian structures allowed by the CCFT
Ward identities.

\subsection{3-point function}
In this section, we consider the Carrollian limit of the Lorentzian CFT 3-point function. We start with the time ordered case. 

\subsubsection{ Time ordered}
\label{Time-ordered}
The 3-point function of three scalar operators with conformal dimensions $\Delta_a\;\; (a=1,2,3)$ in momentum space in the Lorentzian signature can be obtained by  transforming the corresponding expression in position space. The explicit  transform is quite cumbersome \cite{Bautista:2019qxj}. Fortunately, it is possible to work with an integral representation which expresses the Lorentzian momentum space 3-point function as a convolution over the product of three two-point functions
\begin{eqnarray}
   \mathcal{A}_3(p_1,p_2,p_3) &=&\delta^d(p_1+p_2+p_3)\,{\cal N}(\Delta_{1}\,,\Delta_2\,,\Delta_{3})\,\Bigl\langle\hspace*{-.05in}\Bigl\langle {\cal O}_{\Delta_1}(p_1)\,{\cal O}_{\Delta_2}(p_2)\,{\cal O}_{\Delta_3}(p_3)\Bigl\rangle\hspace*{-.05in}\Bigl \rangle
   \end{eqnarray}
  where
 \begin{eqnarray} 
   \Bigl\langle\hspace*{-.05in}\Bigl\langle {\cal O}_{\Delta_1}(p_1)\,{\cal O}_{\Delta_2}(p_2)\,{\cal O}_{\Delta_3}(p_3)\Bigl\rangle\hspace*{-.05in}\Bigl \rangle =\int \frac{d^d k}{(2\pi)^d} \,\frac{1}{(k^2-i\epsilon)^{\frac{d}{2} -\Delta_{12}}\,[( p_2+k)^2-i\epsilon]^{\frac{d}{2}-\Delta_{23}} [(p_1-k)^2-i\epsilon]^{\frac{d}{2} -\Delta_{13}}}\nonumber\\
\label{4.78}
\end{eqnarray}
Here, we have introduced the quantities, $\Delta_{ab}= \frac{\Delta_a+\Delta_b-\Delta_c}{2}$, ($a,\,b,\,c=1,2,3$) and the normalization factor is given by
\begin{eqnarray}
\mathcal{N}(\Delta_{1}\,,\Delta_2\,,\Delta_{3})= \,\prod_{a=1}^3 \left[-i \frac{\pi^{\frac{d}{2}}\,\Gamma[\frac{d}{2} -\Delta_a]}{2^{2\Delta_a -d}\Gamma[\Delta_a]}\right]
\end{eqnarray}
The details of the derivation of this representation of the correlator are given in appendix \ref{3-points}.

\vspace*{.07in}In the above representation of the correlator, the manifest cyclic symmetry under the exchange of the three scalar operators is apparently lost since \eqref{4.78} only depends upon the external momenta $p_1$ and $p_2$. However, the symmetry under all external momenta can be shown by a change of the integration
variable $k$. For example, the shift $k \rightarrow k+p_1$ rewrites, after using  momentum conservation,  the 
correlator in terms of the momenta $(p_1,p_3)$ instead of $(p_1,p_2)$. 

\vspace*{.07in}Since the three-point function is represented as a product of three 2-point functions, the analysis of the Carrollian limit proceeds along the same lines as in the previous section. In what follows, we consider the Carrollian limit for different branches one by one. 

\subsubsection*{Branch with $\delta(\omega_1+\omega_2+\omega_3)$}

For this branch, none of the zeroth components of the external momenta vanish. Hence, we can directly take the limit $c\rightarrow0$. A convenient way to do this is to rescale the loop momentum $k^\mu$ by $p_1$, namely
\[
k^\mu \rightarrow  |p_1|k^\mu\;\;, 
\qquad d^d k \rightarrow  |p_1|^d\, d^d k \, .
\]
where, we have introduced the Lorentzian norm \cite{Bautista:2019qxj}
\be
|p| = \sqrt{|-(p^0)^2+\mathbf{p}^2|}
\ee
With the above rescaling, the correlator factorizes as
\be
\langle\!\langle \, {\cal O}_{\Delta_1}(p_1)\,
{\cal O}_{\Delta_2}(p_2)\,
{\cal O}_{\Delta_3}(p_3)\, \rangle\!\rangle
&=& |p_1|^{\Delta_1+\Delta_2+\Delta_3-2d}\,
f\!\left(\frac{p_2}{|p_1|},\,\Delta_i\right),
\ee
where $f$ is a dimensionless function that depends only on the ratio
$\tfrac{p_2}{|p_1|}$ (with $|p_1|$ being the norm of $d$ dimensional Lorentzian momenta $p_1^\mu$). The Carrollian limit can now be taken by replacing
\begin{eqnarray}
p_1\simeq \frac{\omega_1}{c}, 
\end{eqnarray}
Substituting these scalings, we obtain
\be
\lim_{c\rightarrow 0}c^{2d-\Delta_1-\Delta_2-\Delta_3}\,\langle\!\langle {\cal O}_{\Delta_1}(\omega_1)\,
{\cal O}_{\Delta_2}(\omega_2)\,
{\cal O}_{\Delta_3}(\omega_3)\rangle\!\rangle
= \left(\omega_1^2\right)^{\tfrac{\Delta_1+\Delta_2+\Delta_3 - 2d}{2}}
\, f\!\left(\frac{\omega_2}{\omega_1}\right).\label{5125etyut}
\ee
This result is in agreement with the solution of the Ward identities 
for the branch $F_1$ given in equation \eqref{3.57}.  In contrast with \eqref{3.57},  the function $f\!\left(\frac{\omega_2}{\omega_1}\right)$ in \eqref{5125etyut}, arising in the small-$c$ limit, is not arbitrary. This behaviour is analogous to the ultra-relativistic limit of the two-point function, as discussed at the end of Section \ref{Time-ordered 2point}.

\subsubsection*{Branch with $\delta(\omega_1)\delta(\omega_2+\omega_3)$}

For the 3-point branches involving one or more vanishing energy components, we need to look for additional contributions to the Carrollian limit, scaling with different powers $c^{\alpha}$ of the speed of light. These arise from specific kinematic regions. For \eqref{4.78}, these correspond to 
\begin{eqnarray}
\chi_1 = c k^0 \,\pm\, c|{\bf k}-{\bf p}_1|\simeq 0 \;\;, 
\quad\mbox{or}\qquad 
\chi_2 = -c k^0 \,\pm \, c|{\bf k}+{\bf p}_2| \simeq 0\;\;,
\label{4.80}
\end{eqnarray}
or from configurations in which both the above conditions are simultaneously satisfied. These contributions are proportional to $\delta(\omega_1)$ or $\delta(\omega_2)$ 
(and also to $\delta(\omega_3)$ if one chooses a representation of the 
{CFT correlator} in which the momentum $p_3$ appears explicitly), as well 
as to the product $\delta(\omega_1)\,\delta(\omega_2)\,\delta(\omega_3)$. 
The latter delta function structure originates from the overall energy-conserving 
delta function in the {CFT correlator}, which enforces $\omega_1+\omega_2+\omega_3=0$.

The functions multiplying the delta functions in the energies are fixed by
projecting the CFT correlator onto the corresponding support. For instance,
the coefficient of $\delta(\omega_1)$ is obtained by integrating equation \eqref{4.78}
with respect to $\omega_1$ at fixed $\omega_2 \neq 0$ (so that the $\delta(\omega_2)$ term
does not contribute). We can now fix the coefficient of the $\delta(\omega_1)$ term by integrating
equation \eqref{4.78} with respect to $\omega_1$. A similar analysis can be extended 
to the other branches that emerge in the Carrollian limit. 
The integral over $\omega_1$ is performed in the same way as in 
equation \eqref{4.77} with the result
\begin{eqnarray}
&& \int_{-\infty}^\infty d\omega_1\,
   \frac{1}{\left[(p_1-k)^2-i\epsilon\right]^{\tfrac{d}{2}-\Delta_{13}}}
   \nonumber\\[4pt]
&& \quad =\; 2i \,c^{d-2\Delta_{13}}\,\sin\!\big[\pi\!\left(\tfrac{d}{2}-\Delta_{13}\right)\big]\,
   B\!\left(1-\tfrac{d}{2}+\Delta_{13},\,d-2\Delta_{13}-1\right)\,
   |{\bf p}_1-{\bf k}|^{\,1-d+2\Delta_{13}} .
\end{eqnarray}
The next step is to compute the integral over the integration momenta $k$
\begin{eqnarray}
\int \frac{dk^0\,d^{d-1} k}{(2\pi)^d} \frac{1}{(k^2-i\epsilon)^{\frac{d}{2}-\Delta_{12}}\,[(p_2+k)^2-i\epsilon]^{\frac{d}{2}-\Delta_{23}} |{\bf p}_1-{\bf k}|^{d -2\Delta_{13}-1}}\label{4.82}
 \end{eqnarray}
 \begin{figure}[t!]
	\centering
	\begin{tikzpicture}[scale=1]
	\draw[thick,Blue,->] (-5, 0) -- (5, 0); 
    \draw[thick,Black,->] (-6, 0) -- (6, 0) node[below right] {${\rm Re} (k^0)$};
	\draw[thick,Black,->] (0,-1) -- (0,6) node[left] {${\rm Im} (k^0)$};
	\draw[dashed](0,0.5)--(0,-0.5);
	\draw[thick, blue,->]
	(-5, 0) -- (5, 0) node[midway, below] {};
	\draw[thick, blue] 
	(5,0.0) arc (-0.5:172.5:5) node[midway, above] {$\Gamma$};
	\draw[thick, blue] 
	(-5,0.45) arc (175:180:4.75) node[midway, above] {$$};

	\draw[thick, blue](-4.96,0.7) -- (-0.8,0.7) node[midway, above] {$$};
	\draw[thin, red, decorate, decoration={snake,amplitude=0.6mm, segment length=2mm}]    (-1,0.42) -- (-5,0.42);
    \draw[ thin, red, decorate, decoration={snake,amplitude=0.6mm, segment length=2mm} ] (5, -0.40) -- (1,-0.40);
    \draw[ thin, cyan, decorate, decoration={snake,amplitude=0.6mm, segment length=2mm} ] (5, -0.6) -- (1.5, -0.6);
    \draw[thin, cyan, decorate, decoration={snake,amplitude=0.6mm, segment length=2mm}]    (-1.5,.6) -- (-5,0.6);

	\draw[thick, blue](-0.8,0.5) ++(0.0,-0.2) arc (-90:90:0.2) node[right] {$$};
	
	\draw[thick, blue]
	(-0.8, 0.3) -- (-5, 0.3) node[midway, below] {};

	\node at (-2, -0.3) {$-|{\bf k}|$};
	\node at (-0.9, -0.3) {$-\chi$};
	\node at (1.0, -2) {$\chi=p_2^0+|{\bf p}_2+ {\bf k}|\leq 0$};
	\node at (0.6, -2.5) {$|{\bf k}|>\chi~~,~~p_2^0\leq 0$};
	\fill (-1.5,0) circle (2pt);
	\fill (0,0.5) circle (2pt);
	\node at (0.4, 0.5) {$i\,\epsilon$};
	\fill (0,-0.5) circle (2pt);
	\node at (0.4,- 0.5) {$-i\,\epsilon$};
	\node at (2, 0.3) {$$};
	\node at (0.9, 0.3) {$$};
	\fill (1.5,0) circle (2pt);
	\fill (1,0) circle (2pt);
	\fill (-1,0) circle (2pt);
\node at (-0.5, 0.9) {$C_\epsilon$};
	\end{tikzpicture}
	\caption{The contour choice for performing the $k^0$ integral in equation \eqref{4.82}. There are a total of four branch points described in equation \eqref{5.130ty}. }
	\label{Fig3}
\end{figure}
As in the two-point case, the integration over $k^{0}$ can be carried out by contour method analyzing the analytic structure of the integrand, with particular attention to the branch cuts generated by the non-integer exponents of terms in the denominators. The location of these cuts is determined by the $i\epsilon$ prescription. The branch cuts in the complex $k^{0}$--plane start at 
\begin{equation}
  k^{0}=\pm |{\bf k}|{\mp i\epsilon}, 
  \qquad 
  k^{0}=-p_{2}^{0}\pm |{\bf k}+{\bf p}_{2}|{\mp i\epsilon} \, \label{5.130ty}
\end{equation}
and extend to $\pm\infty {\mp i\epsilon}$.\footnote{If the conformal dimensions of the operators were integers, the corresponding singularity would reduce to an integer-order pole or be absent. In the present discussion we do not consider this special case.} 
Their relative positions depend on the values of the external four-momentum $p_{2}$ and the spatial momentum ${\bf k}$. Two distinct configurations may arise
\begin{eqnarray}
-|{\bf k}| \;<\; -\chi \;\equiv\; -p_{2}^{0}-|{\bf p}_{2}+{\bf k}| 
\;<\; -p_{2}^{0}+|{\bf p}_{2}+{\bf k}| \, , 
\end{eqnarray}
\begin{eqnarray}
\mbox{and};\qquad-|{\bf k}| \;>\; -\chi 
\;\;\;\Longrightarrow\;\;\; 
\chi > |{\bf k}| \, .
\end{eqnarray}
Since the above relations are valid for any value of $\textbf{p}_2$ and ${\bf k}$, in the Carrollian limit ($c\rightarrow 0$) these conditions become 
(for $ c \,(|\textbf{p}_2+{\bf k}|-|{\bf k}|)\simeq  0$)
\begin{eqnarray}
&& \chi \geq |{\bf k}|
   \;\;\Longrightarrow\;\;
   \omega_{2} + c\,|{\bf p}_{2}+{\bf k}| \,\geq\, c\,|{\bf k}|
   \;\;\Longrightarrow\;\;
   \omega_{2} \geq 0 \, , \nonumber \\[4pt]
&& \chi \leq |{\bf k}|
   \;\;\Longrightarrow\;\;
   \omega_{2} + c\,|{\bf p}_{2}+{\bf k}| \,\leq\, c\,|{\bf k}|
   \;\;\Longrightarrow\;\;
   \omega_{2} \leq 0 \, .\label{5.133}
\end{eqnarray}
When the second condition is satisfied, $\omega_{2}$ is negative, while $\chi$ can take either sign. In the Carrollian limit this leads to
\begin{eqnarray}
\omega_{2} \leq 0 \;\;\Rightarrow\;\;
\left\{
\begin{array}{ll}
\chi \geq 0 
& \Longrightarrow \quad 0 \,\geq\, \omega_{2} \,\geq\, -c\,|{\bf p}_{2}+{\bf k}|
    \;=\; -c\,|{\bf k}| \,\simeq\, 0,
    \quad \forall\, |{\bf k}|\in[0,\infty) \, , \\[8pt]
\chi \leq 0 
& \Longrightarrow \quad 0 \,\geq\, \omega_{2} \,\leq\, -c\,|{\bf p}_{2}+{\bf k}|
    \;=\; -c\,|{\bf k}| \,\simeq\, 0,
    \quad \forall\, |{\bf k}|\in[0,\infty) \, .\label{5.134}
\end{array}
\right.
\end{eqnarray}
In the last step we have imposed $c|{\bf k}|\simeq 0$, as appropriate in the
ultra-relativistic (small–$c$) regime, thereby isolating the contribution from
the region singled out by equation \eqref{4.80}. From equations \eqref{5.133} and \eqref{5.134}, the case $|{\bf k}| \geq \chi$ is
compatible with both positive and negative values of $\chi$. However, the case
$\chi \leq 0$ selects only the value $\omega_{2}=0$ in the small-$c$ limit.
We therefore restrict our analysis to the Carrollian limit  with $\chi \geq 0$, which is
compatible with $\omega_{2} \leq 0$. The Carrollian limit for the case
$|{\bf k}| \leq \chi$ with $\omega_{2} \geq 0$ can be treated analogously.

\vspace*{.07in}Now, we consider the $k^0$ integral in \eqref{4.82} along the contour shown in figure \ref{Fig3}. Under the assumption
\begin{eqnarray}
2d-2\Delta_{23}-2\Delta_{12}-1\geq 0
\end{eqnarray}
the contribution from the arc at infinity vanishes. The integral along the semicircle centered around the branch point $-\chi+i\epsilon$ also vanishes. Thus, the desired integral along the real axis is determined by the discontinuities across the cuts 
\begin{eqnarray}
&&{\int \frac{dk^0\,}{2\pi} \frac{1}{(k^2-i\epsilon)^{\frac{d}{2}-\Delta_{12}}\,[(p_2+k)^2-i\epsilon]^{\frac{d}{2}-\Delta_{23}} |{\bf p}_1-{\bf k}|^{d -2\Delta_{13}-1}}}\non\\[.4cm]
 &\equiv& \left[ 2i\sin\pi( d-\Delta_{23} -\Delta_{12})\int_{-|{\bf k}|}^{-\infty}\;\;+\;\; 2\,i\sin\pi\left(\frac{d}{2} -\Delta_{23}\right) \int_{-\chi}^{-|{\bf k}|}\right]\frac{dk^0}{2\pi}\nonumber\\
 &&\times \frac{1}{|k^0-|{\bf k}| |^{\frac{d}{2}-\Delta_{12}}\,|k^0+|{\bf k}| |^{\frac{d}{2}-\Delta_{12}}\,
 |p_2^0+k^0-|{\bf k}+{\bf p}_2| |^{\frac{d}{2} -\Delta_{23}}|p_2^0+k^0+|{\bf k}+{\bf p}_2| |^{\frac{d}{2} -\Delta_{23}}}\nonumber\\[.4cm]
&\equiv& \frac{i}{\pi}\sin\pi( d-\Delta_{23} -\Delta_{12})\,I_1+\,\frac{i}{\pi}\sin\pi\left(\frac{d}{2} -\Delta_{23}\right) I_2
 \end{eqnarray}
 We now evaluate the two contributions separately. After a suitable change of variables, the first integral can be recast in the form
 \begin{eqnarray}
I_1= -(2|{\bf k}|)^{1-2d +2\Delta_{23}+2\Delta_{12}}\int_0^\infty \frac{dt}{ t^{\frac{d}{2}-\Delta_{12}}\,(t+1)^{\frac{d}{2}-\Delta_{12}}\, (t +A_1)^{\frac{d}{2}-\Delta_{23}}\,(t+A_2)^{\frac{d}{2} -\Delta_{23}}} \label{7.96a}
 \end{eqnarray}
 where 
 \begin{eqnarray}
&& A_1
=\frac{-\frac{\omega_2}{c}+|{\bf k}| -|{\bf p}_2 +{\bf k}|}{2|{\bf k}|}=\frac{-\chi +|{\bf k}|}{2\,|{\bf k}|}\geq \frac{-|{\bf k}|+|{\bf k}|}{2|{\bf k}|}\geq 0 
\nonumber\\
 && A_2
 =\frac{-\frac{\omega_2}{c}+|{\bf k}| +|{\bf p}_2 +{\bf k}|}{2|{\bf k}|}
 \end{eqnarray}
In the intermediate steps, the factors in the denominators in \eqref{7.96a} involve the absolute values. However, as shown above, $A_1$ is positive definite. Similarly, since we are considering the case $\omega_{2}\leq 0$, it follows that $A_{2}\geq 0$. Thus, noting that the range of integration is over positive values of $t$, it follows that all the factors in the denominator, in the chosen kinematic region, are positive definite. 
Therefore, we have removed the absolute value from all the terms in the denominator. By performing a further change of variable $t=\frac{y}{1-y}$, we get
\begin{eqnarray}
   I_1&=& -\frac{(2|{\bf k}|)^{1-2d +2\Delta_{23}+2\Delta_{12}}}{(A_1\,A_2)^{\frac{d}{2}-\Delta_{23}}}\int_0^1\,dy\, y^{-\frac{d}{2}+\Delta_{12}}\,(1-y)^{d-2\Delta_{12}-d -2\Delta_{23}-2}\prod_{i=1}^2\left( 1+\frac{1-A_i}{A_i}y\right)^{\Delta_{23} -\frac{d}{2}}\nonumber\\[.4cm]
   &=&-\frac{(2|{\bf k}|)^{1-2d +2\Delta_{23}+2\Delta_{12}}}{(A_1\,A_2)^{\frac{d}{2}-\Delta_{23}}}\frac{\Gamma[\Delta_{12}-\frac{d}{2} +1]\, \Gamma[2 d -2(\Delta_{12}+\Delta_{23})-1]}{\Gamma[ \frac{3d}{2} -2\Delta_{23}-\Delta_{12}]}\nonumber\\[.3cm]
   &&\;F_1\left( \Delta_{12} -\frac{d}{2} +1,\,\frac{d}{2} -\Delta_{23},\,\frac{d}{2} -\Delta_{23}, \frac{3}{2} d-2\Delta_{23}-\Delta_{12},\frac{A_1-1}{A_1},\,\frac{A_2-1}{A_2}\right)\label{4.92}
\end{eqnarray}
where we have used the identity given in equation \eqref{4.123}.

\vspace*{.07in}The second integral gives
 \begin{eqnarray}
 I_2&=&\int^{-|{\bf k}|}_{-\omega} \frac{dk^0}{|k^0-|{\bf k}| |^{\frac{d}{2}-\Delta_{12}}\,|k^0+|{\bf k}| |^{\frac{d}{2}-\Delta_{12}}\,
 |p_2^0+k^0-|{\bf k}+{\bf p}_2| |^{\frac{d}{2} -\Delta_{23}}|p_2^0+k^0+|{\bf k}+{\bf p}_2| |^{\frac{d}{2} -\Delta_{23}}}\nonumber\\
 &=& -(2|{\bf k}|)^{1-2d +2\Delta_{12}+2\Delta_{23}}\int_0^{\frac{-\chi+|{\bf k}|}{2|{\bf k}|}}  \frac{dt}{ |t|^{\frac{d}{2}-\Delta_{12}}\,|t-1|^{\frac{d}{2}-\Delta_{12}}\, |t -A_1|^{\frac{d}{2}-\Delta_{23}}\,|t-A_2|^{\frac{d}{2} -\Delta_{23}}}\nonumber\\
 &=& -(2|{\bf k}|)^{1-2d +2\Delta_{12}+2\Delta_{23}}\int_0^{A_1} \frac{dt}{ t^{\frac{d}{2}-\Delta_{12}}\,(1-t)^{\frac{d}{2}-\Delta_{12}}\, (A_1 -t)^{\frac{d}{2}-\Delta_{23}}\,(A_2-t)^{\frac{d}{2} -\Delta_{23}}}
 \end{eqnarray}
 In going from 1st to 2nd line, we have made the coordinate transformation $k^0=(2t-1)|\textbf{k}|$.
We have also used the condition  $ |{\bf k}|\geq \chi $ which holds in the chosen Kinematic region. In going to the 3rd line, we have removed the absolute values. To see this, we note that $t\in[0, A_1]$ implies $|t-A_1|=(A_1-t)$. Furthermore $A_2\geq A_1$ and since $t\in[0, A_1]$, we have $t\leq A_2$ and therefore $|t-A_2|=(A_2-t)$. We have also $A_1= \frac{1}{2}\left( 1-\frac{\chi}{|{\bf k}|}\right)$, with $0\leq \chi\leq|{\bf k}|$. It follows that $A_1\leq 1$, and therefore $|t-1|=1-t$. 
We now make a change of variable $t=-\frac{z\alpha}{1-z\alpha}$ with $ \alpha=\frac{\chi -|{\bf k}|}{\chi+|{\bf k}|}\leq0$, getting
\begin{eqnarray}
I_2&=&\frac{(-1)^{\frac{d}{2}-\Delta_{12}+1}}{(A_1\,A_2)^{\frac{d}{2} -\Delta_{23}}}\frac{1}{\alpha^{\frac{d}{2}-\Delta_{12}-1}}\int_0^1 dz\,\frac{(2|{\bf k}|)^{2\Delta_{23}+2\Delta_{12}-2d+1}(1-z\,\alpha)^{d-2\Delta_{12}+d-2\Delta_{23}-2}}{z^{\frac{d}{2}-\Delta_{12}}\,(1+\frac{(1-A_1)}{A_1}\,\alpha\,z)^{\frac{d}{2} -\Delta_{23}} \,\,(1+\frac{(1-A_2)}{A_2} \,\alpha\,z)^{\frac{d}{2} -\Delta_{23}}}\nonumber\\[.4cm]
&=&(-1)^{\frac{d}{2}-\Delta_{12}+1}\frac{(2|{\bf k}|)^{2\Delta_{23}+2\Delta_{12}-2d+1}}{(A_1\,A_2)^{\frac{d}{2} -\Delta_{23}}}\frac{1}{\alpha^{\frac{d}{2}-\Delta_{12}-1}(\Delta_{12} -\frac{d}{2} +1)}\;F^{(3)}_D\left[\Delta_{12}-\frac{d}{2}+1,\right.\non\\
&&\left.\left\{-2(d-\Delta_{12}-\Delta_{23}-1),\, \frac{d}{2}-\Delta_{23},\,\frac{d}{2}-\Delta_{23}\right\},\,\Delta_{12} -\frac{d}{2} +2,\,\left\{\alpha,\,\alpha\,\frac{A_1-1}{A_1},\,\alpha\,\frac{A_2-1}{A_2}\right\}\right]\nonumber\\
&&\label{4.96}
 \end{eqnarray}
where $F_D^{(m)}$ denotes the Lauricella hypergeometric function of type~$D$,
defined by the integral representation
\begin{eqnarray}
&&F_D^{(m)}\!\left(a;\,b_1,\dots,b_m;\,c;\,x_1,\dots,x_m\right)
=
\frac{\Gamma(c)}{\Gamma(a)\,\Gamma(c-a)}
\int_{0}^{1}
t^{\,a-1}\,(1-t)^{\,c-a-1}\,
\prod_{j=1}^{m} \bigl(1 - t\,x_j\bigr)^{-b_j}\,dt \nonumber\\
\label{5147}
\end{eqnarray}
with $\mbox{Re}(c)>\mbox{Re}(a)>0$.\footnote{The condition $\mbox{Re}(c)>\mbox{Re}(a)>0$ in our notations are $\Delta_{12} -\frac{d}{2} >-1$. The factor $(k^2-i\epsilon)^{\frac{d}{2} -\Delta_{12}}$ goes to the numerator, when $\Delta_{12}>\frac{d}{2}$. Hence, the identity \eqref{5147} has been used in \eqref{4.96} in the sense of analytic continuation. } 

\vspace*{.07in}We can now take the Carrollian limit $c\rightarrow 0$. In this limit, we first note that 
\begin{eqnarray}
A_{1}\simeq A_{2}\simeq -\frac{\omega_{2}}{2c\,|{\bf k}|},
\qquad
\frac{A_{1}-1}{A_{1}} \;\simeq\; \frac{A_{2}-1}{A_{2}} \;\simeq\; \alpha \;\simeq\; 1 \, .
\end{eqnarray}
With the above results, the integrals in equations \eqref{4.92} and \eqref{4.96} simplify to give
\begin{eqnarray}
I_{2} \;\simeq \; (-1)^{\tfrac{d}{2}-\Delta_{12}}\,I_{1}
\;\simeq\;
c^{\,d-2\Delta_{23}}\,
\frac{(2|{\bf k}|)^{\,2\Delta_{12}-d+1}}{\,(\omega_{2}^{2})^{\,\tfrac{d}{2}-\Delta_{23}}}\,
B\!\left(\Delta_{23}-\tfrac{d}{2}+1,\,d-2\Delta_{23}-1\right)
\end{eqnarray}
where $B(x,y)$ denotes the Euler Beta function. 
Finally, by using the identity 
\begin{eqnarray}
\int \f{d^{d-1}k}{(2\pi)^{d-1}} \f{1}{|\textbf{p}_1-\textbf{k}|^{d-2\beta_2-1}|\textbf{k}|^{d-2\beta_3-1}}= \frac{\pi^{\f{d-1}{2}}\Gamma\left(\f{d-1}{2}-\beta_2-\beta_3\right)}{\Gamma\left(\f{d-1}{2}-\beta_2\right)\Gamma\left(\f{d-1}{2}-\beta_3\right)}B(\beta_2,\beta_3)|\textbf{p}_1|^{2\beta_2+2\beta_3-d+1}\label{4122wea}
\end{eqnarray}
we can now evaluate the last integral in equation \eqref{4.82}, getting
\begin{eqnarray}
&&\hspace*{-.4in}\lim_{c\rightarrow 0}c^{2(\Delta_{23}+\Delta_{13} -d)}
\int_{-\infty}^\infty d\omega_1\langle
\!\langle \, {\cal O}_{\Delta_1}(p_1)\,
{\cal O}_{\Delta_2}(p_2)\,
{\cal O}_{\Delta_3}(p_3)\, \rangle\!\rangle\non\\
&=&-2^{2\Delta_{12} -d+1}\,\pi^{\frac{d-3}{2}}\,\sin\pi\left(\frac{d}{2} -\Delta_{13}\right)\left[ \sin\pi \left(d-\Delta_{23}-\Delta_{12}\right) +(-1)^{\frac{d}{2} -\Delta_{12}}\sin\pi\left( \frac{d}{2} -\Delta_{23}\right) \right]\nonumber\\
&& B\left( \Delta_{13} -\frac{d}{2} +1,\,d-2\Delta_{13} -1\right) B\left( \Delta_{23} -\frac{d}{2} +1,\,d-2\Delta_{23} -1\right) B\left(\Delta_{12} ,\,\Delta_{23}\right)\non\\
&&\frac{\Gamma\left[ \frac{d-1}{2} -\Delta_{12} -\Delta_{23}\right]\,}{\Gamma\left[ \frac{d-1}{2} -\Delta_{12}\right]\,\Gamma\left[\frac{d-1}{2} -\Delta_{23}\right] } \, (\omega_2)^{\Delta_{2}+\Delta_3-\Delta_1 -d}\, |{\bf p}_1|^{\Delta_{1}-d+1}
\end{eqnarray}
The above expression matches with the first term in the $F_3$ branch \eqref{4110fg} with a specific value of the overall constant $c_3$ (which depends upon the spacetime boundary dimension $d$ and the conformal dimension $\Delta_i$). 

\vspace*{.07in}The results of the Carrollian limit for the other branches containing $\delta(\omega_2)$ and $\delta(\omega_3)$ can be obtained in a similar manner and can be written down by permuting the indices in the above result.

\subsubsection*{Branch with $\delta(\omega_1)\delta(\omega_2)\delta(\omega_3)$}

The Carrollian branch containing $\delta(\omega_1)\delta(\omega_2)\delta(\omega_3)$ originates from the region in which both conditions in equation \eqref{4.80} are satisfied simultaneously. In this region, $\omega_1$, $\omega_2$, and consequently $\omega_3$ (due to the presence of overall delta function $\delta(\omega_1+\omega_2+\omega_3)$) vanish giving rise to the factor $\delta(\omega_1)\,\delta(\omega_2)\,\delta(\omega_3)$. In this case, the behavior of the CFT correlator in the Carrollian limit $c \to 0$ is determined by integrating over both $\omega_1$ as well as $\omega_2$. Using the identity \eqref{4.77}, we find
\begin{eqnarray}
&&\hspace*{-.4in} \int_{-\infty}^\infty \prod_{i=1}^2d\omega_i\langle\langle {\cal O}_{\Delta_1}(p_1)\,{\cal O}_{\Delta_2}(p_2)\,{\cal O}_{\Delta_3}(p_3)\rangle \rangle\non\\
&=&\frac{(2i)^3 c^2}{2\pi}\prod_{i<j=1}^3\left[\sin\pi\left( \frac{d}{2} -\Delta_{ij}\right) B\left( 1-\frac{d}{2} +\Delta_{ij},\,d-2\Delta_{ij}-1\right)\right]\nonumber\\
&& \int_{-\infty}^\infty\frac{d^{d-1}k}{(2\pi)^{d-1} }\,|{\bf k}|^{1-d+2\Delta_{12}}|{\bf p}_1-{\bf k}|^{1-d+2\Delta_{13}}\,|{\bf p}_2+{\bf k}|^{1-d + 2\Delta_{23}}
\end{eqnarray}
The momentum-space integral can be evaluated in terms of triple-$K$ integrals. Writing $\delta_t=\delta_1+\delta_2+\delta_3$, one finds~\cite{Bzowski:2013sza}
\begin{eqnarray}
\int_{-\infty}^{\infty} \frac{d^{d-1} k}{(2\pi)^{d-1}}
\frac{1}{{|\bf k}|^{\delta_1}\,|{\bf p}_1-{|\bf k}|^{\delta_2}\,|{\bf p}_2+{\bf k}|^{\delta_3}}= \frac{2^{4\frac{d-1}{2}}}{(4\pi)^{\frac{d-1}{2}}} \frac{I_{\frac{d-1}{2}-1\{\frac{d-1}{2}-\delta_t+\delta_j\}}}{\prod_{a=1}^3\Gamma(\delta_a) \Gamma(d-1-\delta_t)}
\end{eqnarray}
with parameters
\begin{equation}
-\delta_t+\delta_a = \Delta_a - (d-1),\qquad\nu=\tfrac{d-3}{2}, 
\qquad \beta_a = \Delta_a - \frac{(d-1)}{2}
\end{equation}
and the triple-$K$ integral defined by
\begin{eqnarray}
I_{\nu\{\beta_1,\beta_2,\beta_3\}}(k_1,k_2,k_3)
= \int_0^\infty dy \, y^{\nu}\,\prod_{j=1}^3 k_j^{\beta_j} K_{\beta_j}(k_j \,y)\,.
\end{eqnarray}
This result is in agreement with equation \eqref{3.58}, which was obtained by solving the Conformal Carrollian Ward identities.

\subsubsection{ Wightman}
The 3-point CFT Wightman correlator in momentum space momentum in the Lorentzian signature can be expressed in the form \cite{Bautista:2019qxj} (see also \cite{Kravchuk:2021kwe, sachin} for discussions of Wightman functions in CFT)
\be
\mathcal{A}_3 &=& \tilde a(\{\beta_i\}) \int \f{d^dk}{(2\pi)^d} \f{\theta(p_2^0+k^0-|\textbf{p}_2+\textbf{k}|)\theta(p_1^0-k^0-|\textbf{p}_1-\textbf{k}|)\theta(k^0-|\textbf{k}|)}{|p_2+k|^{d-2\beta_1}|p_1-k|^{d-2\beta_2}|k|^{d-2\beta_3}}\label{631wes1}
\ee
where 
\be
\tilde a(\{\beta_i\}) = c_{123} \f{2^{3(d+1)-2\beta_t}\pi^{3(d/2+1)}}{\prod_{j=1}^3\Gamma(\beta_j-d/2+1)\Gamma(\beta_j)}\quad;\qquad \beta_j = \f{\Delta_1+\Delta_2+\Delta_3}{2}-\Delta_j.
\ee
The zeroth component of the momentum vector $p^\mu$ is denoted as $p^0 = \f{\omega}{c}$ and the  norm $|p|$ is defined as\footnote{In \eqref{p}, there is no ambiguity in the sign inside the square root since the Wightman correlator vanishes for spacelike or null like external momenta }
\be
|p| \equiv \sqrt{(p^0)^2 - |\textbf{p}|^2} = \sqrt{\left(\f{\omega}{c}\right)^2 - |\textbf{p}|^2}\label{p}
\ee
Again, the correlator in \eqref{631wes1} is not symmetric under the exchange of the momenta since $(p_3^0,\,{\bf p}_3)$ do not appear. However, with a change of variable
\begin{eqnarray}
k^0\rightarrow k^0-p_2^0~~;~~{\bf k}\rightarrow {\bf k}-{\bf p}_2
\end{eqnarray}
the Wightman correlator in equation \eqref{631wes1} becomes
\be
\mathcal{A}_3 &=& \tilde a(\{\beta_i\}) \int \f{d^dk}{(2\pi)^d} \f{\theta(k^0-|\textbf{k}|)\theta(-p_3^0-k^0-|\textbf{p}_3+\textbf{k}|)\theta(k^0-p_2^0-|\textbf{k}-\textbf{p}_2|)}{|k|^{d-2\beta_1}|p_3+k|^{d-2\beta_2}|k-p_2|^{d-2\beta_3}}
\ee
This depends upon the momenta $p_2$ and $p_3$. Similarly, with the change of variable
\begin{eqnarray}
k^0\rightarrow k^0+p_1^0~~;~~{\bf k}\rightarrow {\bf k}+{\bf p}_1
\end{eqnarray}
the 3-point function can be made to depend upon the momenta $p_1$ and $p_3$. One could again take a sum of the three equivalent expressions divided by $3$ to make apparent the symmetry in the exchange of the three external momenta. 

\vspace*{.07in}The Heaviside step functions in the expression of Wightman 3-point functions imply that $p_1^0\ge 0, p_3^0\le 0$ while $p_2^0$ can be both positive as well as negative. This will be useful below when we consider the Carrollian limit of the above 3-point function.
 In the time-ordered case, we obtained the three branches with two delta functions in Carrollian energy from one of them, by permuting the energy indices. 
The asymmetry in the energy directions introduced by the Heaviside step function in the Wightman CFT correlator, implies that we need to be careful while analysing the Carrollian limit that yields these branches. We will analyse the Carrollian limit yielding each of these branches explicitly. We shall now show how the different branches of Carrollian 3-point functions emerge when we take the limit $c\rightarrow 0$.

\subsubsection*{Branch with $\delta(\omega_1+\omega_2+\omega_3)$}
For this branch, we can again directly take the limit $c\rightarrow 0$ since none of the external zeroth components of the momenta vanish. As in the time ordered case, if we rescale the loop momenta as
\[
k^\mu \rightarrow  |p_1|k^\mu\;\;, 
\qquad d^d k \rightarrow  |p_1|^d\, d^d k \, .
\]
the denominator inside the integral in \eqref{631wes1} rescales as 
\be
\f{1}{|p_2+k|^{d-2\beta_1}|p_1-k|^{d-2\beta_2}|k|^{d-2\beta_3}}\;\;\rightarrow\;\; \f{|p_1|^{\Delta_t-3d}}{\left|\f{p_2}{|p_1|}+k\right|^{d-2\beta_1}\left|\f{p_1}{|p_1|}-k\right|^{d-2\beta_2}|k|^{d-2\beta_3}}
\ee
Next, under this rescaling, the first theta function becomes
\be
\theta(p_2^0+k^0-|\textbf{p}_2+\textbf{k}|)\rightarrow \theta\!\left(|p_1|\left(
\,\frac{p_2^0}{|p_1|}
+k^0- \left|
\frac{\mathbf{p}_2}{|p_1|} + \mathbf{k}
\right|\right)\right)=\theta\!\left(
\,\frac{p_2^0}{|p_1|}
+k^0- \left|
\frac{\mathbf{p}_2}{|p_1|} + \mathbf{k}
\right|\right)
\ee
where, we have used the fact that $|p_1|>0$. Hence, rescaling the theta function by $|p_1|$ will not change its value. Now, in the limit $c\rightarrow 0$, we would have 
\be
\frac{p_2^0}{|p_1|} \simeq  \frac{\omega_2}{|\omega_1|}\quad;\qquad \frac{\mathbf{p}_2}{|p_1|} \simeq c\,\frac{\mathbf{p}_2}{|\omega_1|}\simeq 0
\ee
This shows that the argument of the theta function becomes a function of the ratio $\frac{\omega_2}{|\omega_1|}$. Following a similar manipulation as above, it is easy to see that the other two theta functions become independent of the external momenta in the limit $c\rightarrow 0$. Hence, the Wightman correlator again factorizes and we get the expected result
\be
\lim_{c\rightarrow 0}c^{2d-\Delta_1-\Delta_2-\Delta_3}\,\mathcal{A}_3
= \left(\omega_1^2\right)^{\tfrac{\Delta_1+\Delta_2+\Delta_3 - 2d}{2}}
\, f\!\left(\frac{\omega_2}{\omega_1}\right).\label{i}
\ee
where $f$ is some function which can be explicitly evaluated. Again, unlike \eqref{3.57}, the function $f\!\left(\frac{\omega_2}{\omega_1}\right)$ in \eqref{i}, arising in the Carrollian limit, is not arbitrary.

\subsubsection*{Branch with $\delta(\omega_1)\delta(\omega_2)\delta(\omega_3)$}
The branch with 3 delta functions in the energy direction should take the following form in the Carrollian limit
\be
\lim_{c\rightarrow0}c^\alpha \mathcal{A}_3 = f_1(\textbf{p}_1,\textbf{p}_2,\textbf{p}_3)\delta(\omega_1)\delta(\omega_2)\delta(\omega_3)
\ee
for some $\alpha$. We can determine the function $f_1$ by integrating both sides over $\omega_1$ and $\omega_2$ as follows
\be
&&\hspace*{-.6in}c^\alpha\int_{-\infty}^\infty d\omega_1d\omega_2\;\mathcal{A}_3 \non\\
&=& \;\tilde a(\{\beta_i\})c^\alpha\int_{-\infty}^\infty d\omega_1d\omega_2 \int \f{d^dk}{(2\pi)^d} \f{\theta(p_2^0+k^0-|\textbf{p}_2+\textbf{k}|)\theta(p_1^0-k^0-|\textbf{p}_1-\textbf{k}|)\theta(k^0-|\textbf{k}|)}{|p_2+k|^{d-2\beta_1}|p_1-k|^{d-2\beta_2}|k|^{d-2\beta_3}}\non\\
\label{4105}
\ee
The integration over $\omega_1$ is given by
\be
&&\hspace*{-.7in}\int_{-\infty}^\infty d\omega_1\f{\theta(p_1^0-k^0-|\textbf{p}_1-\textbf{k}|)}{((p_1^0-k^0)^2-|\textbf{p}_1-\textbf{k}|^2)^{\f{d}{2}-\beta_2}}\non\\[.2cm]
&=&\int_{ck^0 +c|\textbf{p}_1-\textbf{k}|}^\infty d\omega_1\f{1}{((\f{\omega_1}{c}-k^0)-|\textbf{p}_1-\textbf{k}|)^{\f{d}{2}-\beta_2}((\f{\omega_1}{c}-k^0)+|\textbf{p}_1-\textbf{k}|)^{\f{d}{2}-\beta_2}}\non\\
&=&\f{2^{2\beta_2-d+1}c}{|\textbf{p}_1-\textbf{k}|^{d-2\beta_2-1}}\int_{0}^\infty dx\f{1}{x^{\f{d}{2}-\beta_2}(x+1)^{\f{d}{2}-\beta_2}}\non\\[.2cm]
&=&\f{2^{2\beta_2-d+1}c}{|\textbf{p}_1-\textbf{k}|^{d-2\beta_2-1}}B\left(\beta_2-\f{d}{2}+1, d-2\beta_2-1\right)\label{4106a}
\ee
In going from the 1st to 2nd equality, we have made a coordinate change 
\be
\omega_1 = (2c|\textbf{p}_1-\textbf{k}|x +ck^0 + c|\textbf{p}_1-\textbf{k}|)\quad\implies\quad d\omega_1 = 2c|\textbf{p}_1-\textbf{k}|\;dx
\ee
In a similar way, the integration over $\omega_2$ is given by 
\be
\int_{-\infty}^\infty d\omega_2\f{\theta(p_2^0+k^0-|\textbf{p}_2+\textbf{k}|)}{|p_2+k|^{d-2\beta_1}}
&=&\f{2^{2\beta_1-d+1}c}{|\textbf{p}_2+\textbf{k}|^{d-2\beta_1-1}}B\left(\beta_1-\f{d}{2}+1, d-2\beta_1-1\right)\label{4108ty}
\ee
With the above integrations, the $k^0$ dependence in \eqref{4105} simplifies and the integration over this can be easily performed
\be
\int \f{dk^0}{(2\pi)}\f{\theta(k^0-|\textbf{k}|)}{|{k}|^{d-2\beta_3}}
&=&\f{1}{2\pi}\f{2^{2\beta_3-d+1}}{|\textbf{k}|^{d-2\beta_3-1}}B\left(\beta_3-\f{d}{2}+1, d-2\beta_3-1\right)
\ee
Combining all the terms, the final expression is given by 
\be
 f_1(\textbf{p}_1,\textbf{p}_2,\textbf{p}_3) &=&c^{2+\alpha}\hat a(\{\beta_i\})\int \f{d^{d-1}k}{(2\pi)^{d-1}}\f{1}{|\textbf{p}_1-\textbf{k}|^{d-2\beta_2-1}|\textbf{p}_2+\textbf{k}|^{d-2\beta_1-1}|\textbf{k}|^{d-2\beta_3-1}}\label{639d}
\ee
Choosing $\alpha=-2$ gives the desired result. In the above expression, we have defined 
\be
\hat a(\{\beta_i\}) &=& \f{\tilde a(\{\beta_i\})}{2\pi}\prod_{i=1}^32^{2\beta_i-d+1}B\left(\beta_i-\f{d}{2}+1, d-2\beta_i-1\right)
\ee
The integral in \eqref{639d} is a triple K integral \cite{Bzowski:2013sza} (see equation \eqref{639df}) giving
\be
f_1(\textbf{p}_1,\textbf{p}_2,\textbf{p}_3)&=& c_{123}\; a(d,\Delta_i)\; I_{\f{d-3}{2}\{\Delta_1-\f{d-1}{2},\Delta_2-\f{d-1}{2},\Delta_3-\f{d-1}{2}\}}(k_1,k_2,k_3)
\ee
where
\be
a(d,\Delta_i)= \f{\pi^{1+d}2^{\f{9+3d}{2}-\Delta_1-\Delta_2-\Delta_3}}{\Gamma\left( \f{\Delta_2+\Delta_3-\Delta_1+d-1}{2}  \right)\Gamma\left( \f{\Delta_1+\Delta_3-\Delta_2+d-1}{2}  \right)\Gamma\left( \f{\Delta_1+\Delta_2-\Delta_3+d-1}{2}  \right)\Gamma\left( \f{\Delta_1+\Delta_2+\Delta_3-d+1}{2}  \right)} 
\ee

\subsubsection*{Branch with $\delta(\omega_1)\delta(\omega_2+\omega_3)$}

In this case, we expect 
\be
\lim_{c\rightarrow0\atop \omega_1\rightarrow 0}c^{\sigma_1}\mathcal{A}_3 =  f_3(\omega_2, \textbf{p}_1)\delta(\omega_1)\delta(\omega_2+\omega_3)
\ee
We first perform the integration over $\omega_1$. This is done above in equation \eqref{4106a} using which we obtain
\be
c^{\sigma_1}\int_{-\infty}^\infty d\omega_1\mathcal{A}_3 &=& c^{\sigma_1+1}a_1 \int \f{d^dk}{(2\pi)^d} \f{\theta(p_2^0+k^0-|\textbf{p}_2+\textbf{k}|)\theta(k^0-|\textbf{k}|)}{|p_2+k|^{d-2\beta_1}|\textbf{p}_1-\textbf{k}|^{d-2\beta_2-1}|k|^{d-2\beta_3}}\label{631wesa}
\ee
where
\be
a_1 = 2^{2\beta_2-d+1}B\left(\beta_2-\f{d}{2}+1, d-2\beta_2-1\right)\tilde a(\beta_i)
\ee
Next, we need to perform the integration over the variable $k^0$. To do this integral, we note that the Heaviside step functions imply 
\be
k^0 \ge |\textbf{k}| \quad\mbox{and}\qquad k^0 \ge -p_2^0+ |\textbf{p}_2+\textbf{k}|
\ee
Both the conditions must be satisfied simultaneously. There are two possibilities for which both the above conditions can be satisfied
\be
|\textbf{k}|  \ge -p_2^0+ |\textbf{p}_2+\textbf{k}|\quad\mbox{or}\qquad |\textbf{k}|  \le -p_2^0+ |\textbf{p}_2+\textbf{k}|\label{5150ytdf}
\ee
A simple analysis shows that the second condition in \eqref{5150ytdf} is not compatible with having $\delta(\omega_1)$. Hence, we discard this option and consider the first one. Thus, the integration over $k^0$ becomes 
\be
&&\hspace*{-.4in} \int_{-\infty}^\infty \f{dk^0}{(2\pi)} \f{\theta(p_2^0+k^0-|\textbf{p}_2+\textbf{k}|)\theta(k^0-|\textbf{k}|)}{|p_2+k|^{d-2\beta_1}|k|^{d-2\beta_3}}\non\\[.3cm]
  &=&\int_{|\textbf{k}|}^\infty \f{dk^0}{(2\pi)} \f{1}{(\f{\omega_2}{c}+k^0-|\textbf{p}_2+\textbf{k}|)^{\f{d}{2}-\beta_1}(\f{\omega_2}{c}+k^0+|\textbf{p}_2+\textbf{k}|)^{\f{d}{2}-\beta_1}(k^0-|\textbf{k}|)^{\f{d}{2}-\beta_3}(k^0+|\textbf{k}|)^{\f{d}{2}-\beta_3}}\non\\[.2cm]
&=&\f{2^{2\beta_3+2\beta_1-2d+1}}{|\textbf{k}|^{2d-2\beta_1-2\beta_3-1}}\int_{0}^\infty \f{dx}{(2\pi)} \f{1}{(x+q)^{\f{d}{2}-\beta_1}(x+r)^{\f{d}{2}-\beta_1}x^{\f{d}{2}-\beta_3}(1+x)^{\f{d}{2}-\beta_3}}\non\\[.2cm]
&=&\f{2^{2\beta_3+2\beta_1-2d+1}}{|\textbf{k}|^{2d-2\beta_1-2\beta_3-1}}\f{1}{(ab)^{\f{d}{2}-\beta_1}}\int_{0}^1 \f{dy}{(2\pi)} \f{y^{\beta_3-\f{d}{2}}(1-y)^{2d-2\beta_1-2\beta_3-2}}{(1-\tilde q y)^{\f{d}{2}-\beta_1}(1-\tilde r y)^{\f{d}{2}-\beta_1}}\non\\[.2cm]
&=&\f{2^{2\beta_3+2\beta_1-2d+1}}{|\textbf{k}|^{2d-2\beta_1-2\beta_3-1}}\f{\Gamma(\beta_3-\f{d}{2}+1)\Gamma(2d-2\beta_1-2\beta_3-1)}{2\pi(qr)^{\f{d}{2}-\beta_1}\Gamma(\f{3d}{2}-2\beta_1-\beta_3)}\non\\
&&F_1\left(\beta_3-\f{d}{2}+1;\f{d}{2}-\beta_1,\f{d}{2}-\beta_1;\f{3d}{2}-2\beta_1-\beta_3;\tilde a,\tilde b\right)
\label{6331}
\ee
In going from 1st to 2nd equality, we have made a coordinate change $k^0 = 2|\textbf{k}|x +|\textbf{k}|$ while in going from 2nd to 3rd equality, we have made the coordinate change $x=\f{y}{1-y}$. We have also defined $\tilde q = \f{q-1}{q}$ and $ \tilde r = \f{r-1}{r}$ with
\be
q \equiv \f{1}{2}+\f{\omega_2}{2c|\textbf{k}|}-\f{|\textbf{p}_2+\textbf{k}|}{2|\textbf{k}|}\quad;\qquad r = \f{1}{2}+\f{\omega_2}{2c|\textbf{k}|}+\f{|\textbf{p}_2+\textbf{k}|}{2|\textbf{k}|}
\ee
In writing the last equality, we have used the Appell hypergeometric identity
\be
\int_0^1 dy \f{y^{\alpha-1}(1-y)^{\gamma-\alpha-1}}{(1-py)^\beta(1-qy)^{\beta'}} = \f{\Gamma(\alpha)\Gamma(\gamma-\alpha)}{\Gamma(\gamma)}F_1(\alpha;\beta,\beta';\gamma;p,q);\quad (\alpha>0; \;\;\;\gamma-\alpha >0)\label{4.123}
\ee
We want to specialize the result in \eqref{6331} in the limit $c\rightarrow0$. In this limit, we have
\be
q\;\simeq\; \f{\omega_2}{2c|\textbf{k}|} \quad;\qquad r\;\simeq\; \f{\omega_2}{2c|\textbf{k}|}\quad;\qquad \tilde q\;\simeq\; 1-\f{2c|\textbf{k}|}{\omega_2}\quad;\qquad \tilde r\;\simeq\; 1-\f{2c|\textbf{k}|}{\omega_2}
\ee
In this limit, we can simplify the expression using the following Appell hypergeometric identity 
\be
\lim_{x\rightarrow 1}F_1(\alpha;\beta,\beta';\gamma;x;x)\; =\; \lim_{x\rightarrow 1}{_2}F_1(\alpha,\beta+\beta';\gamma;x) \; =\; \f{\Gamma(\gamma)\Gamma(\gamma-\alpha-\beta-\beta')}{\Gamma(\gamma-\alpha)\Gamma(\gamma-\beta-\beta')}
\ee
Using the above identity together with
\be
\int \f{d^{d-1}k}{(2\pi)^{d-1}} \f{1}{|\textbf{p}_1-\textbf{k}|^{d-2\beta_2-1}|\textbf{k}|^{d-2\beta_3-1}}= \frac{\pi^{\f{d-1}{2}}\Gamma\left(\f{d-1}{2}-\beta_2-\beta_3\right)}{\Gamma\left(\f{d-1}{2}-\beta_2\right)\Gamma\left(\f{d-1}{2}-\beta_3\right)}B(\beta_2,\beta_3)|\textbf{p}_1|^{2\beta_2+2\beta_3-d+1}\label{4122we}
\ee
the expression in \eqref{6331} gets simplified and we finally get  
\be
c^{\sigma_1}\int_{-\infty}^\infty d\omega_1\mathcal{A}_3 &=& c^{\sigma_1+1+d-2\beta_1}\;c_{123}b_1(d,\Delta_i)\; \omega_2^{2\beta_1-d}|\textbf{p}_1|^{2\beta_2+2\beta_3-d+1}
\ee
where
\be
b_1(d,\Delta_i)
&=&\frac{ \pi ^{2 d+\frac{1}{2}} 2^{3 d-\Delta _1-\Delta _2-\Delta _3} \Gamma \left(\frac{1}{2} \left(d-2 \Delta _1-1\right)\right)}{\Gamma \left(\Delta _1\right) \Gamma \left(\frac{1}{2} \left(-\Delta _1+\Delta _2+\Delta _3\right)\right) \Gamma \left(\frac{1}{2} \left(-d-\Delta _1+\Delta _2+\Delta _3+2\right)\right)}
\ee
By choosing $\sigma_1=2\beta_1-d-1$, we get the desired Carrollian branch
\be
f_3(\omega_2, \textbf{p}_1) = c_{123}b_1(d,\Delta_i)\; \omega_2^{\Delta_2+\Delta_3-\Delta_1-d}|\textbf{p}_1|^{2\Delta_1-d+1} 
\ee

\subsubsection*{Branch with $\delta(\omega_2)\delta(\omega_1+\omega_3)$}

In this case, we expect 
\be
\lim_{c\rightarrow0}c^{\sigma_2}\mathcal{A}_3 =  f_4(\omega_1, |\textbf{p}_2|)\delta(\omega_2)\delta(\omega_1+\omega_3)
\ee
To fix the form of the function $f_4$, we perform the integration over $\omega_2$. The relevant integral was done in equation \eqref{4108ty} which gives
\be
c^{\sigma_2}\int_{-\infty}^\infty d\omega_2\mathcal{A}_3 &=& c^{\sigma_2+1} a_2(\{\beta_i\}) \int \f{d^dk}{(2\pi)^d} \f{\theta(p_1^0-k^0-|\textbf{p}_1-\textbf{k}|)\theta(k^0-|\textbf{k}|)}{|p_1-k|^{d-2\beta_2}|\textbf{p}_2+\textbf{k}|^{d-2\beta_1-1}|k|^{d-2\beta_3}}\label{631wes}
\ee
where
\be
 a_2 = 2^{2\beta_1-d+1}B\left(\beta_1-\f{d}{2}+1, d-2\beta_1-1\right)\tilde a
\ee
Next, we need to perform the integration over $k^0$. Due to the step theta functions, the range of $k^0$ is restricted. The two theta functions imply following conditions
\be
k^0 \ge |\textbf{k}| \quad\mbox{and}\qquad k^0 \le p_1^0- |\textbf{p}_1-\textbf{k}|\label{6357}
\ee
Both the conditions must be satisfied simultaneously. There are two possibilities 
\be
|\textbf{k}|  \ge p_1^0- |\textbf{p}_1-\textbf{k}|\quad\mbox{or}\qquad |\textbf{k}|  \le p_1^0- |\textbf{p}_1-\textbf{k}|
\ee
In the first case, the two conditions of \eqref{6357} are mutually exclusive and can't be satisfied. Hence, the integration range would be null and the integration would be trivially zero. In the second case, the range of $k^0$ integration would be from $|\textbf{k}|$ to $p_1^0-|\textbf{p}_1-\textbf{k}|$. Hence, the integration over $k^0$ gives
\be
&& \int_{-\infty}^\infty \f{dk^0}{(2\pi)} \f{\theta(p_1^0-k^0-|\textbf{p}_1-\textbf{k}|)\theta(k^0-|\textbf{k}|)}{|p_1-k|^{d-2\beta_2}|k|^{d-2\beta_3}}\non\\[.3cm]
  &=&\int^{p_1^0-|\textbf{p}_1-\textbf{k}|}_{|\textbf{k}|}  \f{dk^0}{(2\pi)} \f{1}{(\f{\omega_1}{c}-k^0-|\textbf{p}_1-\textbf{k}|)^{\f{d}{2}-\beta_2}(\f{\omega_1}{c}-k^0+|\textbf{p}_1-\textbf{k}|)^{\f{d}{2}-\beta_2}(k^0-|\textbf{k}|)^{\f{d}{2}-\beta_3}(k^0+|\textbf{k}|)^{\f{d}{2}-\beta_3}}\non\\[.2cm]
   &=&\f{2^{2\beta_2+2\beta_3-2d+1}}{|\textbf{k}|^{2d-2\beta_2-2\beta_3-1}}\int_{0}^a \f{dx}{(2\pi)} \f{1}{x^{\f{d}{2}-\beta_3}(1+x)^{\f{d}{2}-\beta_3}(s-x)^{\f{d}{2}-\beta_2}(t-x)^{\f{d}{2}-\beta_2}}\non\\[.3cm]
   &=&\f{2^{2\beta_2+2\beta_3-2d+1}}{|\textbf{k}|^{2d-2\beta_2-2\beta_3-1}}\f{1}{(st)^{\f{d}{2}-\beta_2}}\int_{0}^{\f{1}{\tilde a}} \f{dy}{(2\pi)} \f{y^{-\f{d}{2}+\beta_3}(1-y)^{2d-2\beta_2-2\beta_3}}{(1-\tilde sy)^{\f{d}{2}-\beta_2}(1-\tilde ty)^{\f{d}{2}-\beta_2}}
\ee
In going from the first to second equality, we have made the coordinate change $k^0 = (2|\textbf{k}|x +|\textbf{k}|)$ whereas in going from the second to third equality, we have made the coordinate transformation $x=\f{y}{1-y}$. We have defined $\tilde s= \f{1+s}{s}$ and $\tilde t= \f{1+t}{t}$, where
\be
s= -\f{1}{2}+\f{\omega_1}{2c|\textbf{k}|}-\f{|\textbf{p}_1-\textbf{k}|}{2|\textbf{k}|}\quad;\quad t= -\f{1}{2}+\f{\omega_1}{2c|\textbf{k}|}+\f{|\textbf{p}_1-\textbf{k}|}{2|\textbf{k}|}
\ee
As before, we only need to know the above integral in the limit $c\rightarrow0$. In this limit, we have
\be
s\;\simeq\; \f{\omega_1}{2c|\textbf{k}|} \quad;\quad t\;\simeq\; \f{\omega_1}{2c|\textbf{k}|}\quad;\qquad \tilde s\;\simeq\; 1+\f{2c|\textbf{k}|}{\omega_1} \quad;\quad \tilde t\;\simeq\; \f{2c|\textbf{k}|}{\omega_1}\non
\ee
At the leading order $\tilde s \simeq \tilde t\simeq 1$ and the integral can be performed as before to obtain
\be
c^{\sigma_2}\int_{-\infty}^\infty d\omega_2\mathcal{A}_3 &=& c^{\sigma_2+1+d-2\beta_2}c_{123} b_2(d,\Delta_i)\omega_1^{2\beta_2-d} |\textbf{p}_2|^{2\beta_1+2\beta_3-d+1}
\ee
where,
\be
b_2(d,\Delta_i)
&=&\frac{ \pi ^{2 d+\frac{1}{2}} 2^{3 d-\Delta _1-\Delta _2-\Delta _3} \Gamma \left(\frac{1}{2} \left(d-2 \Delta _2-1\right)\right)}{\Gamma \left(\Delta _2\right) \Gamma \left(\frac{1}{2} \left(\Delta _1-\Delta _2+\Delta _3\right)\right) \Gamma \left(\frac{1}{2} \left(-d+\Delta _1-\Delta _2+\Delta _3+2\right)\right)}
\ee

By choosing $\sigma_1=2\beta_2-d-1$, we get the desired Carrollian branch
\be
f_4(\omega_1, |\textbf{p}_2|) = c_{123}b_2(d,\Delta_i)\; \omega_1^{\Delta_1-\Delta_2+\Delta_3-d}|\textbf{p}_2|^{2\Delta_2-d+1} 
\ee

\subsubsection*{Branch with $\delta(\omega_3)\delta(\omega_1+\omega_2)$}
In this case, we expect 
\be
\lim_{c\rightarrow0}c^{\sigma_3}\mathcal{A}_3 =  f_5(\omega_2, |\textbf{p}_3|)\delta(\omega_3)\delta(\omega_1+\omega_2)
\ee
For getting this Carrollian branch, we need to consider the representation \eqref{631wes} of the 3-point function. The integration over $\omega_3$ is given by  
\be
\int_{-\infty}^\infty d\omega_3\f{\theta(-p_3^0-k^0-|\textbf{p}_3+\textbf{k}|)}{((p_3^0+k^0)^2-|\textbf{p}_3+\textbf{k}|^2)^{\f{d}{2}-\beta_2}}
&=&\f{2^{2\beta_2-d+1}c}{|\textbf{p}_3+\textbf{k}|^{d-2\beta_2-1}}B\left(\beta_2-\f{d}{2}+1, d-2\beta_2-1\right)
\ee
Next, we need to perform the integration over $k^0$. Using the same procedure as before and making the similar approximation, we find 
\be
\int_{-\infty}^\infty \f{dk^0}{(2\pi)} \f{\theta(k^0-p_2^0-|\textbf{k}-\textbf{p}_2|)\theta(k^0-|\textbf{k}|)}{|k-p_2|^{d-2\beta_3}|k|^{d-2\beta_1}}=\f{2^{2\beta_1-d+1}}{|\textbf{k}|^{d-2\beta_1-1}}\f{\Gamma(\beta_1-\f{d}{2}+1)\Gamma(d-2\beta_1-1)}{2\pi\Gamma(\f{d}{2}-\beta_1)}\left(\f{\omega_2}{c}\right)^{2\beta_3-d}\non\\
\ee
Note that due to the Heaviside theta functions, the integration variable must satisfy the conditions 
\be
k^0\ge |\textbf{k}|\quad\mbox{and}\qquad k^0\ge p_2^0+ |\textbf{k}-\textbf{p}_2|
\ee
Again there are two possibilities 
\be
|\textbf{k}|  \ge p_2^0+ |\textbf{k}-\textbf{p}_2|\quad\mbox{or}\qquad |\textbf{k}|  \le p_2^0+ |\textbf{k}-\textbf{p}_2|
\ee
The second option is not possible in this case since due to the delta functions $\delta(\omega_3)\delta(\omega_1+\omega_2)$, we must have $\omega_2 < 0$ since $\omega_1$ is always positive. 

Finally, using the identity \eqref{4122we}, we obtain
\be
c^{\sigma_3}\int_{-\infty}^\infty d\omega_3\mathcal{A}_3 &=& c^{\sigma_3+1+d-2\beta_3} c_{123}b_3(d,\Delta_a)\omega_2^{2\beta_3-d} |\textbf{p}_3|^{2\beta_1+2\beta_2-d+1}
\ee
where
\be
b_3(d,\Delta_i)
&=&\frac{ \pi ^{2 d+\frac{1}{2}} 2^{3 d-\Delta _1-\Delta _2-\Delta _3} \Gamma \left(\frac{1}{2} \left(d-2 \Delta _3-1\right)\right)}{\Gamma \left(\Delta _3\right) \Gamma \left(\frac{1}{2} \left(\Delta _1+\Delta _2-\Delta _3\right)\right) \Gamma \left(\frac{1}{2} \left(-d+\Delta _1+\Delta _2-\Delta _3+2\right)\right)}
\ee

By choosing $\sigma_3=2\beta_3-d-1$, we get the desired Carrollian branch
\be
f_5(\omega_2, |\textbf{p}_3|) = c_{123}b_3(d,\Delta_a)\; \omega_2^{\Delta_1+\Delta_2-\Delta_3-d}|\textbf{p}_3|^{2\Delta_3-d+1} 
\ee

\section{Discussion}
\label{s4}
In this paper, we have studied the Carrollian conformal theories in the momentum space. We have focused on the scalar operators and only used those aspects of the theories which are dictated by the symmetry. For this, we have used the Ward identities corresponding to the Carrollian conformal symmetry. By taking a  transform of the position space Ward identities, we first obtained these Ward identities in the momentum space and then solved them for the 2 and 3 point functions. Unlike standard conformal field theories, the Ward identities of Carrollian conformal theories admit multiple structures reflecting the fact that the Carrollian manifold has a null direction. The 2-point function has two independent structures known as electric and magnetic branches. On the other hand, the 3-point function admits 5 independent branches. For the 2-point function, both the branches are completely fixed up to some normalization constant. On the other hand, for the 3-point functions, four of the branches are fixed by symmetries whereas one of the branch is only fixed up to a function of ratios of the Carrollian energies.

\vspace*{.07in}We also considered the Carrollian limit of the standard conformal field theory correlators. In the Carrollian limit $c\rightarrow0$, we showed how to recover the different branches of the Carrollian conformal theories in the momentum space. We considered both Wightman as well as time-ordered correlators. If we put the CFT on the AdS boundary, then their Carrollian limit $c\rightarrow 0$ can be interpreted as the flat limit of AdS from the boundary perspective. In the bulk, the flat limit corresponds to taking the large AdS radius limit $R\rightarrow\infty$. By combining (and considering all massless particles) the bulk analysis in momentum space done in \cite{Marotta:2024sce,Farrow:2018yni,Lipstein:2019mpu,spin2} with the Carrollian limit of the boundary CFT developed here, one
would obtain a unified momentum–space description of bulk and boundary
dynamics in flat spacetime. Such a framework could provide new insights into
the emergence of flat Carrollian holography from the flat and ultra-relativistic
limits of AdS/CFT. A detailed presentation of the bulk analysis, including the
flat limit of AdS correlators in momentum space for massless fields, will be
discussed in more detail in \cite{bulk}.

\vspace*{.07in}As mentioned in the introduction, one application of the momentum space analysis considered in this paper is in the study of the analytic properties of Carrollian conformal theories on the flat space null boundaries. The explicit analytic structures of Carrollian conformal correlators in the momentum space would be useful for this purpose. This may also require the regularization of Carrollian conformal correlators for special values of the Carrollian dimensions and a better understanding of the logarithmic terms discussed earlier. We leave these questions for the future work.  

\vspace*{.07in}In this paper, we have not considered the 4 and higher point correlators. However, these correlators must also be a solution of the Ward identities. In the usual CFTs, the higher point functions in the momentum space can be represented geometrically as simplex integrals \cite{Bzowski:2019kwd, Bzowski:2020kfw}. The structure of the higher point Carrollian conformal correlators are expected to be more complicated due to the presence of several branches as well as the fact that symmetries do not completely fix even the 3-point correlators. Hence, we expect more freedom in the higher point Carrollian conformal correlators. 

\vspace*{.07in}We have focused only on the null boundaries in this work. This is relevant for the massless bulk fields. However, for the massive bulk fields, we also need to take into account the timelike and spacelike components of the flat space boundaries \cite{suvrat1,suvrat2,Have:2024dff}. The theories living on these boundary components are still very poorly understood compared to the theories on the null boundaries. It would be interesting to see whether we can use similar momentum space techniques used in this paper to study the theories on the time like and spacelike boundaries. 

\vspace{0.07in}A natural extension of this work is the generalization of our analysis to spinning correlators. In relativistic conformal field theories, tensor structures are organized into a finite set associated with free-field realizations. In the Carrollian case, the appearance of multiple branches already at the scalar level suggests that the space of allowed structures may be significantly enlarged for the spinning correlators.
This should not be viewed as an inconsistency, but rather as a reflection of the fact that Carrollian conformal symmetry imposes weaker constraints compared to its relativistic counterpart. As a consequence, symmetry alone is not expected to fully determine the structure of correlators, and additional consistency conditions, such as operator-product expansions and constraints from four-point functions (including factorization and crossing symmetry), are likely required to further restrict the space of solutions. We emphasize, however, that the scalar analysis presented here does not by itself determine the structure of spinning or stress-tensor correlators.
Understanding how the above mentioned additional conditions constrain spinning correlators, and whether they reduce the apparent freedom to a smaller set of physically realizable structures, remains an important open problem, which we leave for future investigation.

\bigskip

{\bf Acknowledgement:} 
{ We are thankful to Chandramouli Chowdhury, Sitender Kashyap and Ana-Maria Raclariu for discussions. We also thank Kostas Skenderis for the comments on the draft. MV is thankful to INFN Napoli and FZU Prague for hospitality while this work was in progress. MV is also supported in part by the “Young Faculty Research Seed Grant'' (IITI/YFRSG-Dream Lab/2024-25/Phase-IV/02) and its Dream Lab (IITI/YFRSG-Dream Lab/2024-25/Phase-III/02) and PRIUS ( ITII/YFRSG-PRIUS/2023-24/Phase-I/03) components of IIT Indore. AS is supported by the Research Studentship in Mathematical Science, University of Southampton, and funding from the Government of Karnataka, India.}

\appendix

\section{Conventions and useful identities}
\label{sec:appenA}
Throughout this paper, we have worked in Minkowski spacetime of dimension $d+1$. Hence, the boundary Carrollian CFT theory lives on the $d$ dimensional manifold $R\times R^{d-1}$. The position space coordinates of the Carrollian CFT are denoted by $(u,\textbf{x})$. The corresponding momentum space coordinates are denoted by $(\omega,\textbf{p})$. The magnitude of the spatial momenta $\textbf{p}_a$ is denoted by $k_a$. The dimension of the Anti de Sitter spacetime is also taken to be $d+1$ with its dual Lorentzian CFT being $d$ dimensional. 

We have considered Carrollian CFTs in both position as well as momentum spaces. Our convention for the Fourier transform is
\be
f(\omega) = \int_{-\infty}^\infty dx e^{i\omega x}f(x)\qquad\implies\qquad f(x) = \int_{-\infty}^\infty \f{d\omega}{2\pi} e^{-i\omega x}f(\omega)
\ee
and its generalization to higher dimensions. In the above convention, we have following delta function identities
\be
\delta^D(\textbf{k}) = \int \f{d^D\textbf{x}}{(2\pi)^D}e^{i\textbf{k}\cdot\textbf{x}}\qquad\mbox{and}\qquad \delta^D(\textbf{x}) = \int \f{d^D\textbf{x}}{(2\pi)^D}e^{-i\textbf{k}\cdot\textbf{x}}
\ee

For various derivations, we need the Fourier transform of the power law functions. For $D$ dimensional spacelike vectors, the basic identity is given by
\be
\int d^Dx \f{e^{i\textbf{p}\cdot\textbf{x}}}{|\textbf{x}|^{\Delta}}&=& \f{\pi^{D/2}2^{D-\Delta}}{\Gamma(\f{\Delta}{2})}\Gamma\Bigl( \f{D-\Delta}{2} \Bigl)|\textbf{p}|^{\Delta-D} \label{c252}
\ee
This can be inverted to obtain
\be
\f{1}{|\textbf{x}|^\Delta}= \f{(\pi)^{\f{D}{2}}\;2^{D-\Delta}}{\Gamma\Bigl(\f{\Delta}{2}\Bigl)}\Gamma\Bigl(\f{D-\Delta}{2}\Bigl)\int \f{d^D\textbf{k}}{(2\pi)^D}\; e^{-i\textbf{k}\cdot \textbf{x}}|\textbf{k}|^{\Delta-D}\label{a232}
\ee
For 3-point function in momentum space (Carrollian) CFTs, we encounter the so called triple-K integrals $I_{\alpha(\beta_1\beta_2\beta_3)}$. It is defined by
\be
I_{\alpha(\beta_1\beta_2\beta_3)}(k_1,k_2,k_3)&=&
\int_0^\infty dy\; y^{\alpha}\prod_{a=1}^3(k_a)^{\beta_a}K_{\beta_a}(k_ay)\label{triplekwer}
\ee
and $k_i$ denotes the magnitude of the momenta $k_i^\mu$ in some dimension $D$ and $K_\beta$ denotes the modified bessel function of the second kind. A useful integral representation of the triple K integral is obtained by Fourier transforming the position space 3-point function in $D$ dimensions and is given by \cite{Bzowski:2013sza}
\be
\int \f{d^{D}k}{(2\pi)^{D}}\f{1}{|\textbf{p}_2+\textbf{k}|^{2\uptau_1}|\textbf{p}_1-\textbf{k}|^{2\uptau_2}|\textbf{k}|^{2\uptau_3}}=C(\uptau_a)I_{\f{D-2}{2}\Bigl(\f{D}{2}-\uptau_2-\uptau_3,\f{D}{2}-\uptau_1-\uptau_3,\f{D}{2}-\uptau_1-\uptau_2\Bigl)}({k}_1,{k}_2,{k}_3) \label{639df}
\ee
where the constant $C(\uptau_a)$ is given by
\be
C(\uptau_a)&=&\f{\pi^{-\f{D}{2}}2^{4-\f{3D}{2}}\Gamma\left(\f{D}{2}-\uptau_1\right)\Gamma\left(\f{D}{2}-\uptau_2\right)\Gamma\left(\f{D}{2}-\uptau_3\right)}{\Gamma\left(\uptau_1\right)\Gamma\left(\uptau_2\right)\Gamma\left(\uptau_3\right)\Gamma\left(d-\uptau_1\right)\Gamma\left(d-\uptau_2\right)\Gamma\left(d-\uptau_3\right)\Gamma\left(d-\uptau_1-\uptau_2-\uptau_3\right)}
\ee
For more details about these integrals, see, e.g., \cite{Bzowski:2013sza}.  
\section{Lorentzian CFT correlators }
\label{sec:BLor}
\subsection{2-point function}
The two-point function of scalar operators with the same conformal dimension $\Delta$ in Lorentzian space can be obtained from the corresponding Euclidean expression by performing a Wick rotation $t \rightarrow i t$. This prescription is, however, ambiguous and leads to two distinct Lorentzian Wightman functions ($\epsilon>0$)\footnote{We follow the prescription given in Ref.~\cite{Bautista:2019qxj}}:
\begin{eqnarray}
\langle {\cal O}(t_1,{\bf x}_1)\,{\cal O}(t_2,{\bf x}_2)\rangle
&=& \frac{C_2(\Delta)}{\left[-(c\,t_{12}-i\epsilon)^2+|{\bf x}_{12}|^2\right]^\Delta}\,, \nonumber\\[4pt]
\langle {\cal O}(t_2,{\bf x}_2)\,{\cal O}(t_1,{\bf x}_1)\rangle
&=& \frac{C_2(\Delta)}{\left[-(c\,t_{12}+i\epsilon)^2+|{\bf x}_{12}|^2\right]^\Delta}\,.
\end{eqnarray}
The Lorentzian time-ordered two-point function then takes the form (see, e.g.,~\cite{2406.19343})
\begin{eqnarray}
\langle T\big\{ {\cal O}(t_1,{\bf x}_1)\,{\cal O}(t_2,{\bf x}_2)\big\}\rangle
= \frac{C_2(\Delta)}{\left[-c^2 t_{12}^2+|{\bf x}_{12}|^2+i\epsilon \right]^\Delta}\,.
\end{eqnarray}
These correlation functions differ only by the $i\epsilon$ prescription.

\subsection{3-point function}
\label{3-points}

{Upto an overall normalization (which may depend on $d$ and $\Delta_a$), the CFT Wightman 3-point function in the position space is given by the following $i\epsilon$ prescription (see, e.g., \cite{Kravchuk:2018htv})}
\begin{eqnarray}
&&\hspace*{-.4in}\langle {\cal O}(t_1,{\bf x}_1)\,{\cal O}(t_2,{\bf x}_2)\,{\cal O}(t_3,{\bf x}_3)\rangle\non\\
&=& \frac{1}{({ x}_{12}^2+i\epsilon \,t_{12})^{\frac{\Delta_1+\Delta_2-\Delta_3}{2}}({ x}_{23}^2+i\epsilon t_{23})^{\frac{\Delta_2+\Delta_3-\Delta_1}{2}}({ x}_{13}^2+i\epsilon \,t_{13})^{\frac{\Delta_1+\Delta_3-\Delta_2}{2}}}\non
\end{eqnarray}
with $x^2=-t^2+ |{\bf x}|^2$. 

\vspace*{.07in}On the other hand, the time ordered Lorentzian CFT three-point function is given by a different $i\epsilon$ prescription (see, e.g., \cite{Kravchuk:2018htv,2406.19343})
\begin{eqnarray}
\Bigl\langle T\left({\cal O}(x_1)\,{\cal O}(x_2)\,{\cal O}(x_3)\right)\Bigl\rangle =\frac{1}{({x}_{12}^2+i\epsilon )^{\frac{\Delta_1+\Delta_2-\Delta_3}{2}}({ x}_{23}^2+i\epsilon)^{\frac{\Delta_2+\Delta_3-\Delta_1}{2}}({x}_{13}^2+i\epsilon)^{\frac{\Delta_1+\Delta_3-\Delta_2}{2}}}
\label{B.216}
\end{eqnarray}
We want to find the momentum representation of the above time ordered correlator. For our purposes, it will be sufficient to obtain an integral representation of the correlator in the momentum space. This will turn out to be enough for the discussion of Carrollian limit in section \ref{Time-ordered}. To obtain this representation, we first consider the Fourier transform of \eqref{B.216}. 
\begin{eqnarray}
&&\hspace*{-.4in}{\cal A}({ p}_1,{ p}_2,{ p}_3)\non\\
&=&
\int d^d x_1\, d^d x_2, d^d x_3 
\prod_{i=1}^3 e^{-i p_i \cdot x_i},
\bigl\langle T\left({\cal O}(x_1)\,{\cal O}(x_2)\,{\cal O}(x_3)\right)\bigl\rangle 
\nonumber\\
&=& (2\pi)^d \delta^{(d)}(p_1+p_2+p_3)
\int 
\frac{d^d x_{13}\, d^d x_{23}\;\; e^{-i p_1 \cdot x_{13} - i p_2 \cdot x_{23}} }{\big[(x_{13}-x_{23})^2 + i\epsilon\big]^{\Delta_{12}}
(x_{23}^2+i\epsilon)^{\Delta_{23}} (x_{13}^2+i\epsilon)^{\Delta_{13}}}.
\label{A3pt_intermediate}
\end{eqnarray}
We now consider the following representation of unity
\begin{eqnarray}
1 =\int d^d y \, \delta^{(d)}\left[y-(x_{13}-x_{23})\right]
=\int d^d y \int \frac{d^d k}{(2\pi)^d} \,
e^{- i k \cdot \left[y-(x_{13}-x_{23})\right] } \, .
\end{eqnarray}
Inserting this inside the integral in \eqref{A3pt_intermediate} and changing variables \(x_{13}\to x_1\), \(x_{23}\to x_2\), \(y\to x_3\), we obtain
\begin{eqnarray}
&&{\cal A}({ p}_1,{ p}_2,{ p}_3)=(2\pi)^d \delta^d(p_1+p_2+p_3) \int \frac{d^d k}{(2\pi)^d}\int \prod_{a=1}^3 dx_a\,\frac{\;~e^{-i(p_1-k)\cdot x_1-i(p_2+k)\cdot x_2-ik\cdot x_3}}{(x_3^2+i\epsilon)^{\Delta_{12}}(x_{2}^2+i\epsilon)^{\Delta_{23}}(x_{1}^2+i\epsilon)^{\Delta_{13}}}\nonumber\\
\end{eqnarray}
Finally, by using the Fourier-transform identity (see, e.g., equation (2.3) of Ref.\cite{1807.07003})
\begin{eqnarray}
\int d^d x\,\frac{e^{-ip\cdot x}}{[x^2+i\epsilon]^\Delta}=
{\cal N}_\Delta (p^2-i\epsilon)^{\Delta -\frac{d}{2}}\qquad,\qquad {\cal N}_\Delta= -i \frac{\pi^{\frac{d}{2}}\,\Gamma[\frac{d}{2} -\Delta]}{2^{2\Delta -d}\Gamma[\Delta]}
\end{eqnarray}
we obtain the following momentum-space representation
\begin{eqnarray}
   \mathcal{A}_3(p_1,p_2,p_3) &=&\delta^d(p_1+p_2+p_3)\,{\cal N}(\Delta_{1}\,,\Delta_2\,,\Delta_{3})\,\Bigl\langle\hspace*{-.1cm}\Bigl\langle {\cal O}_{\Delta_1}(p_1)\,{\cal O}_{\Delta_2}(p_2)\,{\cal O}_{\Delta_3}(p_3)\Bigl\rangle\hspace*{-.1cm}\Bigl \rangle\label{B.219df}
   \end{eqnarray}
   where we have defined
   \begin{eqnarray}
  \Big\langle\hspace*{-.1cm}\Bigl\langle {\cal O}_{\Delta_1}(p_1)\,{\cal O}_{\Delta_2}(p_2)\,{\cal O}_{\Delta_3}(p_3)\Bigl\rangle\hspace*{-.1cm}\Bigl \rangle  =\int \frac{d^d k}{(2\pi)^d} \,\frac{1}{(k^2-i\epsilon)^{\frac{d}{2} -\Delta_{12}}\,[( p_2+k)^2-i\epsilon]^{\frac{d}{2}-\Delta_{23}} [(p_1-k)^2-i\epsilon]^{\frac{d}{2} -\Delta_{13}}}\nonumber\\
\end{eqnarray}
and
\begin{eqnarray}
\mathcal{N}(\Delta_{1}\,,\Delta_2\,,\Delta_{3})= {\cal N}\,\prod_{i=1}^3 \left[-i \frac{\pi^{\frac{d}{2}}\,\Gamma[\frac{d}{2} -\Delta_i]}{2^{2\Delta_i -d}\Gamma[\Delta_i]}\right]
\end{eqnarray}

\section{Delta distribution identities}
\label{appenc}
In solving the Ward identities, we need to deal with the derivatives of momentum conserving delta functions such as $\delta^d(\sum_{a=1}^n p_a^\mu)$. In this appendix, we describe the method and summarize various identities involving the derivatives of these delta functions. We shall be working in the Carrollian dimension d with indices $\mu, \nu$ running from 0 to $d-1$ whereas $i,j$ running from 1 to $d-1$. We start by defining the \textquotedblleft total\textquotedblright\; momenta 
\be
P^\mu= \sum_{a=1}^n p^\mu_a
\ee
The derivative of the above delta function with respect to a momentum component $p^\mu_a$ is given by
\begin{eqnarray}
\frac {\partial}{\partial p_a^\mu} \delta^{d}(\sum_{b=1}^n p_b^\mu)= \left( \frac{\partial \sum_bp_b^\mu}{\partial p_a^\mu}\right) \frac{\partial}{\partial P^\mu} \delta^{d}(P^\mu)=\frac{\partial}{\partial P^\mu} \delta^{d}(P^\mu)\label{a88} 
\end{eqnarray} 
The spacetime index $\mu$ is not summed over in the above expression. The above identity implies
\begin{eqnarray}
\frac {\partial}{\partial p_a^\mu} \delta^{d}(\sum_{b=1}^n p_b^\mu)=\frac {\partial}{\partial p_c^\mu} \delta^{d}(\sum_{b=1}^n p_b^\mu)\quad;\qquad\forall a,c
\end{eqnarray}
This shows that we can take the derivative of the momentum conserving delta function with respect to the momentum of any field $p_a$ and it will give the same result. This is very useful. E.g., suppose, we have an expression $\sum_{a=1}^np_a^\mu\frac{\partial}{\partial p_a^\mu}$ (where $\mu$ index is summed over) which we want to simplify. We consider an arbitrary test function  
 $f(P)$ and the integral
\begin{eqnarray}
\int d^dP \,f(P)\sum_{a=1}^np_a^\mu\frac{\partial}{\partial p_a^\mu}\,\delta^d(P)&=&\int d^dP \,f(P)\sum_{a=1}^np_a^\mu\frac{\partial}{\partial p_n^\mu}\,\delta^d(P)\nonumber\\
&=&\int d^dP \,f(P)\,P^\mu\,\frac{\partial}{\partial p_n^\mu}\,\delta^d(P)\non\\
&=&\int d^dP \,f(P)\,P^\mu\frac{\partial}{\partial P^\mu}\delta^d(P)\nonumber\\
&=&
-\int d^dP \,P^\mu\,\frac{\partial f(P)}{\partial P^\mu}\,\delta^d(P)-\int d^dP \,f(P)\frac{\partial}{\partial P^\nu}P^\nu\,\delta^d(P)\nonumber\\
&=& - d\int d^dP \,f(P)\,\delta^d(P)\label{319}
\end{eqnarray}
The first term in the second last line vanishes since $\delta^d(P)$ imposes $P^\mu=0$. Equation \eqref{319} gives the delta function identity
\begin{eqnarray}
\sum_{a=1}^np_a^\mu\frac{\partial}{\partial p_a^\mu}\,\delta^d(P)\;=\;-d  \,\delta^d(P)\label{4210}
\end{eqnarray}
Two useful special cases of this identity are 
\begin{eqnarray}
\sum_{a=1}^n\omega_a\frac{\partial}{\partial \omega_a}\delta(\sum_{a=1}^n\omega_a)=-\delta(\sum_{a=1}^n\omega_a)\quad;\qquad\sum_{a=1}^np_a^i\frac{\partial}{\partial p_a^i}\delta^{d-1}(\sum_{a=1}^n\textbf{p}_a)=-(d-1)\delta^{d-1}(\sum_{a=1}^n\textbf{p}_a)\label{ident1}
\end{eqnarray}
For dealing with special conformal Ward identity, we need delta function identities with more than one derivatives. To see how to deal with such identities, consider the following test integral
\begin{eqnarray}
\int d^d P \,f(P) \sum_{a=1}^n  \,\omega_a  \frac{\partial}{\partial p_i^a}\frac{\partial}{\partial p_a^i}  \delta^d(\sum_{b=1}^n p_b)&=&\int d^dP \,f(P)\, P^0 \frac{\partial^2}{\partial P_i\partial P^i} \delta^d(P)\nonumber\\
&=&\int d^dP \,\delta^d(P)\, P^0 \frac{\partial^2}{\partial P_i\partial P^i} f(P)
\end{eqnarray}
In going to the 2nd equality, we have used $\frac{\partial P^0}{\partial P^i}=0$. The last line vanishes since the delta-function imposes $P^0=0$. This gives the delta function identity
\begin{eqnarray}
\sum_{a=1}^n  \,\omega_a  \frac{\partial}{\partial p_i^a}\frac{\partial}{\partial p_a^i}  \delta^d(\sum_{b=1}^n p_b)\;=\;0\label{a94}
\end{eqnarray}
Some other useful identities which can be proved similarly are
\be
\sum_{a=1}^n p_i^a \frac{\partial^2}{\partial p_{ja}\partial p^j_a}\delta^{d-1}({\bf p}_1+\dots +{\bf p}_n)\;=\;-2 \frac{\partial}{\partial P^i}\delta^{d-1}(\textbf{P}) 
\label{C.1}\\[.2cm]
\sum_{a=1}^n p_{aj}\frac{\partial^2}{\partial p_j^a\partial p_a^i}\delta^{d-1}({\bf p}_1+\dots +{\bf p}_n)\;=\;-d \frac{\partial}{\partial P^i}\delta^{d-1}(\textbf{P})\label{C.5}
\ee

\section{Fourier transform of position space Carrollian correlators}
\label{append}
In this appendix, we consider the Fourier transform of the 2 and 3 point position space Carrollian conformal correlators which were reviewed in section \ref{position_rev}. While doing the Fourier transform, we shall consider generic values of $\Delta_a$ for which the Fourier transform exists. For some special values of $\Delta_a$, such as the one discussed at the end of section \ref{sec4:Ward}, there may be divergence issues requiring appropriate regularizations.

\subsection{2-point function}
The 2-point function of scalar operators in the d dimensional Carrollian conformal theories in the position space was given in equation \eqref{2ptposfg}
\be
\left\langle \mathcal{O}(\textbf{x}_1,u_1)\mathcal{O}(\textbf{x}_2,u_2)\right\rangle = \f{C_1}{|\textbf{x}_{12}|^{\Delta_1+\Delta_2}}\delta_{\Delta_1,\Delta_2} +\f{C_2}{|u_{12}|^{\Delta_1+\Delta_2-d+1}}\delta(\textbf{x}_{12})
\ee
The first term is the magnetic branch whereas the second term is electric branch. The Fourier transform of the magnetic branch is given by 
\be
&&\hspace*{-.4in}\Bigl\langle\mathcal{O}_1( \omega_1,\textbf{p}_1)\mathcal{O}_2( \omega_2,\textbf{p}_2 )\Bigl\rangle_M\non\\[.2cm]
&=& \int du_1 du_2 e^{iu_1\omega_1+iu_2\omega_2}\int d^{d-1}x_1d^{d-1}x_2\;  \f{C_1}{|\textbf{x}_{12}|^{\Delta_1+\Delta_2}}\delta_{\Delta_1,\Delta_2}  \; e^{i\textbf{p}_1\cdot \textbf{x}_1+i\textbf{p}_2\cdot \textbf{x}_2 }\non\\
&=& C_1\delta_{\Delta_1,\Delta_2}(2\pi)^2\delta(\omega_1)\delta(\omega_2)(2\pi)^{d-1}\delta^{d-1}(\textbf{p}_1+\textbf{p}_2)\int d^{d-1}y\f{e^{i\textbf{p}_1\cdot \textbf{y}}}{\textbf{y}^{2\Delta_1}}\non\\
&=&C_1\delta_{\Delta_1,\Delta_2}(2\pi)^2(2\pi)^{d-1}\delta(\omega_1)\delta(\omega_2)\delta^{d-1}(\textbf{p}_1+\textbf{p}_2)\f{2^{d-1-2\Delta_1}\pi^{\f{d-1}{2}}\Gamma(\f{d-1}{2}-\Delta_1)}{\Gamma(\Delta_1)}|\textbf{p}_1|^{2\Delta_1-d+1}\non\\
\ee
In going to the 2nd equality, we have made a change of variable $(\textbf{x}_1,\textbf{x}_2)\rightarrow (\textbf{y},\textbf{x}_2)$ where $\textbf{y}=\textbf{x}_1-\textbf{x}_2$. In going to the 3rd equality, we have used the identity \eqref{c252}. 

\vspace*{.07in}Next, the Fourier transform of the electric branch is given by 
\be
&&\hspace*{-.4in}\Bigl\langle\mathcal{O}_1( \omega_1,\textbf{p}_1)\mathcal{O}_2( \omega_2,\textbf{p}_2 )\Bigl\rangle_E\non\\[.2cm]
&=& \int du_1 du_2 e^{iu_1\omega_1+iu_2\omega_2}\f{C_2}{|u_{12}|^{\Delta_1+\Delta_2-d+1}}\int d^{d-1}x_1d^{d-1}x_2\;   \; e^{i\textbf{p}_1\cdot \textbf{x}_1+i\textbf{p}_2\cdot \textbf{x}_2 }\delta^{d-1}(\textbf{x}_{12})\non\\
&=& \int du_1 du_2 e^{iu_1\omega_1+iu_2\omega_2}\f{C_2}{|u_{12}|^{\Delta_1+\Delta_2-d+1}}\int d^{d-1}x_1\;   \; e^{i(\textbf{p}_1+\textbf{p}_2)\cdot \textbf{x}_1}\non\\
&=& C_2(2\pi)^{d-1}\delta^{d-1}(\textbf{p}_1+\textbf{p}_2)\int  du_2e^{i(\omega_1+\omega_2)u_2} \int dv e^{iv\omega_1}\f{1}{|v|^{\Delta_1+\Delta_2-d+1}}\non\\
&=& C_2(2\pi)(2\pi)^{d-1}\delta(\omega_1+\omega_2)\delta^{d-1}(\textbf{p}_1+\textbf{p}_2)2^{d-\Delta_1-\Delta_2}\f{(\pi)^{\f{1}{2}}\Gamma(\f{d-\Delta_1-\Delta_2}{2})}{\Gamma(\f{\Delta_1+\Delta_2-d+1}{2})}|\omega_1|^{\Delta_1+\Delta_2-d}
\ee
In going to the last equality, we have used the integral
\be
\int_{-\infty}^\infty dy e^{iv\omega_1}\f{1}{|v|^{\Delta_1+\Delta_2-d+1}}&=&\int_{-\infty}^0 dv e^{iv\omega_1}\f{1}{|v|^{\Delta_1+\Delta_2-d+1}}+\int_{0}^\infty dv e^{iv\omega_1}\f{1}{|v|^{\Delta_1+\Delta_2-d+1}}\non\\
&=&2\int^{\infty}_0 dv \f{\mbox{cos}(v\omega_1)}{|v|^{\Delta_1+\Delta_2-d+1}}\non\\
&=& 2 \Gamma(d-\Delta_1-\Delta_2)\sin \left(\frac{1}{2} \pi  \left(-d+\Delta _1+\Delta _2+1\right)\right)|\omega_1|^{\Delta_1+\Delta_2-d}\non
\ee
The above integral converges for $0< d-\Delta_1-\Delta_2<1$. The sine factor can be replaced in terms of the Gamma functions by using the identities  
\be
\Gamma(z)\Gamma(1-z)=\f{\pi}{\sin(\pi z)}\quad;\qquad \Gamma(z)\Gamma\Bigl(z+\f{1}{2}\Bigl) = 2^{1-2z}\sqrt{\pi}\,\Gamma(2z)\label{duplication}
\ee

\subsection{3-point function}

The 3-point position space Carrollian conformal correlators were given in equation \eqref{39we}. 
\be
&&\hspace*{-.9in}\bigl\langle \,\mathcal{O}_1(u_1,\textbf{x}_1) \mathcal{O}_2(u_2,\textbf{x}_2)\mathcal{O}_3(u_3,\textbf{x}_3)\bigl\rangle\non\\
&=& G_1(u_1,u_2,u_3)\delta^{d-1}(\textbf{x}_{12})\delta^{d-1}(\textbf{x}_{23})+G_2(\textbf{x}_1,\textbf{x}_2,\textbf{x}_3) +G_3(\textbf{x}_{12},u_{23})\delta^{d-1}(\textbf{x}_{23})\non\\
&&+\;G_4(\textbf{x}_{23},u_{31})\delta^{d-1}(\textbf{x}_{31})+G_5(\textbf{x}_{31},u_{12})\delta^{d-1}(\textbf{x}_{12})\label{39wes}
\ee
where, the form of functions $G_i$ are given in section  \ref{sec:3pointpos}. 
There are a total of 5 branches in the 3-point function. We consider each of them one by one denoting the corresponding Fourier transform by the subscript $G_i$. The Fourier transform of the 1st term in \eqref{39wes} is given by (denoting $\alpha=2d-2-\Delta_t$) 
\be
&&\hspace*{-.4in}\Bigl\langle\mathcal{O}_1( \omega_1,\textbf{p}_1)\mathcal{O}_2( \omega_2,\textbf{p}_2 )\mathcal{O}_3( \omega_3,\textbf{p}_3 )\Bigl\rangle_{G_1}\non\\[.2cm]
&=& \int \prod_{a=1}^3du_ad^{d-1}\textbf{x}_a\; e^{i(\omega_1u_1+\omega_2u_2+\omega_3u_3)}e^{i(\textbf{p}_1\cdot \textbf{x}_1+\textbf{p}_2\cdot \textbf{x}_2+\textbf{p}_3\cdot \textbf{x}_3)}u_{23}^\alpha f\left(\f{u_{13}}{u_{23}}\right)\delta^{d-1}(\textbf{x}_{12})\delta^{d-1}(\textbf{x}_{23})\non\\
&=& \int d^{d-1}\textbf{x}_3\,e^{i(\textbf{p}_1+\textbf{p}_2+\textbf{p}_3)\cdot \textbf{x}_3}\int dU\; e^{i(\omega_1+\omega_2+\omega_3)U} \int drdt\; e^{i(\omega_1t+\omega_2r)}r^\alpha f\left(\f{t}{r}\right)\non\\
&=& (2\pi)^{d-1}\delta^{d-1}(\textbf{p}_1+\textbf{p}_2+\textbf{p}_3)2\pi\delta(\omega_1+\omega_2+\omega_3) \int dr\; r^{\alpha+1}e^{-i\omega_2r} \int dye^{i\omega_1ry}f(y)\non\\
&=&  (2\pi)^{d-1}\delta^{d-1}(\textbf{p}_1+\textbf{p}_2+\textbf{p}_3)(2\pi)\delta(\omega_1+\omega_2+\omega_3) \int dr\; r^{\alpha+1}e^{i\omega_2r} \hat f(\omega_1r)\non\\
&=& (2\pi)^{d-1}\delta^{d-1}(\textbf{p}_1+\textbf{p}_2+\textbf{p}_3)(2\pi)\delta(\omega_1+\omega_2+\omega_3) \omega_1^{-\alpha-2}\int ds\; s^{\alpha+1}e^{i\f{\omega_2}{\omega_1}s} \hat f(s)\non\\
&=& (2\pi)^{d-1}\delta^{d-1}(\textbf{p}_1+\textbf{p}_2+\textbf{p}_3)(2\pi)\delta(\omega_1+\omega_2+\omega_3) \omega_1^{\Delta_t-2d} \hat G_1\left(\f{\omega_2}{\omega_1}\right)\label{b237}
\ee
In going to the second equality, we have performed the $x_1$ and $x_2$ integrals using the delta functions and have made a change of variable $u_3 =U, u_{23} = r $ and $ u_{13} = t$. In going to the 3rd equality, we have made a change of variable $t=yr$. Finally, in going to the 5th equality, we have made a change of variable $\omega_1r=s$.

\vspace*{.07in}The expression \eqref{b237} agrees with \eqref{3.57}. On the other hand, if we consider example \eqref{314we}, its Fourier transform is given by
\be
&&\hspace*{-.4in}\Bigl\langle\mathcal{O}_1( \omega_1,\textbf{p}_1)\mathcal{O}_2( \omega_2,\textbf{p}_2 )\mathcal{O}_3( \omega_3,\textbf{p}_3 )\Bigl\rangle_{G_1}\non\\[.2cm]
&=&(2\pi)^{d-1}\delta^{d-1}(\textbf{p}_1+\textbf{p}_2+\textbf{p}_3) \sum_{p,q,r}B_{pqr}\int  
du_1du_2du_3\f{e^{i(u_1\omega_1+u_{2}\omega_2+u_3\omega_3)}}{|u_{12}|^p|u_{23}|^q|u_{31}|^r}\non\\
&=&(2\pi)^{d-1}\delta^{d-1}(\textbf{p}_1+\textbf{p}_2+\textbf{p}_3)(2\pi)\delta(\omega_1+\omega_2+\omega_3)\sum_{p,q,r}B_{pqr}C_{pqr}\int \f{d\uptau}{2\pi} |\uptau|^{p-1}|\omega_2-\uptau|^{q-1}|\omega_1+\uptau|^{r-1}\non\\
\label{b238we}
\ee
The sum is over values of $p,q,r$ satisfying the condition $p+q+r = \Delta_1+\Delta_2+\Delta_3-2d+2$. In going to the second equality, we have made a change of variable $u_{13} =x\,, u_{23} = y, u_3=z$ and used the identity \eqref{a232} for $D=1$. The constant $C_{pqr}$ is given by
\be
C_{pqr} = \f{(\pi)^{\f{3}{2}}\;2^{3-p-q-r}}{\Gamma\Bigl(\f{p}{2}\Bigl)\Gamma\Bigl(\f{q}{2}\Bigl)\Gamma\Bigl(\f{r}{2}\Bigl)}\Gamma\Bigl(\f{1-p}{2}\Bigl)\Gamma\Bigl(\f{1-q}{2}\Bigl)\Gamma\Bigl(\f{1-r}{2}\Bigl)
\ee
The integral in \eqref{b238we} can be shown to be a triple K integral and the final result is given by
\be
&&\hspace*{-.4in}\Bigl\langle\mathcal{O}_1( \omega_1,\textbf{p}_1)\mathcal{O}_2( \omega_2,\textbf{p}_2 )\mathcal{O}_3( \omega_3,\textbf{p}_3 )\Bigl\rangle_{G_1}\non\\[.3cm]
&=&(2\pi)^{d-1}\delta^{d-1}(\textbf{p}_1+\textbf{p}_2+\textbf{p}_3)(2\pi)\delta(\omega_1+\omega_2+\omega_3) \sum_{p,q,r}\f{\;2^{\f{11}{2}-p-q-r}B_{pqr}}{\Gamma(\f{p}{2})\Gamma(\f{q}{2})\Gamma(\f{r}{2})}\non\\[.2cm]
&&\f{\pi}{\Gamma\Bigl(\f{p+q+r-1}{2}\Bigl)}I_{-\f{1}{2}\{\f{q+r-1}{2},\f{p+r-1}{2},\f{p+q-1}{2}\}}(\omega_1,\omega_2,\omega_3)\non
\ee
 While evaluating the integrals to obtain the above expression, we have assumed that $p,q$ and $r$ are less than 1. Hence, for values of $p,q,r$ not satisfying this condition, the above expression should be taken as analytically continued expression. 

\vspace*{.07in}The Fourier transform of the 2nd term in \eqref{39wes} is given by
\be
&&\hspace*{-.4in}\Bigl\langle\mathcal{O}_1( \omega_1,\textbf{p}_1)\mathcal{O}_2( \omega_2,\textbf{p}_2 )\mathcal{O}_3( \omega_3,\textbf{p}_3 )\Bigl\rangle_{G_2}\non\\[.2cm]
&=& C_{123}\int \prod_{a=1}^3du_ad^{d-1}\textbf{x}_a\; \f{e^{i(\omega_1u_1+\omega_2u_2+\omega_3u_3)}e^{i(\textbf{p}_1\cdot \textbf{x}_1+\textbf{p}_2\cdot \textbf{x}_2+\textbf{p}_3\cdot \textbf{x}_3)}}{|\textbf{x}_{12}|^{\Delta_1+\Delta_2-\Delta_3}|\textbf{x}_{23}|^{\Delta_2+\Delta_3-\Delta_1}|\textbf{x}_{31}|^{\Delta_3+\Delta_1-\Delta_2}}\non\\
&=&C_{123}(2\pi)^3\delta(\omega_1)\delta(\omega_2)\delta(\omega_3)\int \prod_{a=1}^3d^{d-1}\textbf{x}_a\; \f{e^{i(\textbf{p}_1\cdot \textbf{x}_1+\textbf{p}_2\cdot \textbf{x}_2+\textbf{p}_3\cdot \textbf{x}_3)}}{|\textbf{x}_{12}|^{\Delta_1+\Delta_2-\Delta_3}|\textbf{x}_{23}|^{\Delta_2+\Delta_3-\Delta_1}|\textbf{x}_{31}|^{\Delta_3+\Delta_1-\Delta_2}}\non\\
&=&(2\pi)^3\delta(\omega_1)\delta(\omega_2)\delta(\omega_3)C_{123}\f{(\pi)^{\f{3(d-1)}{2}}\;2^{3(d-1)-\Delta_t}}{\Gamma\Bigl(\f{\delta_1}{2}\Bigl)\Gamma\Bigl(\f{\delta_2}{2}\Bigl)\Gamma\Bigl(\f{\delta_3}{2}\Bigl)}\Gamma\Bigl(\f{d-1-\delta_1}{2}\Bigl)\Gamma\Bigl(\f{d-1-\delta_2}{2}\Bigl)\Gamma\Bigl(\f{d-1-\delta_3}{2}\Bigl)\non\\
&&(2\pi)^{d-1}\delta^{d-1}(\textbf{p}_1+\textbf{p}_2+\textbf{p}_3)\int \f{d^{d-1}\textbf{k}}{(2\pi)^{d-1}}\; \f{1}{|\textbf{k}|^{d-1-\delta_3}|\textbf{p}_2+\textbf{k}|^{d-1-\delta_1}|\textbf{p}_1-\textbf{k}|^{d-1-\delta_2}}\label{b247}
\ee
We have defined $\delta_1=\Delta_2+\Delta_3-\Delta_1, \delta_2=\Delta_1+\Delta_3-\Delta_2$ and so on. In going to the 3rd equality, we have first used the translation invariance to extract the $d-1$ dimensional momentum conserving delta function and for the factors in denominator, we have used the identity \eqref{a232} for $D=d-1$. The expression in \eqref{b247} (apart from the factor $\delta(\omega_1)\delta(\omega_2)\delta(\omega_3)$) is basically the momentum space 3-point function of 3 scalar operators in $d-1$ dimensional Euclidean CFT. The integral in \eqref{b247} can again be shown to be proportional to a triple-K integral with the final result given by 
\be
&&\hspace*{-.4in}\Bigl\langle\mathcal{O}_1( \omega_1,\textbf{p}_1)\mathcal{O}_2( \omega_2,\textbf{p}_2 )\mathcal{O}_3( \omega_3,\textbf{p}_3 )\Bigl\rangle_{G_2}\non\\[.2cm]
&=&   \f{(2\pi)^3\delta(\omega_1)\delta(\omega_2)\delta(\omega_3)(2\pi)^{d-1}\delta^{d-1}(\textbf{p}_1+\textbf{p}_2+\textbf{p}_3)C_{123}\pi^{d-1}2^{4+\f{3d-3}{2}-\Delta_t}}{\Gamma\Bigl(\f{\Delta_t-d+1}{2}\Bigl)\Gamma\Bigl(\f{\Delta_1+\Delta_2-\Delta_3}{2}\Bigl)\Gamma\Bigl(\f{\Delta_2+\Delta_3-\Delta_1}{2}\Bigl)\Gamma\Bigl(\f{\Delta_3+\Delta_1-\Delta_2}{2}\Bigl)}\non\\
&&I_{\f{d-3}{2}(\Delta_1-\f{d-1}{2},\Delta_2-\f{d-1}{2},\Delta_3-\f{d-1}{2})}(p_1,p_2,p_3)
\ee
Next, for the 3rd term in \eqref{39wes}, we have 
\be
&&\hspace*{-.4in}\Bigl\langle\mathcal{O}_1( \omega_1,\textbf{p}_1)\mathcal{O}_2( \omega_2,\textbf{p}_2 )\mathcal{O}_3( \omega_3,\textbf{p}_3 )\Bigl\rangle_{G_3}\non\\[.2cm]
&=&  C_3\int \prod_{a=1}^3du_ad^{d-1}\textbf{x}_a\; \f{e^{i(\omega_1u_1+\omega_2u_2+\omega_3u_3)}e^{i(\textbf{p}_1\cdot \textbf{x}_1+\textbf{p}_2\cdot \textbf{x}_2+\textbf{p}_3\cdot \textbf{x}_3)}}{|\textbf{x}_{12}|^{2\Delta_1}|u_{23}|^{\Delta_2+\Delta_3-\Delta_1-d+1}}\delta^{d-1}(\textbf{x}_{23})\non\\[.3cm]
&=&(2\pi)^2\delta(\omega_1)\delta(\omega_2+\omega_3)(2\pi)^{d-1}\delta^{d-1}(\textbf{p}_1+\textbf{p}_2+\textbf{p}_3)C_3 \int d^{d-1}\textbf{z} \f{e^{i\textbf{p}_1\cdot\textbf{z}}}{|\textbf{z}|^{2\Delta_1}}\int dv \f{e^{iv\omega_2}}{|v|^{\Delta_2+\Delta_3-\Delta_1-d+1}}\non\\[.3cm]
&=&(2\pi)^{2}\delta(\omega_1)\delta(\omega_2+\omega_3)(2\pi)^{d-1}\delta^{d-1}(\textbf{p}_1+\textbf{p}_2+\textbf{p}_3)C_3\pi^{\f{d}{2}}2^{2d-1-\Delta_t}\f{\Gamma\Bigl(\f{d-1-2\Delta_1}{2}\Bigl)\Gamma\Bigl(\f{d+\Delta_1-\Delta_2-\Delta_3}{2}\Bigl)}{\Gamma(\Delta_1)\Gamma\Bigl(\f{\Delta_2+\Delta_3-\Delta_1-d+1}{2}\Bigl)}\non\\
&&(p_1)^{2\Delta_1-d+1}(\omega_2)^{\Delta_2+\Delta_3-\Delta_1-d}\label{c251we}
\ee
In going to the 2nd equality, we have performed integrations over $u_1$ and $\textbf{x}_3$ and then made a change of variables $u_3=t, u_{23}=v$ and $\textbf{x}_2= \textbf{y}, \textbf{x}_{12} = \textbf{z}$. In going to the 3rd equality, we have used identity \eqref{c252} for $D=d-1$ and $D=1$. 

\vspace*{.07in}The Fourier transform of 4th and 5th terms in \eqref{39we} is exactly analogous to the previous case and can be written down by permuting the indices in \eqref{c251we}. 

\section{Direct evaluation of equation (5.118)}
\label{appene}
In this appendix we provide an alternative derivation of the integral in equation \eqref{5123ert}. The method involves decomposing the integration range into three distinct regions, identified by the phase of the integrand, and evaluating the contribution of each region separately. This alternative derivation gives an independent consistency check of the result derived in Section \ref{Time-ordered 2point}.

\vspace*{.07in}Following the cuts shown in figure \ref{Fig2}, the integral can be decomposed into three separate contributions as
\begin{eqnarray}
I(\alpha,\,|{\bf p}|)=\left[\int_{-\infty}^{-c|{\bf p}|}+\int_{-c|{\bf p}|}^{c|{\bf p}|}+\int_{c|{\bf p}|}^{\infty}\right] \frac{d\omega}{[(-\omega+c\,|{\bf p}|-i0^+)(\omega+c|{\bf p}| -i0^+)]^\alpha}\;,\quad \alpha= \frac{d}{2}-\Delta
\end{eqnarray}
The integrand between $[-c|{\bf p}|,\,c|{\bf p}|]$ is analytic with no branch cuts and can be evaluated in the standard way
\begin{eqnarray} 
\int_{-c|{\bf p}|}^{c|{\bf p}|} \frac{d\omega}{[(-\omega+c\,|{\bf p}|)(\omega+c|{\bf p}|)]^\alpha}&=&
 (2c|{\bf p}|)^{1-2\alpha}\int_0^1 dt\, t^{-\alpha}\,(1-t)^{-\alpha}\nonumber\\
&=&(2c|{\bf p}|)^{1-2\alpha}\,B(1-\alpha,\,1-\alpha)\nonumber\\
\end{eqnarray}
The two tails of the integral contribute with two distinct phase factors along the real axis
\begin{eqnarray}
I_{\rm tails}(\alpha,|{\bf p}|)&\equiv&\Big[\int_{-\infty}^{-c|{\bf p}|}+\int_{c|{\bf p}|}^{\infty}\Big] \frac{d\omega}{[(-\omega+c\,|{\bf p}|-i0^+)(\omega+c|{\bf p}| -i0^+)]^\alpha}\nonumber\\
&=&
e^{i\pi \alpha}\int_{-\infty}^{-c|{\bf p}|}\, d\omega\,|-\omega^2+c^2|{\bf p}|^2|^{-\alpha}+e^{i\pi\alpha} \int_{c|{\bf p}|}^{\infty} d\omega\,|-\omega^2+c^2|{\bf p}|^2|^{-\alpha}\nonumber\\
&=&2 e^{i\pi\alpha}  \int_{c|{\bf p}|}^{\infty} d\omega\,[\omega^2-c^2|{\bf p}|^2]^{-\alpha}\non\\
&=&2\, e^{i\pi\alpha}\,\, (2c|{\bf p}|)^{1-2\alpha}B(1-\alpha, 2\alpha-1)\label{e239}
\end{eqnarray}
where, in going to the last equality, we have made a change of variable $x = \omega - c|{\bf p}|$ and then rescaled the $x$ variable as $x = 2c|{\bf p}|\, t$. 

\vspace*{.07in}Now, using the reflection formula and the duplication formula for the Gamma functions given in equation \eqref{duplication} and the trigonometric relation $\sin\!\bigl(\pi(\alpha-\tfrac{1}{2})\bigr) = -\cos(\pi\alpha)$, we can obtain the following identity 
\be
B(1-\alpha, 2\alpha-1)&=& -\frac{1 }{2\cos\pi\alpha}\,B(1-\alpha,1-\alpha)
\ee
Using the above identity in \eqref{e239} and combining all the contributions, we find the desired result
\begin{eqnarray}
I(\alpha,\,|{\bf p}|)&=&\Big[2 e^{i\pi\alpha} -2\cos\pi\alpha\Big]\,(2c|{\bf p}|)^{1-2\alpha}\,B(1-\alpha,\,2\alpha-1)\non\\
&=& 2i\sin\pi\alpha\,(2c|{\bf p}|)^{1-2\alpha}\,B(1-\alpha,\,2\alpha-1)
\end{eqnarray}
This agrees with the result obtained in equation \eqref{4.77}.

\end{document}